\documentclass[longauth]{aa}  
\usepackage[nottoc]{tocbibind}
\usepackage{enumerate}
\usepackage{float}
\usepackage{titlesec}
\usepackage{afterpage}
\usepackage{amssymb,amsmath}
\usepackage[]{graphicx} \graphicspath{ {Fig/} }
\usepackage{caption}
\usepackage{subfig}
\usepackage[]{xcolor}
\usepackage{booktabs, multirow}
\usepackage[colorlinks,linkcolor=black,urlcolor=blue,citecolor=blue]{hyperref}
\makeatletter
\renewcommand*\aa@pageof{, page \thepage{} of \pageref*{LastPage}}
\makeatother\usepackage{natbib}

\usepackage{url}
\usepackage[utf8]{inputenc}
\usepackage{titlesec}
\usepackage{amssymb,amsmath}
\usepackage{verbatim}
\usepackage{graphicx}
\usepackage{enumitem}
\usepackage{color}
\usepackage{sidecap}
\usepackage{graphicx, caption}
\usepackage{caption}
\usepackage{cuted}
\renewcommand{\theenumi}{(\roman{enumi})}%

\usepackage{txfonts}
\begin{document}

   \title{The imprint of cosmic voids from the DESI Legacy Survey DR9 LRGs in the \textit{Planck} 2018 lensing map through \\spectroscopically calibrated mocks}

   \author{S. Sartori\inst{1}, P. Vielzeuf\inst{1}, S. Escoffier\inst{1}, M. C. Cousinou\inst{1}, A. Kovács\inst{2\and3}, J. DeRose\inst{4}, S. Ahlen\inst{5}, D. Bianchi\inst{6}, D. Brooks\inst{7},  E.~Burtin\inst{8}, T.~Claybaugh\inst{9}, A.~de la Macorra\inst{10}, J.~E.~Forero-Romero\inst{11\and12}, J. Garcia-Bellido\inst{13}, S.~Gontcho A Gontcho\inst{9}, G.~Gutierrez\inst{14}, K.~Honscheid\inst{15\and16\and17}, R.~Kehoe\inst{18}, D.~Kirkby\inst{19}, T.~Kisner\inst{9}, M.~Landriau\inst{9}, M.~E.~Levi\inst{9}, 
   A.~Meisner\inst{20}, R.~Miquel\inst{21\and22}, J.~Moustakas\inst{23}, J. A. Newman\inst{24}, N.~Palanque-Delabrouille\inst{8\and9}, I.~P\'erez-R\`afols\inst{25}, F.~Prada\inst{26}, G.~Rossi\inst{27}, E.~Sanchez\inst{28}, D.~Sprayberry\inst{20}, G.~Tarl\'{e}\inst{29}, B.~A.~Weaver\inst{20} }

   \institute{Aix Marseille Univ, CNRS/IN2P3, CPPM, Marseille, France 
   \and MTA-CSFK Lend\"ulet "Momentum" Large-Scale Structure (LSS) Research Group, 1121 Budapest, Konkoly Thege Mikl\'os \'ut 15-17, Hungary
   \and Konkoly Observatory, HUN-REN CSFK, MTA Centre of Excellence, Budapest, Konkoly Thege Mikl\'os {\'u}t 15-17. H-1121 Hungary
   \and Brookhaven National Laboratory, Physics Department, Upton, NY 11973, USA
   \and Physics Dept., Boston University, 590 Commonwealth Avenue, Boston, MA 02215, USA
   \and Dipartimento di Fisica ``Aldo Pontremoli'', Universit\`a degli Studi di Milano, Via Celoria 16, I-20133 Milano, Italy
   \and Department of Physics \& Astronomy, University College London, Gower Street, London, WC1E 6BT, UK
   \and IRFU, CEA, Universit\'{e} Paris-Saclay, F-91191 Gif-sur-Yvette, France
   \and Lawrence Berkeley National Laboratory, 1 Cyclotron Road, Berkeley, CA 94720, USA
   \and Instituto de F\'{\i}sica, Universidad Nacional Aut\'{o}noma de M\'{e}xico,  Circuito de la Investigaci\'{o}n Cient\'{\i}fica, Ciudad Universitaria, Cd. de M\'{e}xico  C.~P.~04510,  M\'{e}xico
   \and Departamento de F\'isica, Universidad de los Andes, Cra. 1 No. 18A-10, Edificio Ip, CP 111711, Bogot\'a, Colombia
   \and Observatorio Astron\'omico, Universidad de los Andes, Cra. 1 No. 18A-10, Edificio H, CP 111711 Bogot\'a, Colombia
   \and Instituto de Fisica Teorica IFT-UAM/CSIC, Universidad Autonoma de Madrid, Cantoblanco 28049 Madrid, Spain
   \and Fermi National Accelerator Laboratory, PO Box 500, Batavia, IL 60510, USA
   \and Center for Cosmology and AstroParticle Physics, The Ohio State University, 191 West Woodruff Avenue, Columbus, OH 43210, USA
   \and Department of Physics, The Ohio State University, 191 West Woodruff Avenue, Columbus, OH 43210, USA
   \and The Ohio State University, Columbus, 43210 OH, USA
   \and Department of Physics, Southern Methodist University, 3215 Daniel Avenue, Dallas, TX 75275, USA
   \and Department of Physics and Astronomy, University of California, Irvine, 92697, USA
   \and NSF NOIRLab, 950 N. Cherry Ave., Tucson, AZ 85719, USA
   \and Instituci\'{o} Catalana de Recerca i Estudis Avan\c{c}ats, Passeig de Llu\'{\i}s Companys, 23, 08010 Barcelona, Spain
   \and Institut de F\'{i}sica d'Altes Energies (IFAE), The Barcelona Institute of Science and Technology, Campus UAB, 08193 Bellaterra Barcelona, Spain
   \and Department of Physics and Astronomy, Siena College, 515 Loudon Road, Loudonville, NY 12211, USA
   \and Department of Physics \& Astronomy and Pittsburgh Particle Physics, Astrophysics, and Cosmology Center (PITT PACC), University of Pittsburgh, 3941 O’Hara Street, Pitts- burgh, PA 15260, USA
   \and Departament de F\'isica, EEBE, Universitat Polit\`ecnica de Catalunya, c/Eduard Maristany 10, 08930 Barcelona, Spain
   \and Instituto de Astrof\'{i}sica de Andaluc\'{i}a (CSIC), Glorieta de la Astronom\'{i}a, s/n, E-18008 Granada, Spain
   \and Department of Physics and Astronomy, Sejong University, Seoul, 143-747, Korea
   \and CIEMAT, Avenida Complutense 40, E-28040 Madrid, Spain
   \and University of Michigan, Ann Arbor, MI 48109, USA}


\abstract{The cross-correlation of cosmic voids with the lensing convergence ($\kappa$) map of the Cosmic Microwave Background (CMB) fluctuations provides a powerful tool to refine our understanding of the current cosmological model. However, several studies have reported a moderate tension (up to $\sim2\sigma$) between the lensing imprint of cosmic voids on the observed CMB and the $\mathrm{\Lambda}$CDM signal predicted by simulations. To address this "lensing-is-low" tension and to obtain new, precise measurements of the signal, we exploit the large DESI Legacy Survey Luminous Red Galaxy (LRG) dataset, covering approximately 19,500 $\deg^2$ of the sky and including about 10 million LRGs at $z < 1.05$. Our $\mathrm{\Lambda}$CDM template was created using the Buzzard mocks, which we specifically calibrated to match the clustering properties of the observed galaxy sample by exploiting more than one million DESI spectra. We identified our catalogs of 3D voids in the range $0.35 < z < 0.95$ and cross-correlated them through a stacking methodology, dividing the sample into bins according to the redshift and $\lambda_\mathrm{v}$ values of the voids. For the full void sample, we report a 14$\sigma$ detection of the lensing signal, with $A_\kappa = 1.016 \pm 0.054$, which increases to 17$\sigma$ when considering the \textit{void-in-void} ($A_\kappa = 0.944 \pm 0.064$) and the \textit{void-in-cloud} ($A_\kappa = 0.975 \pm 0.060$) populations individually, the highest detection significance for studies of this kind. We observe a full agreement between the observations and $\mathrm{\Lambda}$CDM predictions across all redshift bins, sky regions, and void populations considered. 
In addition to these findings, our analysis highlights the importance of accurately matching sparseness and redshift error distributions between mocks and observations, as well as the role of $\lambda_\mathrm{v}$ in enhancing the signal-to-noise ratio through void population discrimination.}


   \authorrunning{S. Sartori et al.}
   \titlerunning{The lensing CMB imprint of cosmic voids from the DESI Legacy Survey}
   \maketitle
%

\section{INTRODUCTION}
The past 25 years have radically transformed our perception and our ability to study the Cosmos. On one hand, the observation of distant Type Ia supernovae (\citealt{Riess1998}, \citealt{Schmidt1998}, \citealt{Perlmutter1999}), the observations of large-scale structures and of the anisotropies of the cosmic microwave background (CMB; \citealt{Eisenstein2005bao}, \citealt{Komatsu2011wmap}, \citealt{Bennet2013wmap}, \citealt{Planck2020}) have introduced and confirmed the model of an accelerating Universe. On the other hand, the remarkable technological advancements have enabled the development of a new generation of wide and deep-field observational surveys (\citealt{SDSS}, \citealt{thedarkenergysurveycollaboration2005dark}  as key examples of the first major surveys of this kind), complemented by cosmological simulations of unprecedented scale and resolution.

These cutting-edge observational techniques have enabled the mapping of the large-scale distribution of matter, confirming theoretical predictions \citep{Peebles1980} of a Universe composed of overdense sheets and filaments interspersed with vast underdensities, known as cosmic voids. Collectively, these intricate structures form what is widely referred to as the cosmic web.

Cosmic voids dominate the Late Universe in terms of volume. While lacking a universally agreed-upon definition, these structures can be characterized as vast, underdense regions of space that, when isolated, evolve over time by transferring matter from their central regions to the surrounding filaments and overdense walls.

Thanks to their unique environmental characteristics, voids have emerged as fundamental resources for assessing the validity of the Lambda-Cold Dark Matter ($\mathrm{\Lambda}$CDM) model and exploring scenarios of modified gravity (\citealt{Clampitt2013}, \citealt{Spolyar2013}, \citealt{Cai2015}, \citealt{Zivick2016}, \citealt{Wilson2023}, \citealt{Mauland2023}), massive neutrinos (\citealt{Massara2015}, \citealt{Banerjee2016}, \citealt{Kreisch2019}, \citealt{Schuster2019}, \citealt{Mauland2023}, \citealt{Thiele2023}, \citealt{Vielzeuf2023}), primordial non-gaussianities \citep{Chan2019} and more in general phenomena beyond the standard model of particle Physics (see \citealt{pisani2019cosmic} and \citealt{Moresco2022} for extensive reviews).

The constraining power of cosmic voids emerges from a variety of probes. Examples include the void size function (\citealt{Jennings2013}, \citealt{Pisani2015}, \citealt{Contarini2019}, \citealt{Ronconi2019}, \citealt{Correa2022a}, \citealt{Contarini2023}, \citealt{Pelliciari2023}, \citealt{Song2024}, \citealt{Verza2024}), which represents the number of voids as a function of their radius, and the void-galaxy cross-correlation function (\citealt{Hamaus2014a}, \citealt{Nadathur2019}, \citealt{Aubert2022}, \citealt{Correa2022b}, \citealt{Mauland2023}, \citealt{Radinovic2023}, \citealt{Schuster2023}, \citealt{Schuster2024}), which provides information on the density profile of voids and the distribution of mass within them. Additional and more recent probes come from the velocity profiles (\citealt{Paz2013}, \citealt{zivick2015}, \citealt{Hamaus2016}, \citealt{Massara2022}, \citealt{Wilson2023}), void auto-correlation function (\citealt{Chan2014}, \citealt{Hamaus2014b}, \citealt{Clampitt2016}) and power-spectrum \citep{Bonici2023}, weak lensing (\citealt{Melchior2014}, \citealt{Barreira2015}, \citealt{Sanchez2017}, \citealt{Baker2018}, \citealt{Hossen2022}, \citealt{boschetti2023}), modelling of dynamical and geometrical distortions (\citealt{Ryden1995}, \citealt{Lavaux2012}, \citealt{Sutter2012}, \citealt{Paz2013}, \citealt{Hamaus2014b}, \citealt{Sutter2014}, \citealt{Hamaus2015}, \citealt{Hamaus2022}, \citealt{Correa2022c}, \citealt{Contarini2024}) as well as shape and ellipticity of voids (\citealt{Lee2009}, \citealt{zivick2015}, \citealt{Rezaei2020}, \citealt{Schuster2023}).

The large-scale matter distribution in the Universe alters the CMB and introduces anisotropies therein (\citealt{ISW1967}, \citealt{SZ1970}, \citealt{Blanchard1987}, \citealt{Kashlinsky1988}, \citealt{Cole1989}, \citealt{Sasaki1989}, \citealt{Tomita1989}, \citealt{Linder1990}). 
Analogously to their overdense counterparts, cosmic voids have been observed to induce various effects on the observed CMB. These effects have a unique value as additional probes to test the cosmological model.  While the presence of gas modifies the energy of CMB photons (thermal Sunyaev-Zel'dovich effect, tSZ, \citealt{SZ1970}), the temporal variation of the gravitational potential due to the accelerated expansion of the Universe introduces temperature anisotropies that can be used to test the effects of dark energy (Integrated Sachs-Wolfe effect, ISW, \citealt{ISW1967}). For illustrative examples of the ISW effect within cosmic voids, see \citet{Nasathur2012}, \citet{Ilic2013}, \citet{Cai2014}, \citet{Kovacs2018}, and \citet{Naidoo2024}. Similarly, for examples of the tSZ effect within voids, see \citet{Alonso2018}, \citet{Li2020}, and \citet{Li2024}.

The imprint of cosmic voids on the CMB lensing signal provides a unique probe of the underlying total matter distribution. Unlike clusters and filaments, cosmic voids induce a de-magnification effect on the CMB, resulting from the deflection of photons as they cross these underdense regions. This effect is observed as a negative signal in the CMB convergence ($\kappa$) maps.
Despite the complexity of obtaining a measurement of the lensing signal from an individual void due to the amplitude of noise in CMB lensing maps, which is dominant at typical scales of these cosmic structures, numerous stacked measurements have been conducted in recent years, reporting significance values ranging between $\sim 3\sigma$ and $\sim 13\sigma$ (\citealt{Cai2014}, \citealt{Raghunathan2020}, \citealt{Vielzeuf2021}, \citealt{Hang2021}, \citealt{Kovacs2022}, \citealt{Camacho-Ciurana2023}, \citealt{Demirbozan2024}), highlighting the effectiveness of stacking techniques in isolating the CMB lensing signal induced by voids from the background noise.

However, several recent studies have reported moderate tensions, up to $\sim 3\sigma$, between the observed signal amplitudes and predictions from $\mathrm{\Lambda}$CDM simulations (\citealt{Vielzeuf2021}, \citealt{Hang2021}, \citealt{Kovacs2022}, \citealt{Camacho-Ciurana2023}). These discrepancies depend on factors such as the void identification strategies employed, the degree of smoothing applied to the CMB lensing maps, and the specific sub-populations of voids considered. In light of these tensions, and furthermore considering recent analyses of the weak lensing of the DESI LRG sample \citep{chen2024}, which have addressed some of the existing lensing tensions through an innovative treatment of systematics, it is crucial to thoroughly understand the nature of these disagreements with the $\mathrm{\Lambda}$CDM predictions for properly evaluating the systematics associated with voids. This may, in turn, provide further insight into the discrepancies observed within the $\mathrm{\Lambda}$CDM model.

The aim of this paper is to exploit the large catalog of voids identified in the photometric Luminous Red Galaxies (LRG) population of the Dark Energy Spectroscopic Instrument (DESI, \citealt{DESI1}, \citealt{DESI2}) Legacy Imaging Surveys \citep{Legacy} to investigate the current tensions in the literature, measuring the cross-correlation with the lensing of the CMB from \textit{Planck} 2018 and comparing the amplitude of the signal with those obtained from the Buzzard mocks \citep{buzzard}, designed and adapted to match our observations by analyzing and correcting a number of possible systematics. The paper is organized as follows. In Section \ref{sec2}, we introduce our observed and simulated datasets. Section \ref{sec3} describes how our mocks are adapted and validated and, moreover, provides a description of our stacking methodology and error analysis. Finally, we present and discuss the main observational results of this paper in Section \ref{sec4}, followed by Section \ref{sec5}, a summary of our conclusions. 

\section{DATASETS}
\label{sec2}
\subsection{DESI Legacy Surveys DR9 Luminous Red Galaxies}
We identify our cosmic voids sample exploiting the LRG population extracted from the ninth Data Release (DR9; \citealt{SchlegelDR9}) of the DESI Legacy Imaging Surveys (see \citealt{Zhou2023} for further details on the LRG catalog and on the selection process). 

\begin{figure*}[h]
\centering
  \includegraphics[scale=0.33]{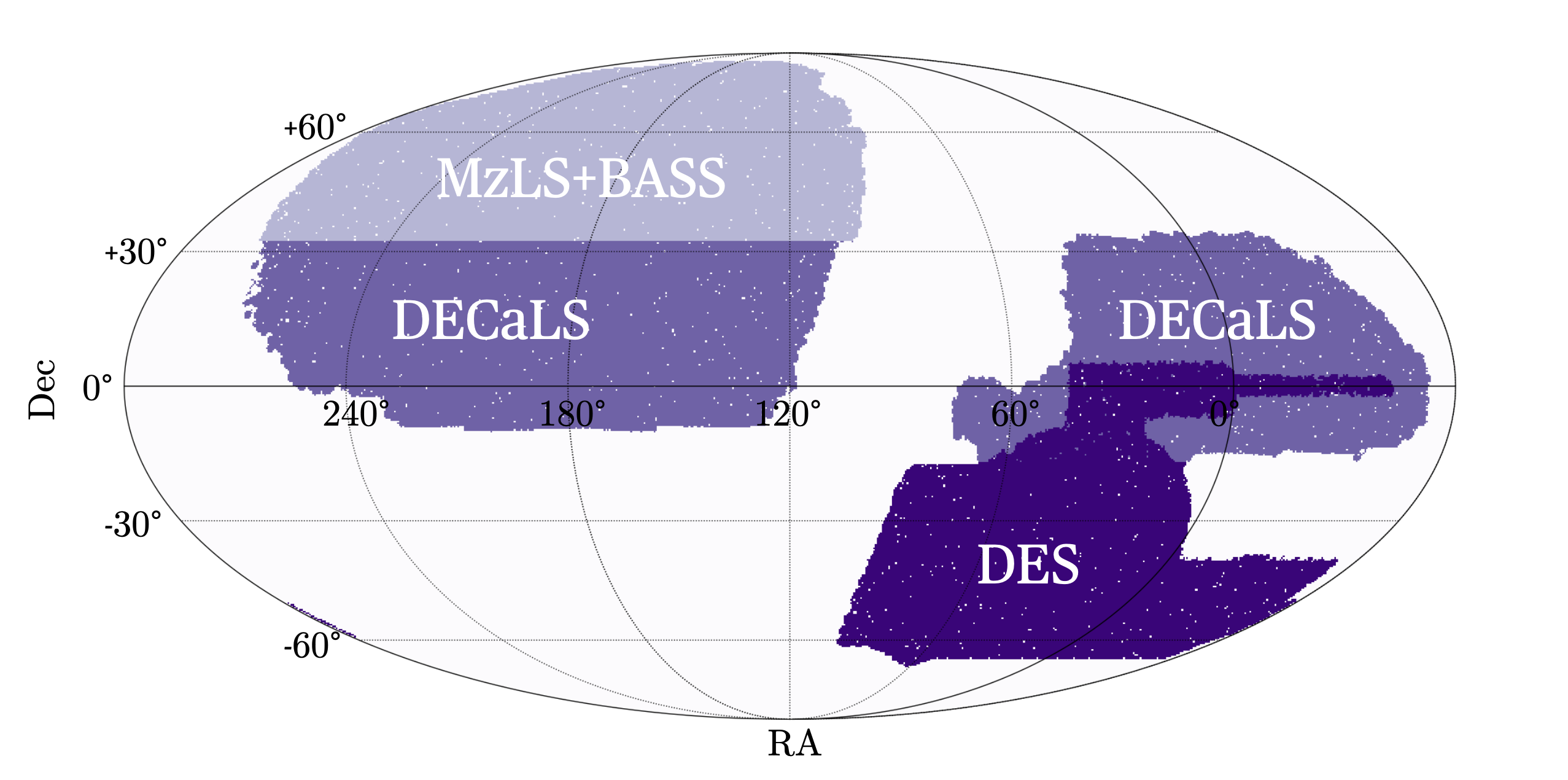}
  \caption{The different footprints of the MzLS+BASS, DECaLS, and DES surveys composing the DESI Legacy Survey, covering a total observed sky area of $19,573\,\deg^2$.}
  \label{fig:LegacyFootprint}
\end{figure*}

The Legacy Surveys are motivated by the need of a target selection for the DESI Survey and consist in a mosaic of three different observative projects, as shown in Fig. \ref{fig:LegacyFootprint}. The Northern hemisphere is mapped by a combination of different surveys: the sky area with Dec $\ge+32.735^\circ$ (J2000 coordinates) is observed by the Mayall z-band Legacy Survey (MzLS; \citealt{MzLS}), using the MOSAIC-3 camera at the prime focus of the 4-meter Mayall telescope at Kitt Peak National Observatory, and the Beijing–Arizona Sky Survey (BASS; \citealt{BASS}), that exploits the 90Prime camera at the prime focus of the Bok 2.3-m telescope. The Dark Energy Camera (DECam; \citealt{Flaugher2015}) of the Blanco 4m telescope, located at the Cerro Tololo Inter-American Observatory, provides the observation of the remaining two-third of the survey footprint. In particular, the Dark Energy Camera Legacy Survey (DECaLS; \citealt{DECALS}) completes the observations of the North Galactic Cap observing the region at Dec $\le+32.375^\circ$ and provides the observation of the equatorial regions of the South Galactic Cap, while the remaining part of the Southern hemisphere is integrated by the Dark Energy Survey (DES; \citealt{DES}) who previously exploited the same instrument to map around $5,000\, \deg^2$ of the South Hemisphere sky.
The Legacy Survey DR9, moreover, includes fluxes from the all-sky
Wide-Field Infrared Survey Explorer (WISE, \citealt{WISE}) at the locations of Legacy Surveys optical sources.
The exploited LRG sample, selected using g, r, z, and W1 photometry from the DESI Legacy Imaging Surveys, is highly robust against imaging systematics. With a comoving number density of $5\times10^{-4}\, h^3 \mathrm{Mpc}^{-3}$ in $0.4<z<0.8$, this sample proved to have a significantly higher density than previous LRG surveys, such as SDSS Legacy survey \citep{SDSS1999}, BOSS \citep{BOSS}, and eBOSS \citep{eBOSS2016}.
The total observed sky area of $19,573 \, \deg^2$, in combination with the high comoving number density, allows us to exploit a total of 10,402,494 LRGs in the redshift range $0.30<z_{\mathrm{ph}}<1.05$.

The selected sample is integrated with 1,005,750 spectroscopic redshifts from DESI, in particular from the first phase of Survey Validation (SV1), the one-percent survey (SV3; covering 1\% of  the final area) and the first two months of the Main Survey. 

\subsection{CMB convergence map from \textit{Planck} mission DR2018}
The convergence $\kappa$, our observable, is defined as the ratio
\begin{equation}
    \kappa(\theta) = \frac{\Sigma(\theta)}{\Sigma_\mathrm{cr}} \, ,
    \label{convergence}
\end{equation}
where $\Sigma$ is the projected density along the line of sight $\theta$. The critical projected density $\Sigma_\mathrm{cr}$, for the specific case of a lens located between the observer and the last scattering surface of the CMB, is defined as
\begin{equation}
    \Sigma_{cr} = \frac{c^2}{4\pi G} \frac{r_\mathrm{cmb}}{(r_\mathrm{cmb}-r)r} \, ,
    \label{sigmaCrit}
\end{equation}
assuming $r_{cmb}$ the distance between the observer and the CMB and $r$ as the distance between the observer and the lens.

Assuming the classical Poisson Equation for the gravitational potential $\Phi(r,\theta)$
\begin{equation}
    \nabla^2\Phi(r,\theta) = 4\pi G\rho(r,\theta) \, ,
\end{equation}
\begin{strip}
\centering
  \includegraphics[scale=0.33]{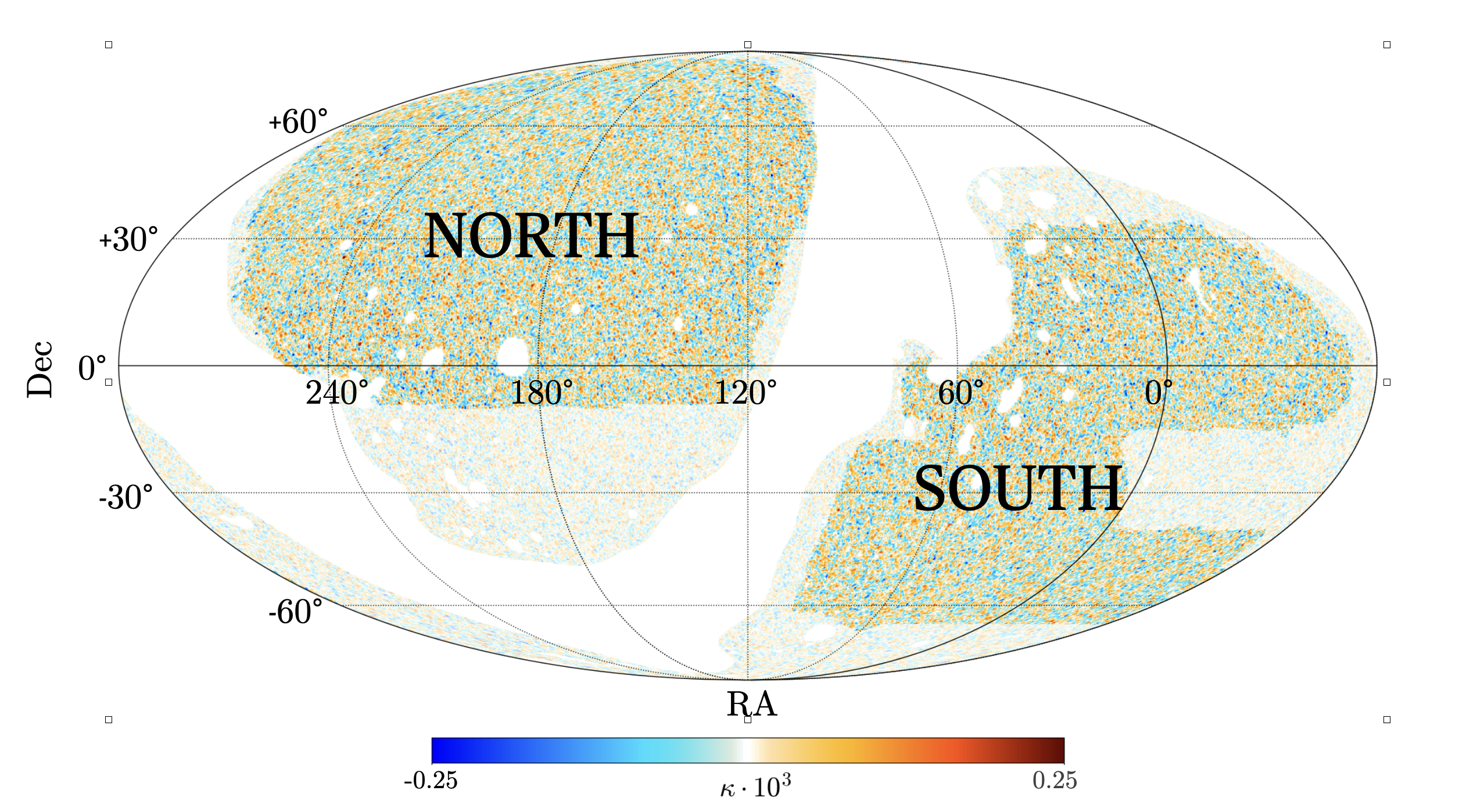}
  \captionof{figure}{\textit{Planck} CMB lensing map, smoothed with a Gaussian filter of FWHM=$0.5^\circ$. The DESI Legacy Survey footprint is overlaid, indicating the region where voids are identified and cross-correlations are performed. The footprint is divided into North and South regions, corresponding to the North and South Galactic Caps, respectively.}
  \label{fig:planck2018}
\end{strip}
it is straightforward to obtain a formulation for $\Sigma$:
\begin{equation}
    \Sigma(\theta) = \frac{1}{4\pi G}\int_{0}^{r_\mathrm{cmb}} \nabla\Phi(r,\theta) dr \, .
    \label{sigma}
\end{equation}

Replacing Eqs. \ref{sigmaCrit} and \ref{sigma} in Eq. \ref{convergence}, 
we find that 
\begin{equation}
    \kappa(\theta) = \frac{1}{c^2}\int_{0}^{r_\mathrm{cmb}} \frac{(r_\mathrm{cmb}-r)r}{r_\mathrm{cmb}} \nabla^2\Phi(r,\theta) dr \, .
\end{equation}

Now, knowing that the cosmological formulation of the Poisson equation for the gravitational potential can be written as
\begin{equation}
    \nabla^2\Phi(r,\theta) = \frac{3}{2}\frac{H_0^2\Omega_\mathrm{m}}{a} \delta(r,\theta) \, ,
\end{equation}
where $\delta(r,\theta)$ is the local density contrast.
We can immediately recover the equation for the convergence ($\kappa(\theta)$), estimated from the matter density field $\delta(r,\theta)$, in the Born approximation, with the Hubble constant $H_0$ and matter density parameter $\Omega_\mathrm{m}$ made explicit.: 
\begin{equation}
    \kappa(\theta) = \frac{3}{2}\frac{H_0^2\Omega_\mathrm{m}}{c^2} \int_{0}^{r_\mathrm{cmb}} \frac{(r_\mathrm{cmb}-r)r}{r_\mathrm{cmb}}\frac{\delta(r,\theta)}{a(t)} dr \, .
\end{equation}

For our analysis, we exploit the publicly available reconstructed CMB lensing convergence ($\kappa$) map provided by the \textit{Planck} collaboration (data release 2018; see \citealt{PlanckCMB2020}). The map is reconstructed using a minimum-variance (MV) quadratic estimator \citep{HuOkamoto2002}, which combines CMB temperature and polarization data, and is provided in the form of harmonic coefficients $\kappa_{lm}$ up to $l_\mathrm{max},m_\mathrm{max}=2048$.

Despite the potential for higher resolution, we generated a \texttt{healpy} map with ${N_\mathrm{side}}=512$ and $l_\mathrm{max},m_\mathrm{max}=1536$. The choice of ${N_\mathrm{side}}$ ensures adequate resolution for the angular dimensions of the structures analyzed, while the $l_\mathrm{max},m_\mathrm{max}$ values are selected to maximize the precision in generating the random CMB maps used for error estimation (see Sec. \ref{error} for further details).

A Gaussian smoothing is finally applied to the map to suppress the noise at small scales and improve the signal-to-noise ratio. Unlike some previous works (\citealt{Vielzeuf2021}, \citealt{Kovacs2022}, \citealt{Camacho-Ciurana2023}) that employed a Gaussian smoothing with $\mathrm{FWHM}=1.0^\circ$, we opted for a more conservative filter with $\mathrm{FWHM}=0.5^\circ$. This approach ensures the maximization of the measured signal while preserving the integrity of the smaller voids imprint on the CMB, whose angular radii often measure less than $1.0^\circ$.

The CMB lensing map, along with the extent of the North and South regions, is shown in Fig. \ref{fig:planck2018}.

\subsection{BUZZARD Mocks: simulated galaxies and CMB convergence map}
The Buzzard mocks are a suite of quarter sky galaxy catalogs constructed from $N$-body lightcone simulations using  \texttt{Addgals} \citep{Wechsler2021}, an algorithm designed to reproduce the clustering properties of subhalo abundance matching models \citep{DeRose2021} in low-resolution lightcone simulations. The underlying $N$-body simulations were run with \texttt{L-Gadget2}, a streamlined version of \texttt{Gadget2} \citep{Springel05} designed to run large $N$-body simulations and initialized with second-order Lagrangian perturbation theory at $z=49$ using \texttt{2LPTIC} \citep{Crocce2005}, with a linear power spectrum produced using CAMB \citep{Lewis2002} assuming a $\Lambda$CDM cosmology with $\Omega_m=0.286$, $\Omega_b=0.046$, $n_s=0.96$, $h=0.7$, $\sigma_8=0.82$, $\Omega_r=0$, and $\Omega_{\nu}=0$. Each lightcone simulates a quarter of the sky to $z=2.34$ and is constructed from three different simulation boxes, with $L_{\rm box}=\{1050, 2600, 4000\}\, h^{-1} \rm Mpc$ and $1400^3$, $2048^3$ and $2048^3$ particles respectively for $z=[0.0,0.34),[0.34,0.9), [0.9,2.34)$. Each realization assumes the same cosmological parameters, but varies the white noise field used to initialize the simulations.

Galaxy catalogs are generated using \texttt{Addgals}, which assigns each galaxy to a particle in the lightcone simulations, embuing them with positions, velocities, and spectral energy distributions which can be integrated to provide broadband photometry. DESI-like LRG catalogs are then selected from these galaxy catalogs using slightly modified versions of the color cuts applied to the DESI Legacy Survey data \citep{DeRose2024}:
\begin{align}
\centering
    & z_\mathrm{fiber} < 21.60 \label{eq:lrg_cuts} \nonumber \\
    & z-W1 > 0.8\times(r-z) - 0.6 \nonumber\\
    & (g-W1>2.9) \ |\  (r-W1>1.8) \\
   \begin{split}
    & ((r-W1 > 1.8\times(W1-17.14) + c_1) \ \& \\
    & (r-W1 > W1-16.33 )) \ | \ (r-W1>3.3 + c_2) \nonumber
    \end{split}
\end{align}
\noindent where $c_1=0.4$ and $c_2=0.275$ are the differences in color cuts from the DESI targeting cuts, adjusted to roughly match the angular target density of the DESI LRG sample.

Photometric redshifts are then generated by fitting a model for $\sigma(z_{\rm phot}|z_{\rm spec})$ to the DESI Legacy Survey data, and using this to apply Gaussian errors to the true redshifts of each simulated LRG. Finally, CMB lensing maps are produced by applying the Born approximation to $50\, h^{-1}\rm Mpc$ thick density shells measured from the lightcone simulations.

This work exploit 4 realizations of the Buzzard mocks.

\section{METHODS}

\label{sec3}
\subsection{Mocks validation and calibration}
\label{MockVal}
To evaluate the lensing imprint on the CMB of our cosmic void sample and any potential deviations from the $\mathrm{\Lambda}$CDM predictions, our goal is to compare through a template-fitting process the observed cross-correlation signal with a $\mathrm{\Lambda}$CDM template derived from mocks. The lensing imprint we measure is strictly dependent on our void population, which, in turn, is inherently linked to the characteristics of the galaxy sample. Consequently, it is crucial that the simulated galaxy populations replicate the clustering properties of the observed sample.
To address potential systematic errors arising from discrepancies between the simulated and observed LRG datasets, we evaluated two key factors affecting void identification: \textit{photometric error} and \textit{sparseness}. Comparing two equally biased galaxy populations with identical photometric properties and mass distributions but differing in either sparseness or redshift error distributions can lead to markedly different void populations. Identical galaxy samples with differing photometric redshift ($z_\mathrm{phot}$) errors result in voids with varying degrees of contamination from galaxies whose redshifts have been inaccurately determined. This contamination particularly affects the estimation of the inner density contrast, thereby impacting both the identification and size of the voids. Conversely, different levels of sparseness significantly influence the resolution of the void-finding method, thereby affecting the void size function. It is evident that distinct void populations could exhibit different lensing properties.

\subsubsection{Photometric Redshift correction}
\label{PhotCorrection}
As a first step of our analyses, we examined the distribution of the photometric redshift errors. We prioritized this aspect because modifications to the photometric redshift distribution in the mocks would subsequently alter the sparseness. We utilized both photometric and spectroscopic redshift ($z_\mathrm{spec}$) data, available for the mock populations and approximately 10\% of the observed galaxies, to estimate photometric errors, with spectroscopic redshifts considered as the true redshift values. The matching between the spectroscopic and the photometric redshifts of our sources give us the opportunity to correct the photometric redshift associated at the galaxies of the mocks, estimating a new and more precise photometric redshift distribution. To account for potential variations in the error distribution as a function of true redshift, we divided our sample into 75 bins with $\Delta z = 0.01$. For each redshift bin, we calculated the distribution of the redshift error $z_\mathrm{err} = z_\mathrm{spec} - z_\mathrm{phot}$, and then determined the cumulative distribution function (CDF), normalized between 0 and 1. This CDF provides the probability that $z_\mathrm{err,i} < Z$, where $Z$ represents a specific value within the distribution.

\begin{figure}[h!]
    \centering
    \includegraphics[scale=0.46]{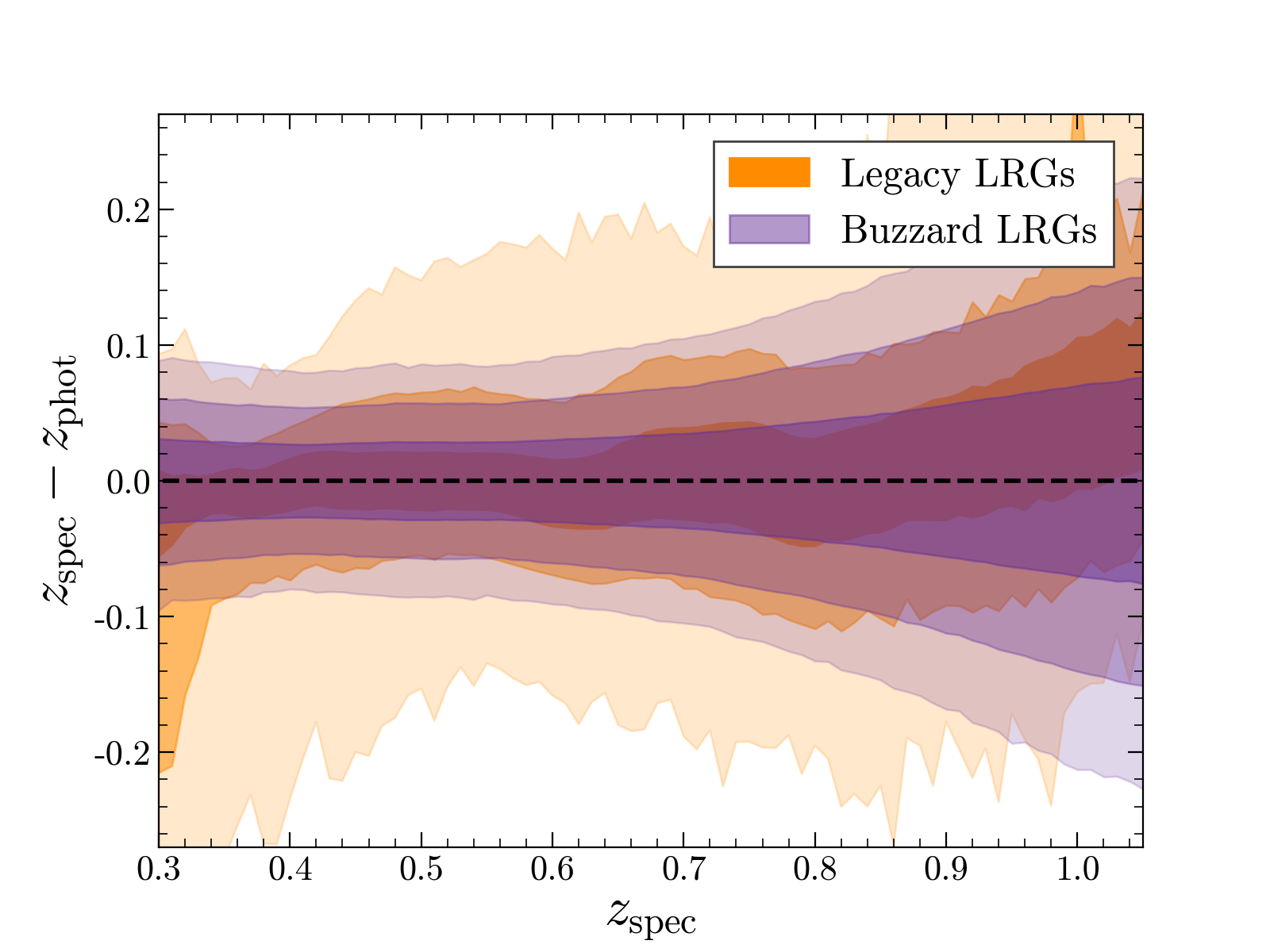}
    \caption{Redshift error distributions for the observed and one simulated LRG datasets, shown in orange and purple, respectively, at the 68.27, 95.45, and 99.73 percentiles. Darker shades represent smaller percentile ranges. The two distributions exhibit distinct trends and deviations from zero. These differences must be corrected to ensure proper matching between observations and mocks.}
    \label{zerrDist}
\end{figure}
At this point, we extract a set of random numbers from a uniform distribution between 0 and 1, matching the size of our LRG catalog for each mock realization. Each mock galaxy is then assigned a redshift error $z_\mathrm{err} = \mathrm{CDF}^{-1}(\mathrm{rand})$. The new photometric redshift for the mock galaxies is then straightforwardly recovered as $z_\mathrm{phot,i} = z_\mathrm{spec,i} - z_\mathrm{err,i}$.

Figure \ref{zerrDist} shows the redshift error distributions for the observed dataset and one Buzzard realization at the 68.27, 95.45, and 99.73 percentiles. It is evident that not only the two distributions exhibit significantly different standard deviations, but they also diverge substantially in their overall trends.

\begin{figure}[h!]
    \centering
    \includegraphics[scale=0.46]{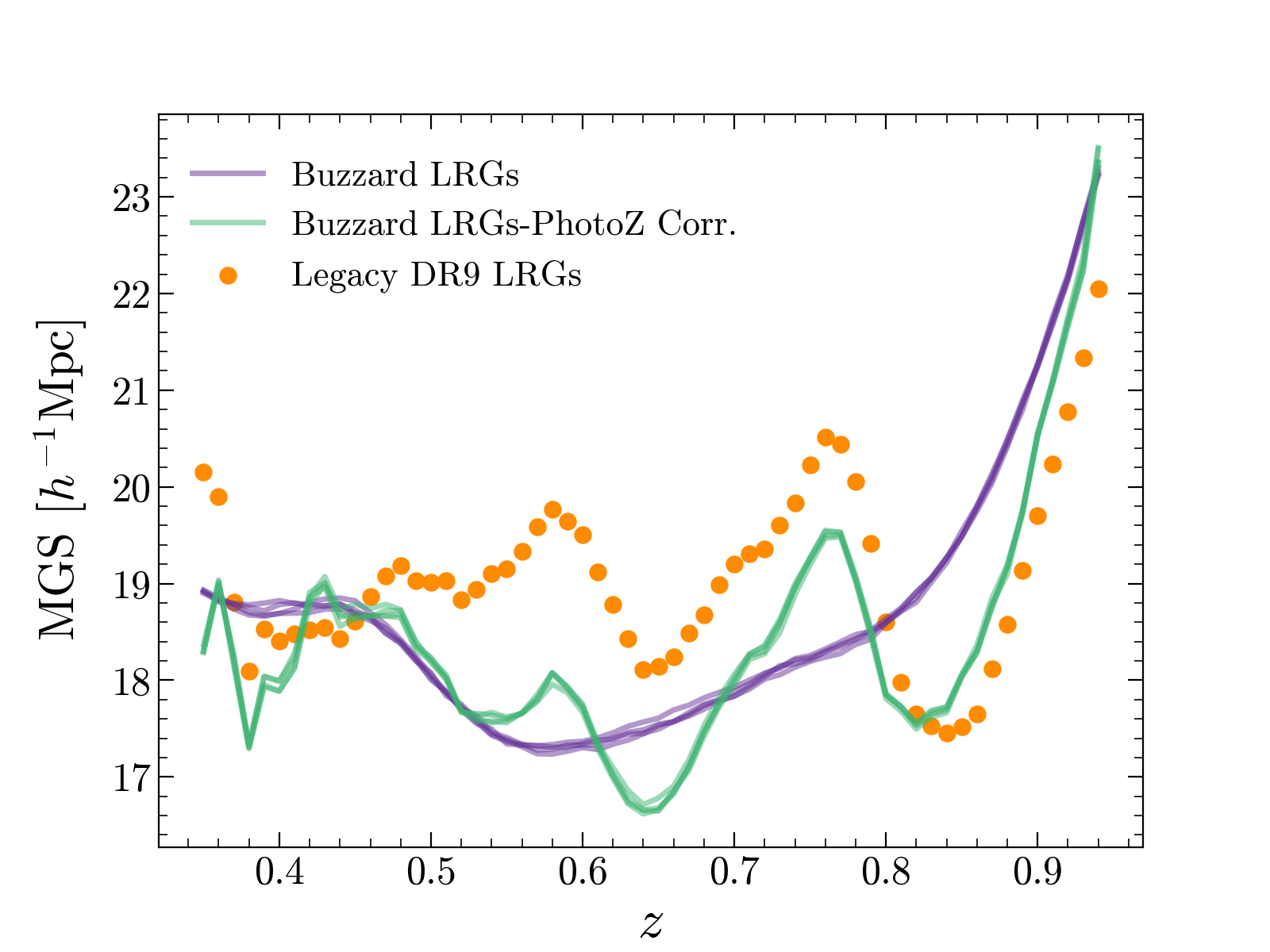}
    \caption{Mean Galaxy Separation (MGS) redshift evolution observed in the Legacy Survey (orange dots) compared with the MGSs of the four Buzzard mock realizations before (purple lines) and after (green lines) the photometric redshift correction. The shape change after correction, now matching the MGS trend of the Legacy Survey galaxy sample, is a strong indicator of the accuracy of the redshift calibration procedure.}
    \label{sparseness}
\end{figure}

Figure \ref{sparseness} shows the Mean Galaxy Separation (MGS), indicator of the sparseness of the sample, defined as:
\begin{equation}
    \mathrm{MGS} = n^{-1/3} \, ,
\end{equation}
where $n$ is the galaxy number density, of the observed and simulated sample. Moreover, it shows how the MGS changes after the $z_\mathrm{phot}$ correction of the mocks. The post-correction shape matching of the two MGS distributions is a strong indication of the success of the procedure. 

\subsubsection{Sparseness matching}
\label{sparsenessMatch}
As a second step in the mock validation and adaptation process, we corrected the sparseness of the simulated datasets to align with the observations. To account for the slight differences in sparseness between the different regions of the survey, we subdivided the observed dataset into two regions, as shown in Fig. \ref{fig:planck2018}, and handled the North and South Galactic Caps separately.

Similar to the approach used for photometric redshift correction, we divided our samples into bins with $\Delta z = 0.01$. Within each bin, we randomly sampled our mock galaxy populations to match the number of objects in both the simulated and observed catalogs. This procedure is feasible only for redshift ranges where our mock datasets are denser than the observed ones. However, this condition is not consistently met for $0.35<z<0.50$ and $z>0.80$ (see Fig. \ref{sparseness}). Despite the inability to fully match the level of sparseness in these redshift ranges, the density discrepancies between the different populations remain below $5\%$. This minimal discrepancy ensures that the impact of varying sparseness on void identification systematics is almost negligible.

We applied this procedure to our four mock realizations, both for the North and South observed catalogs, resulting in eight corrected simulated datasets (four for the North Galactic Cap and four for the South Galactic Cap). 

\begin{figure}[h!]
    \centering
    \includegraphics[scale=0.44]{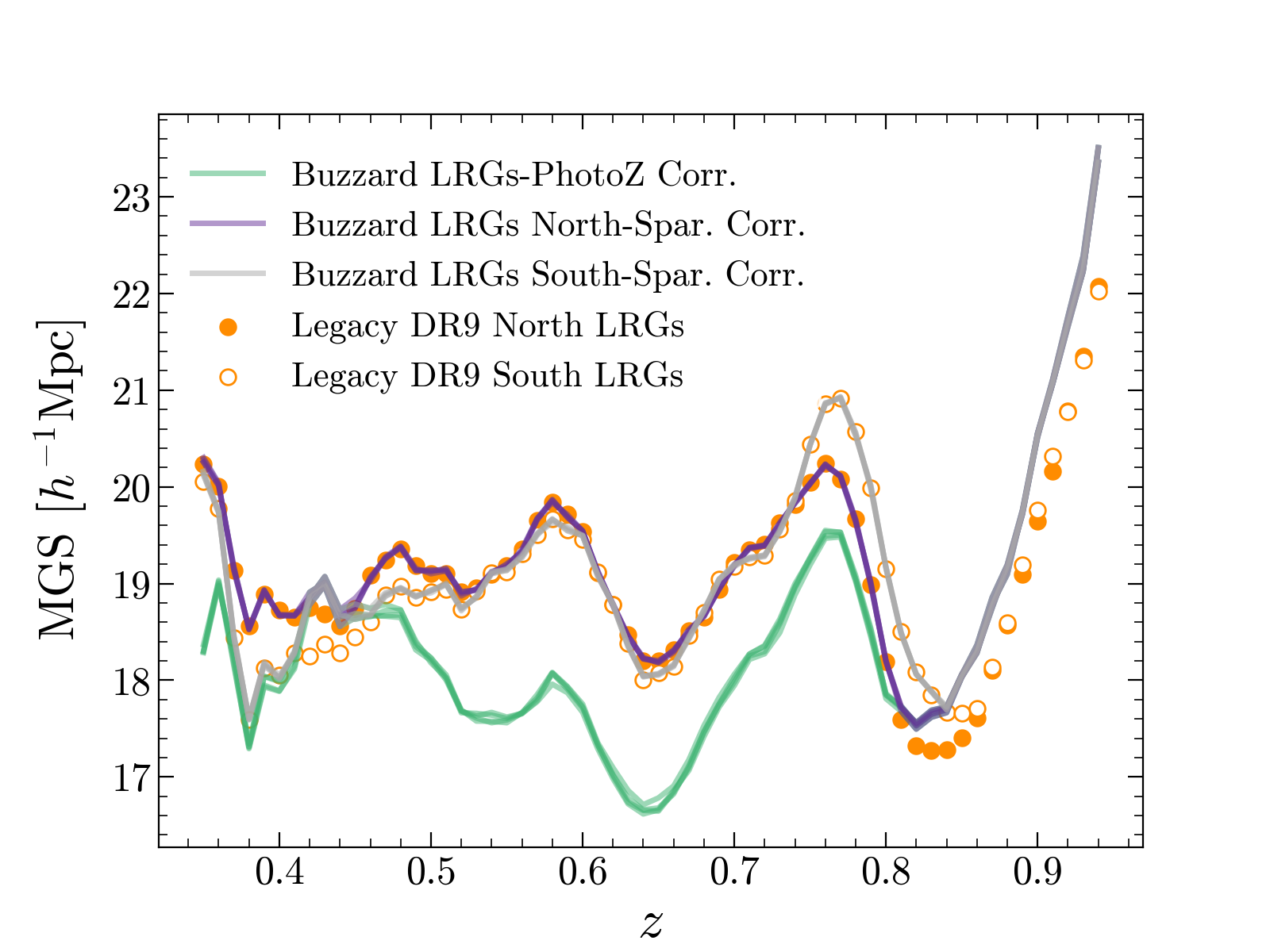}
    \caption{Mean Galaxy Separation (MGS) redshift evolution observed in the Legacy Survey, measured in the North (Orange dots) and South (white dots) regions. The slight difference in sparseness between the two regions is due to random fluctuations and minor observational differences between the surveys. The observed MGS values are compared with those from the four redshift calibrated Buzzard mock realizations (purple lines) and then with the MGS values after the sparseness correction, performed independently to match the North (green lines) and South (grey lines) regions. We observed an almost complete match between the observations and simulations, with only minor differences at $z\sim0.4$ and $z>0.8$, where the mocks appeared more sparse than the observations, making subsampling infeasible.}
    \label{sparsenessCorr}
\end{figure}

Figure \ref{sparsenessCorr} shows the final effect of the combination of the two correction procedures, for both the North and South datasets, as an almost perfect MGS matching in the whole redshift range considered by this cosmological analysis. 
\subsection{Void Finding}
\label{voidFinding}
Over recent decades, the absence of a precise theoretical definition of cosmic voids has led to the development of various void-finding algorithms. According to \citet{LavauxWandelt2010}, these methods can be classified into three main categories: density-based, geometry-based, and dynamics-based, depending on how the algorithms exploit the mass tracer field (see \citealt{Colberg2008} for an overview on the void finding methods and their systematics). Furthermore, cosmic voids can be identified in either two or three dimensions (2D or 3D). Although 2D void identification has been slightly more effective in maximizing the signal-to-noise ratio (S/N) for lensing analysis (see, e.g., \citealt{Cautun2018}), we opted for a 3D void catalog in our study. This choice allows us to address tensions observed in previous works, particularly those involving 3D voids (see, i.e, \citealt{Vielzeuf2021} and \citealt{Camacho-Ciurana2023}), and helps mitigate issues related to the intrinsic alignment of voids that can produce spurious lensing signals.
\begin{figure*}
\centering
  \includegraphics[width=1.\linewidth]{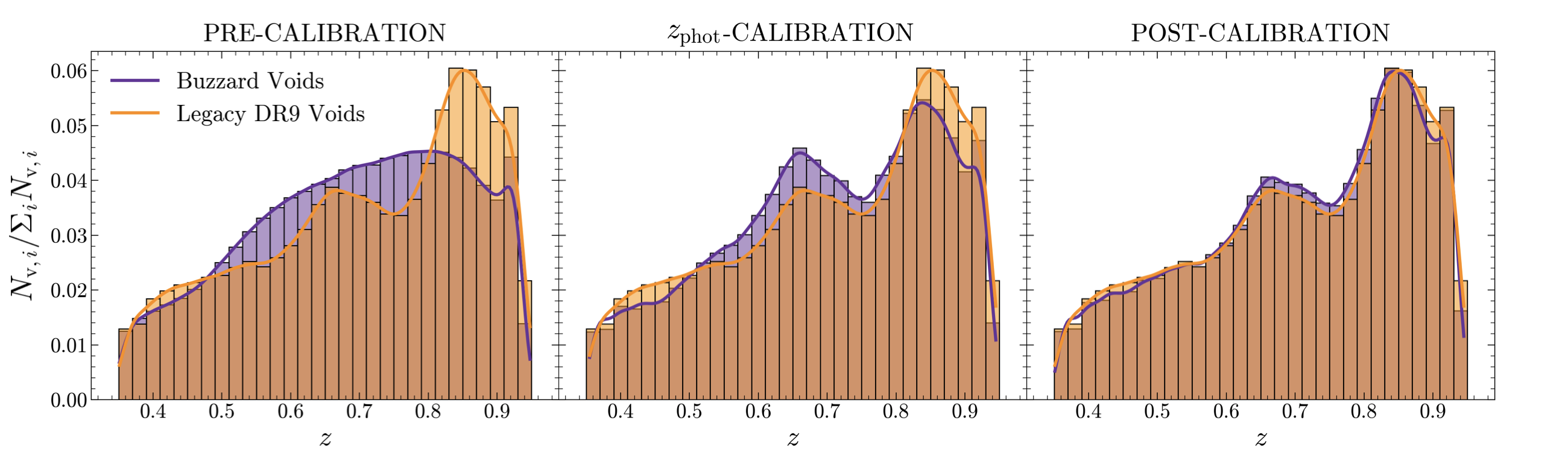}
  \captionof{figure}{Probability function for the number of voids as a function of redshift for the full-sky sample, illustrating the effects of mock calibration. \textit{Left}: Void distributions prior to mock calibration. \textit{Center}: Void distributions following the photometric redshift calibration of the mock galaxies. \textit{Right}: Void distributions following both photometric redshift and sparseness calibrations.}
  \label{fig:voidDist}
\end{figure*}
For identifying our 3D voids, we used the \texttt{REVOLVER}\footnote{\url{https://github.com/seshnadathur/Revolver}} (REal-space VOid Locations from surVEy Reconstruction) void-finding code \citep{Nadathur2019b}, based on a modified version of the \texttt{ZOBOV} algorithm \citep{zobov2008}. \texttt{ZOBOV} identifies cosmic voids in the minima of the reconstructed density field, a reconstruction achieved through a Voronoi tessellation. The identification process is fully described in \citet{Nadathur2019b}, but can be summarized as follows:
\begin{enumerate}
  \item The three-dimensional space covered by the tracer catalog is subdivided through a Voronoi tessellation. This process assigns a unique Voronoi cell to each tracer particle, defining the region that is closer to it than to any other particle. The density at any point within each cell is then determined by taking the inverse of the cell volume. The effect of the mask is taken in account overflowing forbidden regions with mock tracers, simulating overdense section of space. 
  \item Density minima in the Voronoi density field are identified. A Voronoi cell is defined as a density minimum when its volume is bigger than the volume of its neighbouring cells. These minima constitute the initial seeds of the voids. 
  \item From each density minimum, the algorithm merges adjacent cells with increasing density until no higher adjacent density cells are found. These merged regions, or basins, are referred to as zones, indicating local depressions in the density field.
  \item A watershed transform is applied to merge zones into larger voids. To preserve the underdense environment of voids, a merging condition is applied: the ridge between any two zones must be less than 20 percent of the average tracer density. Finally, a hierarchy of voids and subvoids is created. A full discussion of the watershed methods is provided in \citet{platen2007}.
  \item The center of each void is assigned as the circumcenter of the positions of the lowest-density galaxy in the void and its three lowest-density mutually adjacent neighbours. This is equivalent to defining the centre of the largest empty sphere that can be inscribed in the void.
  \item Finally, the void radii are assigned. From total volume of each void, calculated as the sum of the volumes of its constituent Voronoi cells $V_\mathrm{v} = \sum_i V_i$, an effective void radius $R_\mathrm{v}$ is assigned. $R_\mathrm{v}$ is simply the radius of an equivalent sphere of the volume $V_\mathrm{v}$,
  \begin{equation}
      R_\mathrm{v} = \left(\frac{3}{4\pi} V_\mathrm{v}\right)^{1/3} .
  \end{equation}
  
The code provides, moreover, a set of properties of the voids, as the central density contrast $\delta_\mathrm{v}$, the average density contrast $\overline{\delta}_\mathrm{v} = \frac{1}{V}\int_V\delta_\mathrm{v} \mathrm{d}^3\Vec{x}$, or the parameter $\lambda_\mathrm{v}$, a useful proxy of the gravitational potential in the center of the voids \citep{nadathur2017}, defined as
\begin{equation}
    \lambda_\mathrm{v} \equiv \overline{\delta}_\mathrm{v} \left( \frac{R_\mathrm{v}}{1\, h^{-1}\mathrm{Mpc}}\right)^{1.2} \, .
\end{equation}
\end{enumerate}
The parameter $\lambda_\mathrm{v}$ plays a fundamental role in distinguishing void populations with varying internal potential values and, consequently, different lensing properties. Negative $\lambda_\mathrm{v}$ voids are generally large structures that evolve within underdense regions, often identifiable as \textit{void-in-voids}. In contrast, positive $\lambda_\mathrm{v}$ voids are typically smaller structures evolving within overdense regions, commonly classified as \textit{void-in-clouds} (for an extensive discussion about \textit{void-in-voids} and \textit{void-in-clouds} formation and evolution see \citealt{Sheth2004}). The lensing imprint of these two void populations is expected to differ significantly: the convergence profile of \textit{void-in-voids} is anticipated to be strongly negative, approaching zero only at large radii. By comparison, the convergence profile of \textit{void-in-clouds} should turn positive already in the outer regions of the structures, reflecting the intrinsic overdensity of the surrounding environments. To disentangle the combined lensing signal from these two populations, the discrimination based on $\lambda_\mathrm{v}$ becomes crucial.
We further introduce the measurement of the lensing imprint for a third void population, with $\lambda_\mathrm{v}$ values around zero ($\lambda_\mathrm{v} \in (-5,5]$). These voids are less likely to be part of larger structures; isolating them from the other two populations enables a more precise separation between the \textit{void-in-voids} and \textit{void-in-clouds} samples. This distinction also provides an additional independent measurement of the lensing signal, enhancing the robustness of our analysis.
\begin{figure}
    \centering
    \label{lambdaDist}
    \includegraphics[width=1.\linewidth]{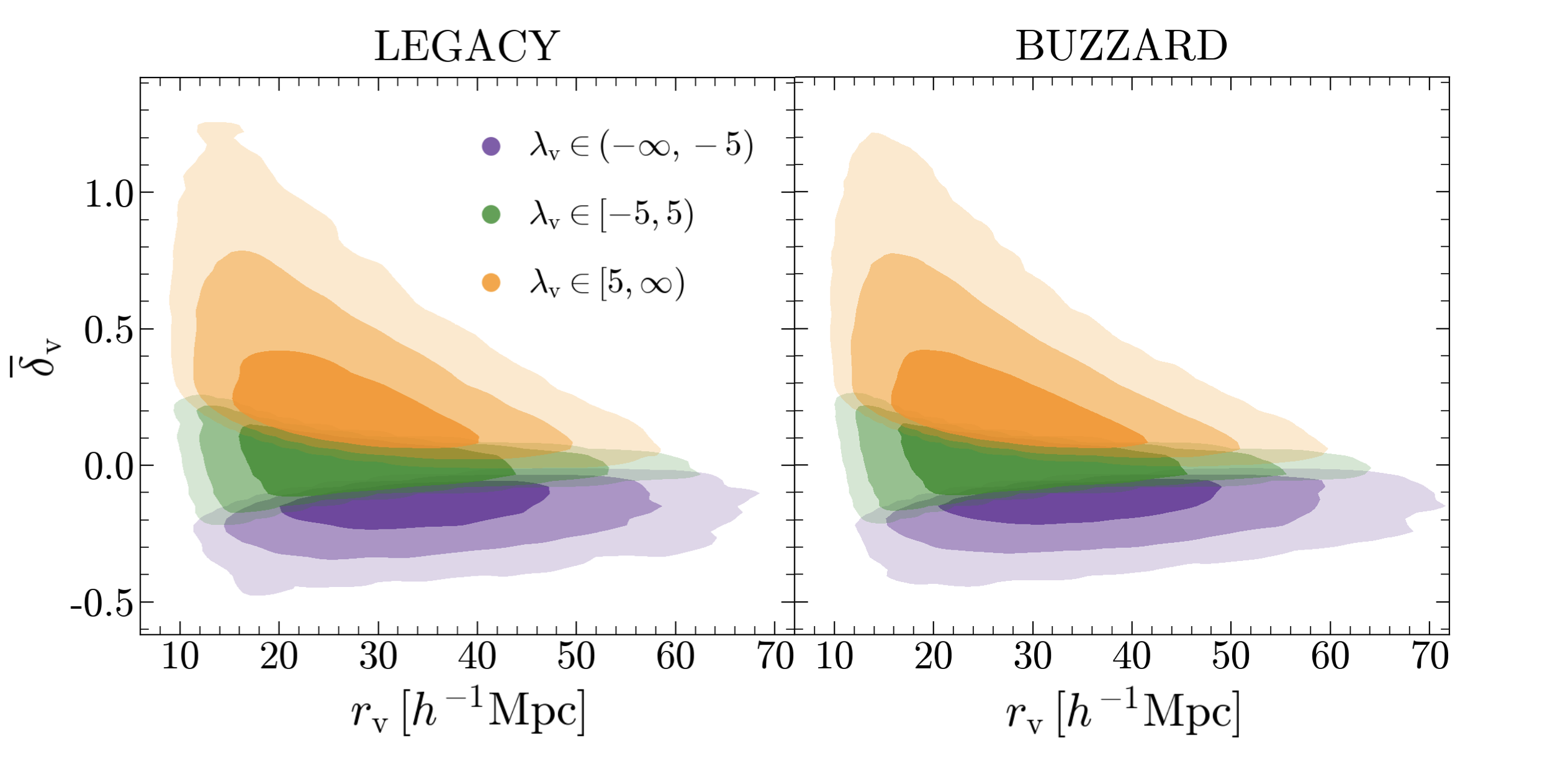}
    \captionof{figure}{Contours in the $r_\mathrm{v}-\bar{\delta_\mathrm{v}}$ space for the full observed (\textit{left}) and simulated (\textit{right}) void samples. The voids are divided into three populations according to their $\lambda_\mathrm{v}$ values, following the binning strategy described in Sec. \ref{voidCat}. It can be seen that the $\lambda_\mathrm{v}$ subdivision is primarily driven by the $\bar{\delta_\mathrm{v}}$ values: voids with $\lambda_\mathrm{v} \in (5, \infty)$ (orange) exhibit positive $\bar{\delta_\mathrm{v}}$ values, voids with $\lambda_\mathrm{v} \in (-\infty, -5]$ (purple) show negative $\bar{\delta_\mathrm{v}}$, and voids in the range $\lambda_\mathrm{v} \in (-5,5]$ (green) tend to have $\bar{\delta_\mathrm{v}} \sim 0$, forming a transition zone between the two populations. Negative-$\lambda_\mathrm{v}$ voids are also more likely to have larger $r_\mathrm{v}$ values compared to positive-$\lambda_\mathrm{v}$ voids. 
    It is evident that voids in different $\lambda_\mathrm{v}$ bins are physically distinct and undergo significantly different formation processes and evolutions.}
    \label{lambda}
\end{figure}
\begin{figure*}
    \centering
    \includegraphics[scale=0.56]{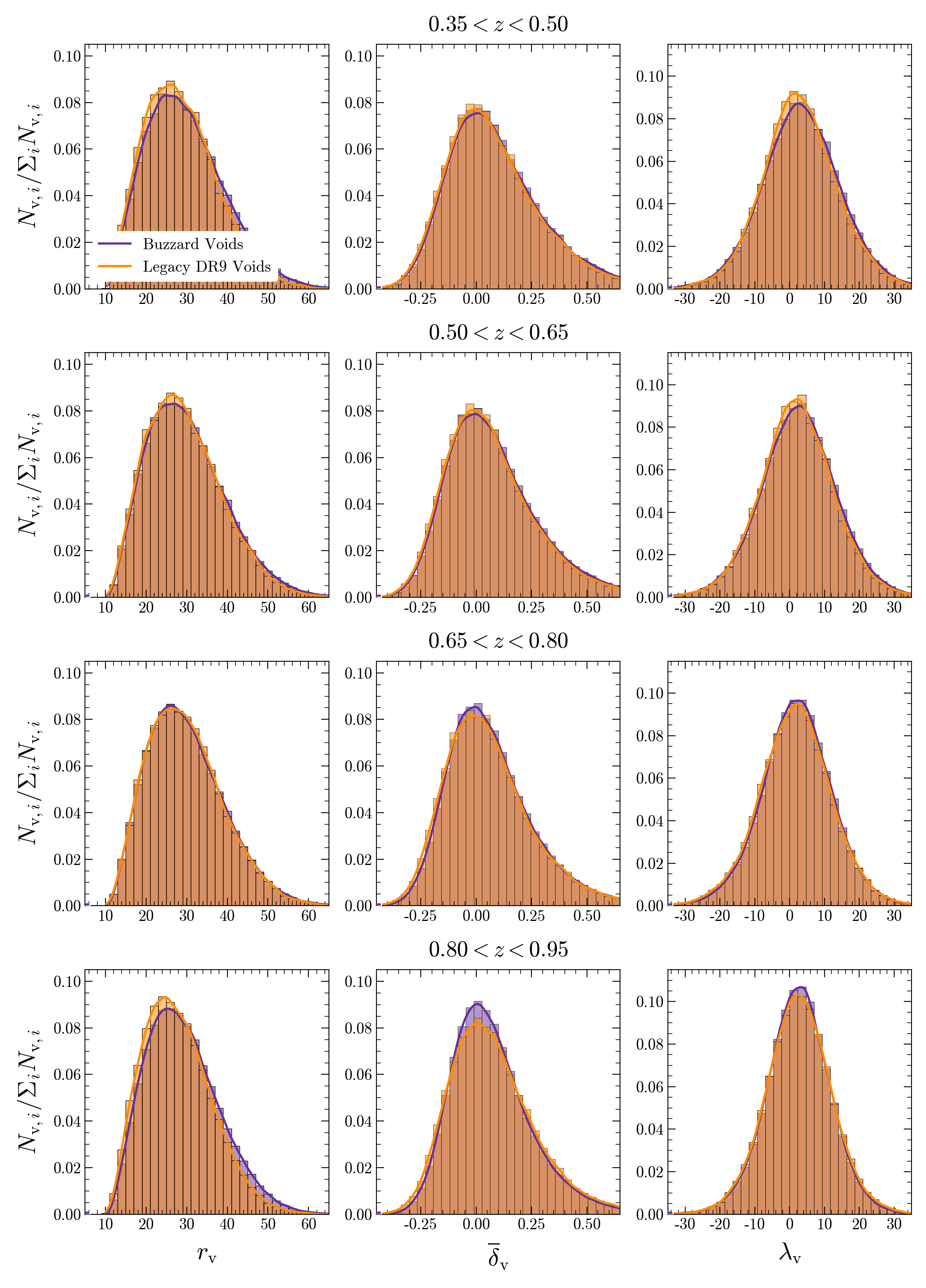}
    \captionof{figure}{Probability distributions for $r_\mathrm{v}$ (first column), $\bar{\delta_\mathrm{v}}$ (second column), and $\lambda_\mathrm{v}$ (third column) for voids identified in the calibrated Buzzard mocks (purple) and in the Legacy Survey (Orange). The void samples are subdivided into four equispaced redshift bins with $\mathrm{d}z=0.15$, with the redshift increasing for each row from top to bottom. The excellent agreement between the properties of the observed and simulated voids ensures effective control of systematics related to potential mismatches between observations and simulations.}
    \label{voidProp}
\end{figure*}
\subsubsection{Void catalogs}
\label{voidCat}

We identified cosmic voids by partitioning, again, the observed dataset into the North and South Galactic Caps, thereby accounting for the varying degrees of sparseness in these regions. Additionally, we build a void catalog for each of the eight (four for the North region, four for the South region) mock realizations presented in Sec. \ref{MockVal}. For both the observed and simulated datasets, we excluded voids whose center is identified outside the redshift range $0.35 < z < 0.95$. Maintaining a spatial buffer near the tracer catalog borders helps to mitigate issues related to the fragmentation of border voids introduced by the identification algorithm.
To reproduce the full-sky observation, we combined the two observed void catalogs into a single dataset. To match this full-sky catalog, we generated 12 mock full-sky void samples by combining the four North and four South mock void catalogs, ensuring that no single realization was used in both regions to avoid covariance issues. The normalized void count distributions are shown in Fig. \ref{fig:voidDist} for the North-South combined void catalogs, before and after the mock calibrations.
\begin{table*}
    \centering
    \caption{Number of voids in each sample used in this analysis, both observed and simulated, for the North and South regions as well as the full-sky. The table presents the number of voids for the entire dataset and for each redshift bin, further differentiated by the three $\lambda_\mathrm{v}$ bins considered. For the Buzzard mocks, the number of voids is the average across the four realizations. Note that the number of voids for the Buzzard full-sky sample is not simply the sum of the North and South realizations; instead, it is weighted to account for the different areas of the two regions of the sky.}
    \begin{tabular}{ccccccc}
        \multicolumn{7}{c}{\textbf{Full sample}} \\
        \specialrule{.15em}{.2em}{0.2em} 
	    \specialrule{.05em}{.05em}{0.5em} 
        \textit{catalog} & Legacy North & Legacy South & Legacy Full-Sky & Buzzard North & Buzzard South & Buzzard Full-Sky \\
        \specialrule{.1em}{.2em}{0.8em} 
        All $\lambda_\mathrm{v}$ & 78,231 & 62,481 & 140,712 & 71,041 & 70,555 & 127,765 \\[0.6em]
        Negative $\lambda_\mathrm{v}$ & 17,687 & 14,035 & 31,722 & 14,560 & 14,516 & 26,449 \\[0.6em]
        Zero $\lambda_\mathrm{v}$ & 30,603 & 24,018 & 54,621 & 28,112 & 27,741 & 50,394 \\[0.6em]
        Positive $\lambda_\mathrm{v}$ & 29,941 & 24,428 & 54,369 & 28,369 & 28,298 & 50,922 \\[0.8em]
        \specialrule{.1em}{.2em}{1em} 
        \multicolumn{7}{c}{\boldmath{$0.35\leq z<0.50$}} \\
        \specialrule{.15em}{.2em}{0.2em} 
	    \specialrule{.05em}{.05em}{0.5em} 
        \textit{catalog} & Legacy North & Legacy South & Legacy Full-Sky & Buzzard North & Buzzard South & Buzzard Full-Sky \\
        \specialrule{.1em}{.2em}{0.8em} 
        All $\lambda_\mathrm{v}$ & 10,480 & 9,343 & 19,823 & 9,211 & 9,676 & 17,008 \\[0.6em]
        Negative $\lambda_\mathrm{v}$ & 2,338 & 2,017 & 4,355 & 2,012 & 2,086 & 3,748 \\[0.6em]
        Zero $\lambda_\mathrm{v}$ & 3,973 & 3,371 & 7,344 & 3,187 & 3,330 & 5,860 \\[0.6em]
        Positive $\lambda_\mathrm{v}$ & 4,169 & 3,955 & 8,124 & 4,012 & 4,260 & 7,400 \\[0.8em]
        \specialrule{.1em}{.2em}{1em} 
        \multicolumn{7}{c}{\boldmath{$0.50\leq z<0.65$}} \\
        \specialrule{.15em}{.2em}{0.2em} 
	    \specialrule{.05em}{.05em}{0.5em} 
        \textit{catalog} & Legacy North & Legacy South & Legacy Full-Sky & Buzzard North & Buzzard South & Buzzard Full-Sky \\
        \specialrule{.1em}{.2em}{0.8em} 
        All $\lambda_\mathrm{v}$ & 15,868 & 12,937 & 28,805 & 14,560 & 14,861 & 26,498 \\[0.6em]
        Negative $\lambda_\mathrm{v}$ & 3,770 & 3,132 & 6,902 & 3,348 & 3,425 & 6,146 \\[0.6em]
        Zero $\lambda_\mathrm{v}$ & 6,120 & 4,770 & 10,890 & 5,215 & 5,380 & 9,540 \\[0.6em]
        Positive $\lambda_\mathrm{v}$ & 5,978 & 5,035 & 11,013 & 5,998 & 6,055 & 10,811 \\[0.8em]
        \specialrule{.1em}{.2em}{1em} 
        \multicolumn{7}{c}{\boldmath{$0.65\leq z<0.80$}} \\
        \specialrule{.15em}{.2em}{0.2em} 
	    \specialrule{.05em}{.05em}{0.5em} 
        \textit{catalog} & Legacy North & Legacy South & Legacy Full-Sky & Buzzard North & Buzzard South & Buzzard Full-Sky \\
        \specialrule{.1em}{.2em}{0.8em} 
        All $\lambda_\mathrm{v}$ & 21,640 & 16,763 & 38,403 & 20,714 & 20,080 & 36,860 \\[0.6em]
        Negative $\lambda_\mathrm{v}$ & 5,439 & 4,234 & 9,673 & 4,743 & 4,598 & 8,500 \\[0.6em]
        Zero $\lambda_\mathrm{v}$ & 8,368 & 6,321 & 14,689 & 8,308 & 7,940 & 14,691 \\[0.6em]
        Positive $\lambda_\mathrm{v}$ & 7,833 & 6,208 & 14,041 & 7,664 & 7,542 & 13,669 \\[0.8em]
        \specialrule{.1em}{.2em}{1em} 
        \multicolumn{7}{c}{\boldmath{$0.80\leq z \leq 0.95$}} \\
        \specialrule{.15em}{.2em}{0.2em} 
	    \specialrule{.05em}{.05em}{0.5em} 
        \textit{catalog} & Legacy North & Legacy South & Legacy Full-Sky & Buzzard North & Buzzard South & Buzzard Full-Sky \\
        \specialrule{.1em}{.2em}{0.8em} 
        All $\lambda_\mathrm{v}$ & 29,853 & 23,118 & 52,971 & 26,170 & 25,565 & 46,710 \\[0.6em]
        Negative $\lambda_\mathrm{v}$ & 6,045 & 4,574 & 10,619 & 4,381 & 4,321 & 7,907 \\[0.6em]
        Zero $\lambda_\mathrm{v}$ & 12,008 & 9,439 & 21,447 & 11,250 & 10,958 & 20,042 \\[0.6em]
        Positive $\lambda_\mathrm{v}$ & 11,800 & 9,105 & 20,905 & 10,539 & 10,286 & 18,760 \\[0.8em]
    \end{tabular}
    \label{voidNumber}
\end{table*}
We subdivided our void samples into four equispaced redshift bins of $\Delta z = 0.15$. Additionally, for both the total void sample and each redshift bin, we distinguished three different void populations based on their $\lambda_\mathrm{v}$ values: negative-$\lambda_\mathrm{v}$ ($\lambda_\mathrm{v} \in (-\infty, -5]$), zero-$\lambda_\mathrm{v}$ ($\lambda_\mathrm{v} \in (-5,5]$), and positive-$\lambda_\mathrm{v}$ ($\lambda_\mathrm{v} \in (5,\infty)$) voids. Figure \ref{lambda} illustrates how the value of $\lambda_\mathrm{v}$ is predominantly driven by the mean density inside the voids; however, voids with more positive $\lambda_\mathrm{v}$ tend to be smaller than those with more negative values, and vice versa. Table \ref{voidNumber} summarizes the number of voids for each case considered, demonstrating that the void bins are not equally populated. These specific $\lambda_\mathrm{v}$-binning values were chosen to approximately maximize the signal-to-noise ratio. A more refined selection will be explored in future work. Moreover, Fig. \ref{voidProp} shows the comparison between the probability distributions for a set of properties of the mock and observed voids, $r_\mathrm{v}$, $\bar{\delta}_\mathrm{v}$ and $\lambda_\mathrm{v}$, are shown for each redshift bin.
The general agreement between the properties of the simulated and observed void datasets allows us to mitigate potential systematics arising from discrepancies between the two void populations. We highlight that the minor discrepancies observed in the void size functions (\textit{left} panel of Fig. \ref{voidProp}) at both lower and higher redshift bins are reasonably due to the slight differences in sparseness of the two tracer catalogs, which enhance the sensitivity of voids to this parameter.  

\subsection{Cross-correlation measurement}
\label{ccmesurament}
Despite the difficulty in detecting the lensing signal from a single void \citep{Krause2013}, due to the weak and noisy nature of the signal in current observations, measuring a stacked signal from a substantial number of voids can yield a significant detection of the void-lensing cross-correlation. Our approach follows to established methods in the literature (see, for instance, \citealt{Camacho-Ciurana2023} and references therein), which are generally consistent with one another. The procedure can be summarized as follows:
\begin{enumerate}
    \item We initialize two empty $512\times512$ pixel matrices: one to store the final stacked $\kappa$ values from our CMB lensing map, and the other to account for the impact of the CMB mask. The mask applied is the \textit{Planck} 2018 mask for the observational data, and a corresponding mask covering the populated quarter of the sky for the Buzzard mocks. The weight for each pixel is progressively accumulated, ranging from 0 (indicating an unobserved pixel) to 1 (indicating a fully observed pixel).
    \item For each void, we extract a square CMB patch centred at the same sky position as the void center, with the side length set to five times the angular diameter of the void. To perform this extraction, we use the \texttt{healpy} Python package, specifically its \textit{gnomonic projection function}. This function identifies the patch by performing an appropriate coordinate rotation and rescales the CMB pixels to match the resolution required by the angular diameter of the void. The same process is applied to the CMB mask.
    \item The CMB and mask patches are then added to their respective 512×512 matrices.
    \item Once all voids have been processed, the stacked CMB patch is normalized by dividing it by the weight matrix. This step yields the average $\kappa$ value for each pixel, relative to its distance from the center of the voids, while properly accounting for the effects of the CMB mask.
\end{enumerate}

To accurately compare different measurements, we must account for the bias introduced by the incomplete observation of the CMB which results in a mean convergence over the observed area of the sky that deviates from zero. To account for this effect, simply subtracting the mean convergence of the observed region from our stacked patch is insufficient, as the voids do not uniformly sample the sky. Some regions are oversampled due to void overlap, which introduces additional complexity to the measurement. We correct our CMB lensing map by subtracting the bias, which is calculated by averaging all the pixels within the circular CMB patches behind the voids with a radius of $R=5r_\mathrm{v}$. Each pixel is weighted according to the number of times it is selected by the overlapping patches.

To compute the cross-correlation, we measured the radial profile of the final stacked patch. We extended our measurement to $R=5r_\mathrm{v}$ (where the subscript \textit{v} denotes the \textit{stacked void}) to capture the influence of the surrounding environment. The radial profile was calculated using 25 bins with a width of $\mathrm{d}R=0.2r_\mathrm{v}$, ensuring an optimal balance between noise reduction and the fitting accuracy of the profile.

\subsection{Error estimation and template-fitting analysis}
\label{errorEst}
\begin{figure}[H]
    \centering
    \includegraphics[width=1.\linewidth]{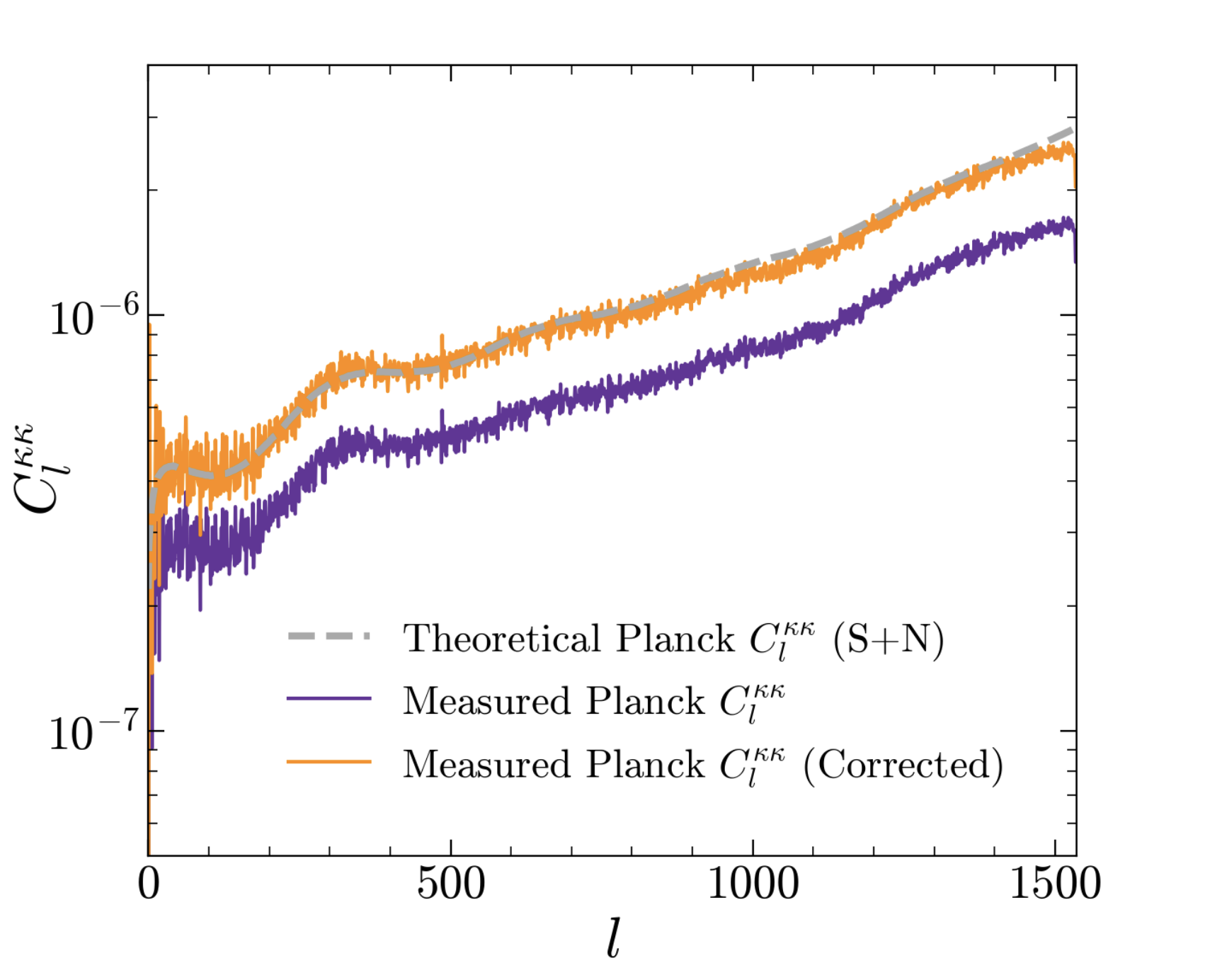}
    \caption{Theoretical (dashed grey line) and observed power spectra, shown before (purple line) and after (Orange line) the linear correction accounting for the mask effect. The theoretical prediction is derived from the best-fit cosmology of the \textit{Planck} 2018 results.}
    \label{powerSpectrum}
\end{figure}

The statistical uncertainty in the measurement of the void-CMB lensing cross-correlation is dominated by the instrumental noise in the \textit{Planck} CMB temperature measurements. Other relevant sources of uncertainty include cosmic and sample variances, as well as, to a smaller degree, the error associated with the reconstruction of the convergence maps. The combination of the wide coverage of the sky provided by both \textit{Planck} (which offers the broadest sky coverage for CMB observations to date) and DESI Legacy observations, along with the size of the void dataset identifiable in the DESI Legacy LRG sample, enables us to stack our signal from a wide range of independent regions of the Universe and uncorrelated areas of the sky, resulting in an unprecedented reduction of the uncertainty associated with our measurements. To account for these systematic effects, we employed the following strategy for each void sample, based on \citet{Vielzeuf2021}:
\begin{enumerate}
    \item Using the \texttt{healpy} function \texttt{anafast}, we firstly compute the auto Power Spectrum of the anisotropies of the \textit{Planck} CMB lensing map \citep{PlanckCMB2020}, $C^{kk}_l$. We then generated, through the \texttt{healpy} \texttt{synfast} function, 1,000 random maps, with $n_\mathrm{side}=512$ and $l_\mathrm{max}=n_\mathrm{side}*3-1$,  based on the measured Power Spectrum. To account for the effects of the CMB mask, a first-order correction to the power spectrum is applied: $C_{l,\mathrm{real}}^{\kappa\kappa} = C_{l,\mathrm{masked}}^{\kappa\kappa}/f_\mathrm{sky}$, where $f_\mathrm{sky}$ represents the ratio between the observed and total sky areas. Although $f_\mathrm{sky}$ varies between 0 and 1, this correction is only reliable when $f_\mathrm{sky}$ is close to 1. In our case, $f_\mathrm{sky} \sim 0.66$, ensuring an adequate correction of the power spectrum in the contest of our analysis. We note that for future studies aiming to constrain cosmological parameters or working with smaller sky regions, a mode-coupling Monte Carlo correction should be considered (see Appendix C of \citealt{sailer2024} for a detailed description of this methodology). Figure \ref{powerSpectrum} shows the comparison between the observed \textit{Planck} $C^{kk}_l$, before and after the correction, with the theoretical Power Spectrum derived from the 2018 \textit{Planck} best-fit cosmology provided by the \textit{Planck} Collaboration. Finally, monopole and dipole of the generated maps are removed. 
    \item Similarly to the procedure applied to the real CMB lensing map, we perform a Gaussian smoothing with a FWHM of $0.5^\circ$ on each random realization. 
    \item \textit{Planck} CMB lensing mask is applied to the random realization in order to conserve the amount of information provided by every map. 
    \item Following the procedure outlined in Sec. \ref{ccmesurament}, we cross-correlated each randomly generated map with our voids. It is essential to maintain the positions of the void centers to avoid losing information on void clustering, which could result in an overestimation of the errors. To assess the impact of cosmic variance, which arises from working with a single realization of our observable Universe, we performed a Jackknife analysis, with $N_\mathrm{JK} = N_\mathrm{v}$ for each bin. This procedure generates a jackknife sample of the same size as the void sample in the considered bin. The standard error associated with each bin is then computed as
    \begin{equation}
        \sigma_{JK} = \sqrt{\frac{N_\mathrm{v}-1}{N_\mathrm{v}}\sum (\theta_i-\bar{\theta})^2} \; .
    \end{equation}
    The impact of the jackknife on the error associated with our signal is less than 1.5\% for each bin and it is therefore considered negligible.
    \item We estimate the elements $i,j$ of the covariance matrix $\hat{C}$ as: 
    \begin{equation}
        C_{ij} = \frac{1}{N-1} \sum_k^N (x_i^k - \bar{x}_i)(x_j^k - \bar{x}_j) \, ,
    \end{equation}
    where $N=1000$ is the size of the random lensing CMB map-voids cross-correlation sample, $x_i^k$ is the measurement of $i$-th data component of the $k$-th cross-correlation, and $\bar{x}_i$ is the mean measurement of the $i$-th component. 
    \item The error bars associated with our measurements are then given by the square root of the diagonal elements of the covariance matrices.
\end{enumerate}
A similar procedure is applied to each CMB mock realization, allowing us to compute the errors associated at the simulated cross-correlations. It is important to note that, in this case, the contribution from instrumental noise is zero. However, we should still account for cosmic and sample variances, which contribution to the error budget is not negligible. The average value of the mock cross-correlation signal is obtained from the mean of the four realizations, while the elements of the associated covariance matrix are computed as:
\begin{equation}
C^B_{ij} = \frac{1}{N} \sum_{n=1}^N C_{i,k}^{(n)} + \frac{1}{N-1}\sum_{n=1}^N(M_{n,i} - \bar{M}_i)(M_{n,j} - \bar{M}_j) \, ,
\end{equation}
where $N$ is the number of realizations and $M$ refers to the value of the cross-correlation signal. The first term represents the average covariance matrix between the various realizations, while the second term accounts for the variance arising from the deviation of each simulated signal from the average template.

At this stage, we can quantify the consistency of our observations with the $\mathrm{\Lambda}$CDM prediction from the Buzzard mocks. Following the approach outlined in \citet{Vielzeuf2021}, we assess the level of consistency by estimating the best-fitting CMB lensing amplitude parameter, $A_\kappa = \kappa_{\mathrm{Legacy}}/\kappa_\mathrm{Buzzard}$ along with its corresponding uncertainty $\sigma_{A_\kappa}$. 

The best-fitting CMB lensing amplitude parameter, $A_\kappa$, is constrained through a $\chi^2$ minimization, with the $\chi^2$ statistic assuming the form: 
\begin{equation}
    \chi^2 = \sum_{ij} (x_i^\mathrm{L}-A_k\cdot x_i^\mathrm{B})
    ((C_{ij}^{\mathrm{L}})^{-1}+A^2_k\cdot (C^{\mathrm{B}}_{ij})^{-1})(x_j^\mathrm{L}-A_k\cdot x_j^\mathrm{B}) \, ,
\end{equation}
where $x_i$ is the average CMB lensing signal in a radius bin $i$ and $C$ is the associated covariance matrix. The exponents L and B denote respectively elements from Legacy Survey and Buzzard Mocks. Each inverse covariance matrix, following the Anderson-Hartlap procedure \citep{Hartlap2007}, is corrected by the factor 
\begin{equation}
        \alpha = \frac{(N_\mathrm{randoms} - N_\mathrm{bins} - 2)}{(N_\mathrm{randoms} - 1)} \, ,
\end{equation}
where $N_\mathrm{randoms}=1000$ and $N_\mathrm{bins}=25$. This correction decrease of $\sim 2.6\%$ the covariance value, having, in our specific case, a small impact on the error measurement. 
In this statistics, we take into account the error estimated for the simulated cross-correlations including its relative covariance, re-scaled for the factor $A_k^2$. The $A_\kappa$ corresponding uncertainty, $\sigma_{A_\kappa}$, is estimated as the range given by the two values assumed by the parameter $A_\kappa$ when $\chi^2_{\mathrm{min}}$ is increased by 1. 

The signal-to-noise ratio associated with our measurement is calculated based on the maximum detected anisotropy, or, in other words, as
\begin{equation}
    \frac{S}{N} = \max_j\left(\frac{M_j}{\sigma_j}\right) \, ,
\end{equation}
where $M_j$ and $\sigma_j$ denote the measured amplitude of the signal in each bin and its corresponding relative error, respectively.
\label{error}

\section{RESULTS \& DISCUSSION}
\label{sec4}
\begin{figure*}
    \centering
    \includegraphics[width=1.0\linewidth]{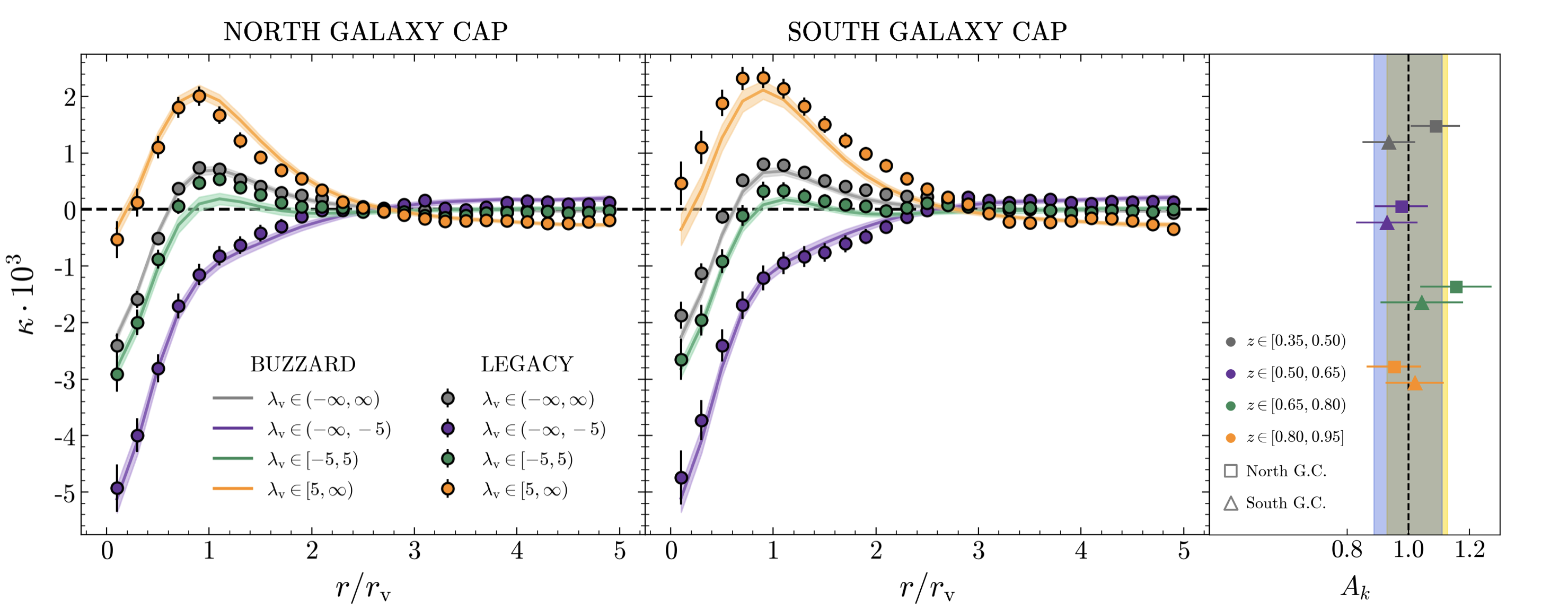}
    \captionof{figure}{Cross-correlation signals for the voids identified in the North (\textit{left}) and South (\textit{center}) regions of DESI Legacy Survey and the corresponding calibrated Buzzard mock realizations. The measurements are provided for the full void samples and the three different $\lambda_\mathrm{v}$ bins considered. The \textit{right} panel summarizes the $A_\kappa$ values for the various measurements (see  \ref{NSsignificances}). The yellow and blue vertical bands represent the combined measurements for the $\lambda_\mathrm{v}$ bins, with associated errors, for the North and South regions, respectively. These two estimates yield CMB lensing amplitude parameter values of $A_{\kappa,\mathrm{North}} = 1.028 \pm 0.099$ and $A_{\kappa,\mathrm{South}} = 0.998 \pm 0.112$, in perfect agreement with the full sample measurements and the simulated $\mathrm{\Lambda}$CDM predictions.} 
    \label{CCNS} 
\end{figure*}
Following the procedures outlined in the previous sections, we are now able to measure the cross-correlation between the voids identified in the LRGs from the DESI Legacy Survey DR9 and the CMB lensing map from \textit{Planck}. Additionally, we estimate the deviation from the $\mathrm{\Lambda}$CDM predictions based on the analysis of the spectroscopically calibrated LRG population of the Buzzard Mocks. 

We chose to conduct our analysis by dividing the dataset into two distinct regions corresponding to the North and South Galactic Caps, matching each mock realization to the respective dataset in order to account for the slight differences in the galaxy catalogs between the two regions. Subsequently, we performed a combined analysis by merging the two regions to account for the full-sky observation. The detailed procedure is outlined in Sec. \ref{PhotCorrection} and Sec. \ref{sparsenessMatch}.

\subsection{North and South Galaxy Caps}
In our initial analysis, we separately cross-correlated the 78,231 DESI Legacy voids in the North Galaxy Cap and the 62,481 voids in the South Galaxy Cap with the \textit{Planck} CMB lensing map. To account for the different environments and evolution of our voids, following the procedure illustrated in \citet{nadathur2017}, we further subdivided the datasets according to the $\lambda_\mathrm{v}$ value of the voids (see \citealt{Raghunathan2020}, \citealt{Camacho-Ciurana2023} and \citealt{Demirbozan2024} for previous works exploiting the $\lambda_\mathrm{v}$-binning methodology), more precisely creating three non-equipopulated bins with $\lambda_\mathrm{v} \in (-\infty, -5)$, $\lambda_\mathrm{v} \in [-5, 5)$, and $\lambda_\mathrm{v} \in [5, \infty)$.
All the observed cross-correlations are then compared with the $\mathrm{\Lambda}$CDM templates calculated from the Buzzard Mocks in order to estimate the best-fitting CMB lensing amplitude parameter, $A_\kappa$, and its corresponding uncertainty, $\sigma_{A_\kappa}$, providing an estimate of the deviation from the simulated $\mathrm{\Lambda}$CDM cosmology.

\begin{figure*}
    \centering
    \includegraphics[width=1.0\linewidth]{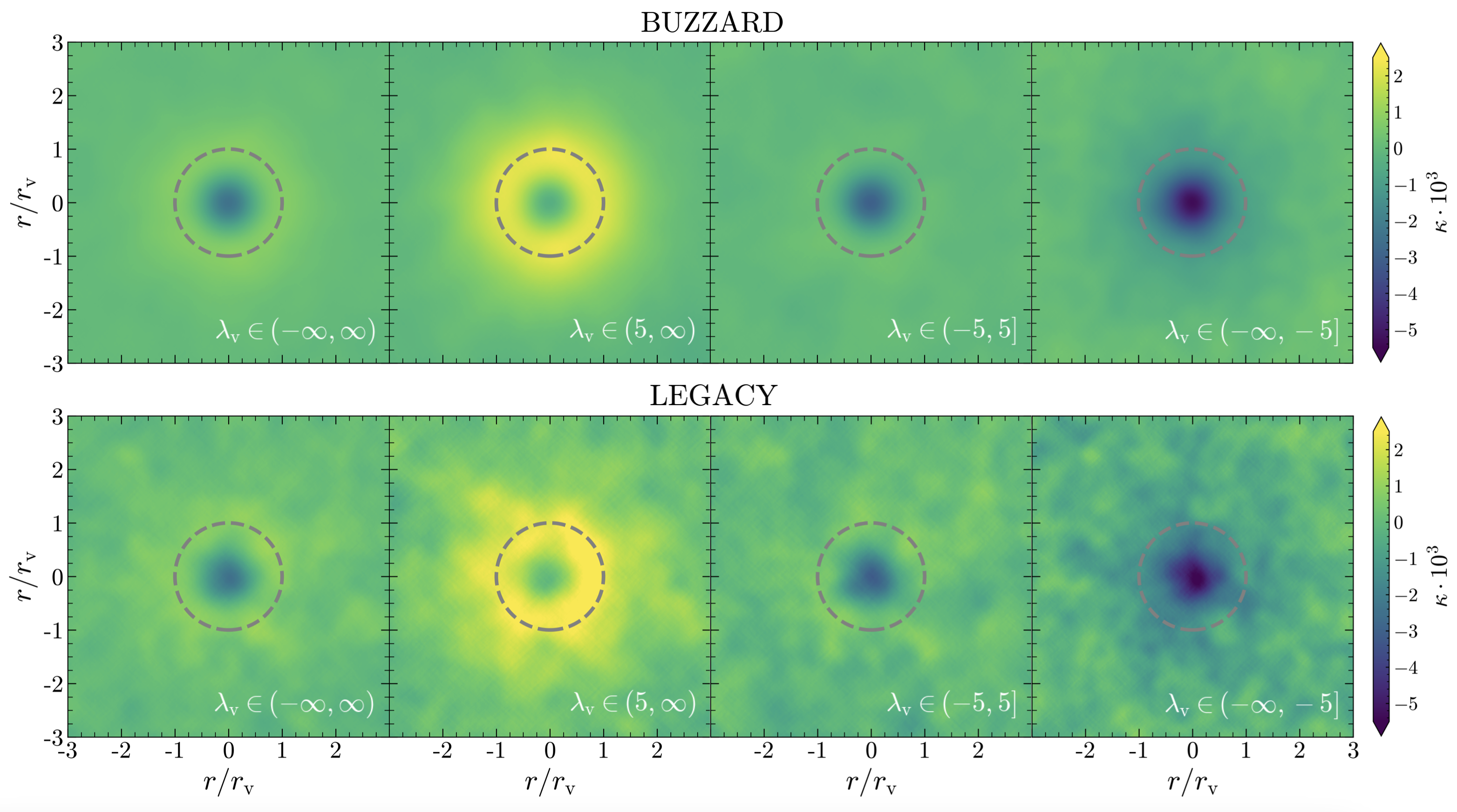}
    \caption{Stacked images of the simulated cross-correlation from the Buzzard mocks (\textit{top}) and the observed (\textit{bottom}) cross-correlation between the full-sky void dataset from the Legacy Survey and the \textit{Planck} CMB convergence map. The first column displays the stacked images for the full void population, while the subsequent three columns correspond to void bins categorized by $\lambda_\mathrm{v}$, ranging from positive to negative values. Despite the presence of instrumental noise in the observed stacked CMB patches, the images exhibit strong visual agreement between simulations and observations, clearly illustrating the distinct properties of the different void populations analyzed.}
    \label{CCimage}
\end{figure*}

The main results of the North/South Galaxy Caps analysis are summarized in Fig. \ref{CCNS}, while an illustrative example of the stacked images from which the cross-correlation profile is derived is presented in Fig. \ref{CCimage}, referred in this case to the full-sky void sample. For both the full dataset and the three $\lambda_\mathrm{v}$ bins, we present the observed cross-correlation alongside the simulated $\mathrm{\Lambda}$CDM signal up to $5r_\mathrm{v}$. In the right panel, the $A_\kappa$ values for each measurement are displayed.
Despite appearing nearly identical, the simulated signals for the North and South regions exhibit slight differences, primarily due to the varying tracer sparseness between the two regions (see Fig. \ref{sparsenessCorr}) which leads to small differences in the void populations identified. Both regions display a similar behaviour in the observed cross-correlations, with all measurements being consistent with the $\mathrm{\Lambda}$CDM expectations, with fluctuation from $A_\kappa=1$ not exceeding $\sim 1 \sigma$. 

When considering the full void sample, we measured $A_\kappa=1.088 \pm 0.081$ (S/N = 11.47) for the North region and $A_\kappa=0.936 \pm 0.087$ (S/N = 7.83) for the South region. Despite the similarity in the measured signals, the differences in the signal-to-noise ratio can be attributed to the varying number of voids and sky coverage in the two regions. As expected, the full sample signals reveal a profile consistent with the projected density profile of the stacked void. Typically, the density contrast profile of voids is negative in their inner part, increasing toward the outer regions in the so-called \textit{compensation wall}, which corresponds to the surrounding walls and filaments and exhibits positive density contrast values. The cross-correlation profiles analogously show a negative convergence in the inner regions, increasing to a positive maximum around $r/r_\mathrm{v} \sim 1$, corresponding to the overdense compensation wall of the stacked void, and then decreasing to zero at high radius values.
\begin{table}
    \centering
    \caption{Summary of the estimated $A_\kappa$ parameters, along with their relative uncertainties and signal-to-noise ratios, for the various measurements performed in the North and South regions of the sky.}
    \begin{tabular}{ccc}
        \multicolumn{3}{c}{\textbf{North Galaxy Cap}} \\
        \specialrule{.15em}{.2em}{0.2em} 
	    \specialrule{.05em}{.05em}{0.5em} 
        Void Sample & S/N & $A_\kappa\pm \sigma_{A_\kappa}$ \\
        \specialrule{.1em}{.2em}{0.8em} 
        $\lambda_\mathrm{v} \in (-\infty,\infty)$ & 11.47 & $1.088\pm0.081$ \\[0.5em]
        $\lambda_\mathrm{v} \in (-\infty,-5)$ & 16.06 & $0.976\pm 0.087$ \\[0.5em]
        $\lambda_\mathrm{v} \in [-5, 5)$ & 8.92 & $1.155\pm0.117$ \\[0.5em]
        $\lambda_\mathrm{v} \in [5, 1000)$ & 11.67 & $0.953\pm0.090$ \\[0.7em]
        \specialrule{.1em}{.2em}{0.8em} 
        \multicolumn{3}{c}{\textbf{South Galaxy Cap}} \\
        \specialrule{.15em}{.2em}{0.2em} 
	    \specialrule{.05em}{.05em}{0.5em} 
        Void Sample & S/N & $A_\kappa\pm \sigma_{A_\kappa}$ \\
        \specialrule{.1em}{.2em}{0.8em} 
        $\lambda_\mathrm{v} \in (-\infty,\infty)$ & 7.83 & $0.936\pm0.087$ \\[0.5em]
        $\lambda_\mathrm{v} \in (-\infty,-5)$ & 12.71 & $0.930\pm0.101$ \\[0.5em]
        $\lambda_\mathrm{v} \in [-5, 5)$ & 6.85 & $1.044\pm0.135$ \\[0.5em]
        $\lambda_\mathrm{v} \in [5, 1000)$ & 12.41 & $1.021\pm0.095$ \\[0.7em]
        \specialrule{.1em}{.2em}{1em} 
    \end{tabular}
\label{NSsignificances}
\end{table}
When considering the three $\lambda_\mathrm{v}$ bins, it becomes evident that the void sample is split into three distinct populations, each exhibiting markedly different lensing characteristics. To understand this behaviour, it is important to recall that voids with very negative $\lambda_\mathrm{v}$ are, by definition, highly underdense, often corresponding to large voids and/or structures likely classified as \textit{void-in-voids}. On the other hand, voids with very positive $\lambda_\mathrm{v}$ are defined as slightly underdense structures evolving in overdense environments, commonly referred to as \textit{void-in-clouds}. The distinct nature of these two populations leads to significant differences in their respective lensing signals, resulting in the original cross-correlation splitting into three distinct signals. This effect can be exploited to enhance the signal-to-noise ratio (For example, the signal-to-noise ratio is enhanced to S/N = 16.06 for the negative $\lambda_\mathrm{v}$ void sample in the North region, compared to S/N = 11.47 for the full void sample in the same Galaxy Cap), thereby yielding observations with higher significance. The significance levels of the various measurements, along with the values of the CMB lensing amplitude parameter $A_\kappa$ and their associated uncertainties, are reported in Table \ref{NSsignificances}, while the number of voids considered in each sample is listed in Table \ref{voidNumber}.

\subsection{Full-Sky analysis}
As the second step of our analysis, we combined the void samples from the North and South Galaxy Caps to reconstruct the full sky void catalog, consisting of 140,712  voids. We then compared the observed cross-correlations with the simulated $\mathrm{\Lambda}$CDM template derived from the 12 North/South Buzzard mock combinations. For details on 
the full creation procedure of the mock catalogs, refer to Sec. \ref{voidCat}.

Figure \ref{CCimage} presents a detailed view of the stacked CMB patches resulting from the cross-correlation between the full-sky void dataset and the CMB lensing map, shown for both the observed data and a single mock realization, as well as for the different $\lambda_\mathrm{v}$ bins. The first column shows the stacked image for the full void sample, where a distinct negative convergence region is visible at the center of the patches, illustrating the de-magnification effect induced by the voids. Additionally, a positive ring around $r \sim 1r_\mathrm{v}$ highlights the magnification effect of the compensation wall. It is evident that the stacked image derived from the DESI Legacy voids appears noisier than that from the simulations, an effect attributable to residual instrumental noise. The subsequent three columns, from left to right, show the void bins ordered by decreasing positive to negative $\lambda_\mathrm{v}$ values. The convergence values of the stacked voids clearly distinguish the different environments in which the various void populations evolve.

\begin{figure*}
    \centering
    \includegraphics[width=1.0\linewidth]{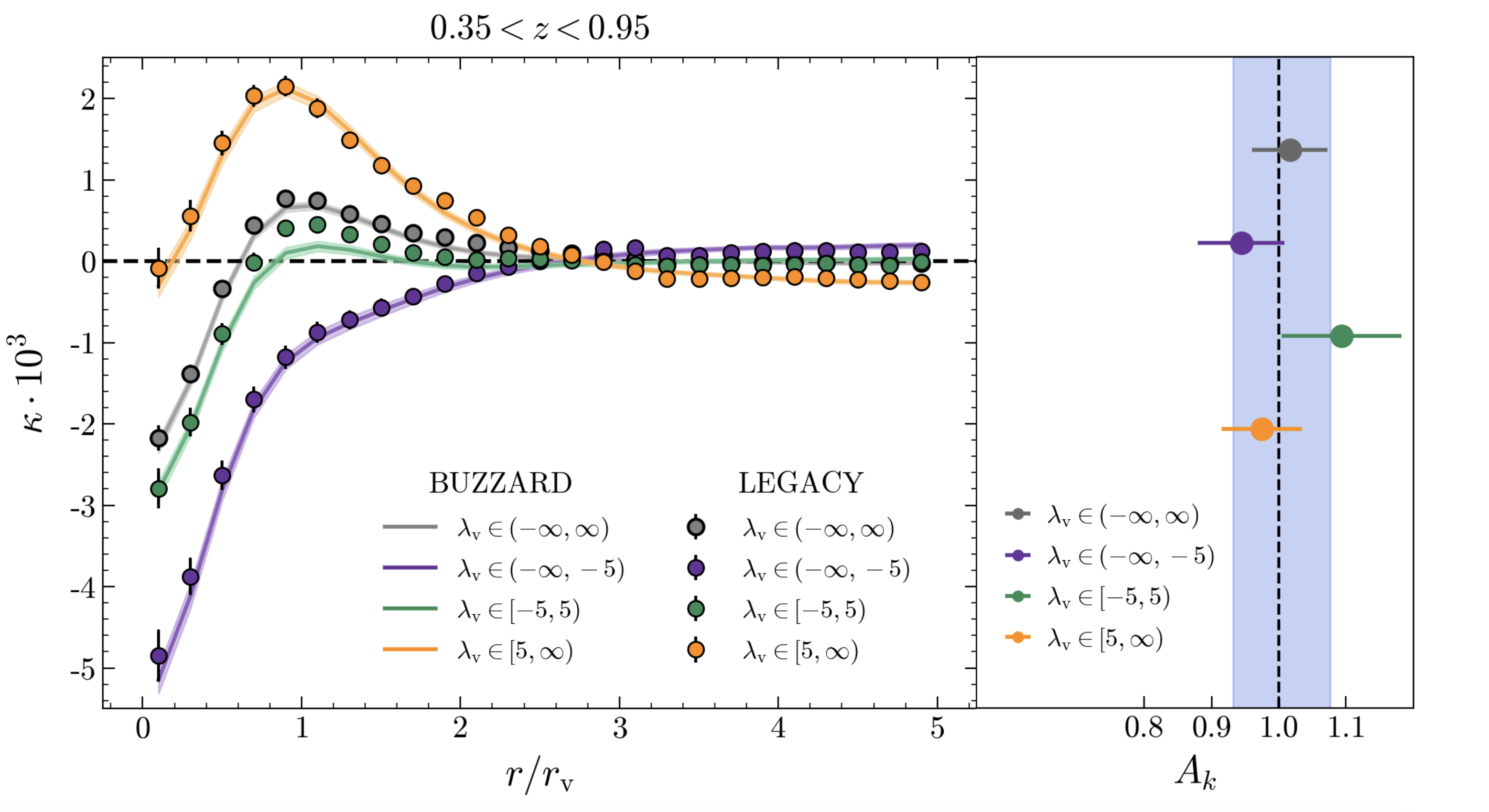}
    \caption{\textit{Left}: Cross-correlation signals for the voids identified in the full-sky DESI Legacy Survey and the corresponding calibrated Buzzard mock realizations. The measurements are provided for the full void samples and the three different $\lambda_\mathrm{v}$ bins considered. \textit{Right}: Summarized $A_\kappa$ values for the various measurements (see  \ref{ALLsignificances}). The blue vertical band represents the combined measurements for the $\lambda_\mathrm{v}$ bins, with associated error, providing an estimate of $A_{\kappa,\mathrm{full-sky}} = 1.004 \pm 0.072$, in perfect agreement with the full sample measurement and the simulated $\mathrm{\Lambda}$CDM predictions.}
    \label{CCall}
\end{figure*}
\begin{figure*}
    \centering
    \includegraphics[width=1.0\linewidth]{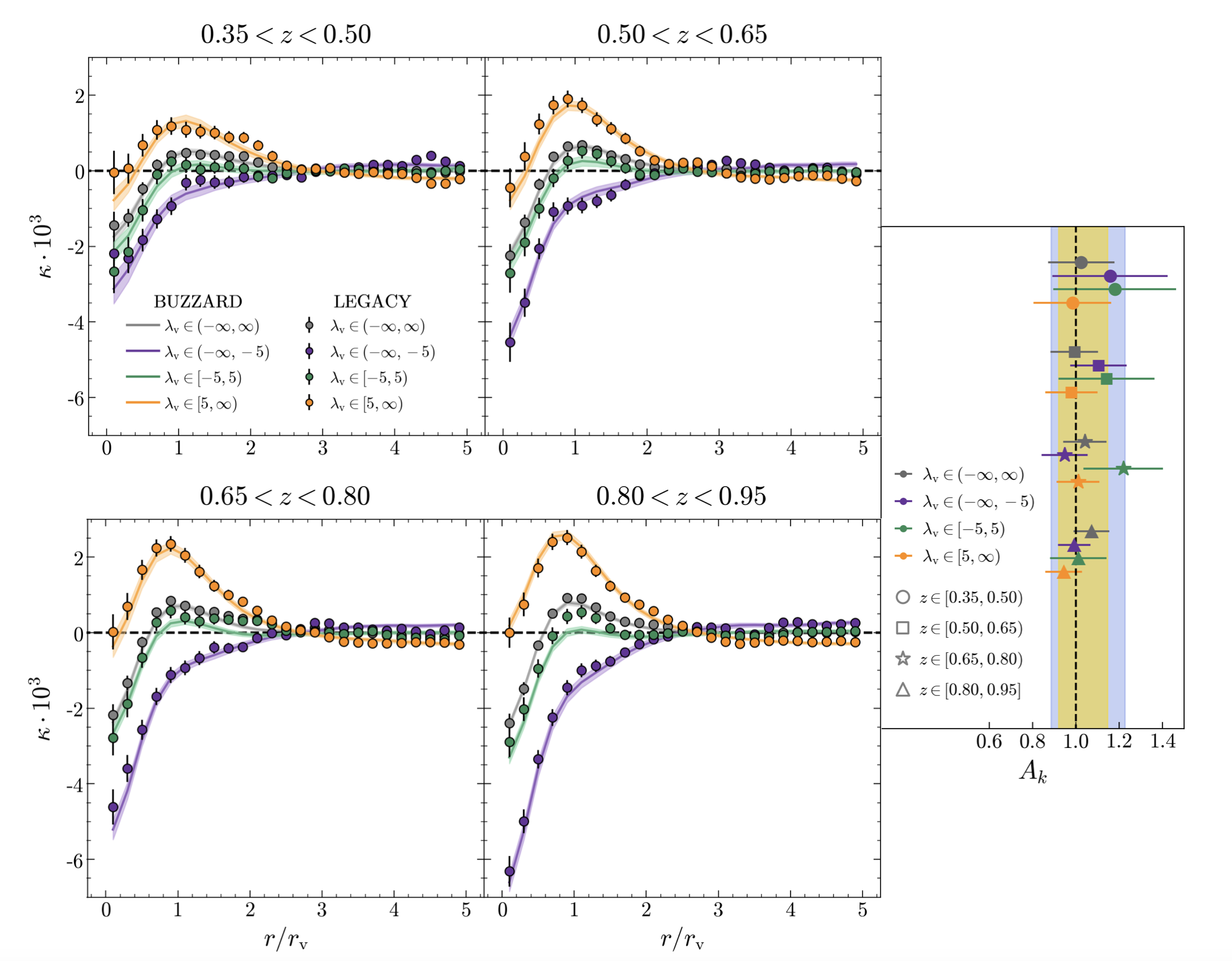}
    \caption{\textit{Left}: Tomographic cross-correlation signals for the voids identified in the full-sky DESI Legacy Survey and the corresponding calibrated Buzzard mock realizations. The measurements are provided for the full void samples and the three different $\lambda_\mathrm{v}$ bins, analyzed in four equispaced redshift bins with $\mathrm{d}z=0.15$. \textit{Right}: Summarized $A_\kappa$ values for the various measurements (see  \ref{CCamplBins}). The blue and yellow vertical bands represent the combined measurements for the $\lambda_\mathrm{v}$ bins and the full void samples at different redshifts, with associated errors. The estimates provide CMB lensing amplitude parameters of $A_{\kappa,\lambda_\mathrm{v}} = 1.056 \pm 0.170$ and $A_{\kappa,\lambda_\mathrm{v}} = 1.033 \pm 0.114$, in perfect agreement with the individual measurements and the simulated $\mathrm{\Lambda}$CDM predictions.}
    \label{CCbins}
\end{figure*}
As in the previous analysis, we first cross-correlated the entire void sample and then subdivided the void catalog into three distinct $\lambda_\mathrm{v}$ bins. The results of the full sky analysis are summarized in Fig. \ref{CCall}, along with the best-fitting CMB lensing amplitude parameters for the signals. We measured $A_\kappa = 1.016 \pm 0.054$ (S/N = 14.06) for the whole void sample. For the three $\lambda_\mathrm{v}$ bins, the measured parameters are $A_\kappa = 0.944 \pm 0.060$ (S/N = 16.94) for the negative $\lambda_\mathrm{v}$ bin, $A_\kappa = 1.093 \pm 0.090$ (S/N = 11.33) for the zero $\lambda_\mathrm{v}$ bin, and $A_\kappa = 0.975 \pm 0.060$ (S/N = 17.02) for the positive $\lambda_\mathrm{v}$ bin. These value are also tabulated in Table ~\ref{ALLsignificances}. Once again, all measurements are consistent with the $\mathrm{\Lambda}$CDM expectations and demonstrate the enhanced signal-to-noise ratio achieved by differentiating between the various void populations. This behaviour highlights that improving the signal-to-noise ratio is not solely dependent on the number of voids but also on the type and quality of the void catalog selected. We note that future analyses could further improve the signal-to-noise ratio through optimal fine-tuning of the $\lambda_\mathrm{v}$ binning.
\begin{table}
    \centering
    \caption{Summary of the estimated $A_\kappa$ parameters, along with their relative uncertainties and signal-to-noise ratios, for the various measurements performed in the full-sky sample.}
    \begin{tabular}{ccc}
        \multicolumn{3}{c}{\textbf{Full Sample}} \\
        \specialrule{.15em}{.2em}{0.2em} 
	    \specialrule{.05em}{.05em}{0.5em} 
        Void Sample & S/N & $A_\kappa\pm \sigma_{A_\kappa}$ \\
        \specialrule{.1em}{.2em}{0.8em} 
        $\lambda_\mathrm{v} \in (-\infty,\infty)$ & 14.06 & $1.016\pm0.054$ \\[0.5em]
        $\lambda_\mathrm{v} \in (-\infty,-5)$ & 16.94 & $0.944\pm 0.064$ \\[0.5em]
        $\lambda_\mathrm{v} \in [-5, 5)$ & 11.33 & $1.093\pm0.090$ \\[0.5em]
        $\lambda_\mathrm{v} \in [5, 1000)$ & 17.02 & $0.975\pm0.060$ \\[0.7em]
        \specialrule{.1em}{.2em}{0.8em} 
    \end{tabular}
\label{ALLsignificances}
\end{table}

\begin{table*}
    \centering
    \caption{Summary of the estimated $A_\kappa$ parameters, along with their relative uncertainties and signal-to-noise ratios, for the various tomographic measurements performed in the full-sky sample across the different redshift bins.}
    \begin{tabular}{ccccccccc}
        \multicolumn{1}{c}{} & \multicolumn{2}{c}{\boldmath{$0.35\leq z < 0.50$}} & \multicolumn{2}{c}{\boldmath{$0.50\leq z < 0.65$}} & \multicolumn{2}{c}{\boldmath{$0.65\leq z < 0.80$}} & \multicolumn{2}{c}{\boldmath{$0.80\leq z \leq0.95$}} \\
        \specialrule{.15em}{.2em}{0.2em} 
	    \specialrule{.05em}{.05em}{0.5em} 
        Void Sample & S/N & $A_\kappa\pm \sigma_{A_\kappa}$ & S/N & $A_\kappa\pm \sigma_{A_\kappa}$ & S/N & $A_\kappa\pm \sigma_{A_\kappa}$ & S/N & $A_\kappa\pm \sigma_{A_\kappa}$ \\
        \specialrule{.1em}{.2em}{0.8em} 
        $\lambda_\mathrm{v} \in (-\infty,\infty)$ & 5.20 & $1.025\pm0.153$ & 7.33 & $0.993\pm0.109$ & 7.94 & $1.041\pm0.099$ & 10.34 & $1.072\pm0.083$ \\[0.5em]
        $\lambda_\mathrm{v} \in (-\infty,-5)$ & 6.13 & $1.158\pm 0.266$ & 9.27 & $1.104\pm0.131$ & 10.10 & $0.947\pm0.107$ & 15.92 & $0.992\pm0.075$\\[0.5em]
        $\lambda_\mathrm{v} \in [-5, 5)$ & 5.48 & $1.180\pm0.283$ & 5.22 & $1.142\pm0.222$ & 5.93 & $1.219\pm0.184$ & 7.17 & $1.011\pm0.130$\\[0.5em]
        $\lambda_\mathrm{v} \in [5, 1000)$ & 5.71 & $0.984\pm0.179$ & 8.47 & $0.979\pm0.120$ & 11.28 & $1.010\pm0.100$ & 12.55 & $0.943\pm0.085$\\[0.7em]
        \specialrule{.1em}{.2em}{0.8em} 
    \end{tabular}
    \label{CCamplBins}
\end{table*}
In recent years, \citet{Kovacs2022} and \citet{Camacho-Ciurana2023} have identified a redshift-dependent tension with the $\mathrm{\Lambda}$CDM prediction in the cross-correlation, with the tension becoming more pronounced at lower redshifts. To address these discrepancies, we extended our analysis by exploiting the large void sample available to conduct a tomographic study of the cross-correlation. We partitioned the redshift range into four equally spaced bins with $\mathrm{d}z = 0.15$. Additionally, as in previous studies, we further divided the void sample within each redshift bin into three distinct populations according to their $\lambda_\mathrm{v}$ values. The number of voids in each bin is reported in Table \ref{voidNumber}.

The results of the tomographic analysis are shown in Fig. \ref{CCbins}, alongside the summary plot of the various estimates of the CMB lensing amplitude parameter. Considering the full void samples for each redshift bin, we measure, from lower to higher redshift, $A_\kappa = 1.025 \pm 0.153$ (S/N = 5.20), $A_\kappa = 0.993 \pm 0.109$ (S/N = 7.33), $A_\kappa = 1.041 \pm 0.099$ (S/N = 7.94), and $A_\kappa = 1.072 \pm 0.083$ (S/N = 10.34). We can observe that for the full void sample, all measurements are in agreement with the $\mathrm{\Lambda}$CDM prediction, with the significance level increasing as the number of voids grows. Furthermore, the cross-correlations exhibit similar amplitudes across the different redshift bins, whereas a stronger signal at higher redshifts would be expected due to the magnification effect of the CMB lensing kernel, which is zero at $z=0$ and peaks at $z\sim1.7-1.8$. To better understand these behaviours, it is useful to separate the void sample based on the $\lambda_\mathrm{v}$ values of the voids. This procedure allows us to distinguish between different void populations and avoid the signal flattening effect caused by the combination of voids with varying physical and lensing properties. It is observed that, when analyzing the signals from the negative and positive $\lambda_\mathrm{v}$ voids separately, both demonstrate a coherent evolution, with signal amplitude increasing from low to high redshift. 

All the measurements are compatible with the $\mathrm{\Lambda}$CDM simulated prediction at $\sim1\sigma$ level. The significance levels of the various measurements, along with the $A_\kappa$ values and their associated uncertainties, are reported in Table \ref{CCamplBins}. 

We observe that the zero-$\lambda_\mathrm{v}$ bin, for all redshift bins except the highest, consistently yields an $A_\kappa$ value systematically above one, at approximately $1\sigma$ level. This bin has limited cosmological significance due to its intrinsic superposition of various void populations, which results in a cross-correlation amplitude close to zero and a lower significance compared to the other two populations; although, it serves as an important indicator of the alignment between the properties of observed and simulated voids. This void population is likely to represent voids with an average internal density $\bar{\delta}_\mathrm{v} \sim 0$. As shown in Fig. \ref{voidProp}, these voids are located in the region of the peak of the $\bar{\delta}_\mathrm{v}$ distribution. Consequently, even minor mismatches in $\bar{\delta}_\mathrm{v}$ between observed and simulated void populations (due to differences in sparseness, error distributions, or other systematics) can lead to $A_\kappa$ values deviating from one, creating apparent discrepancies with the $\mathrm{\Lambda}$CDM predictions. Despite the apparent trend in our measurements, it cannot be ruled out that this result is due to random fluctuations. Indeed, considering the excellent overlap of the two void populations, as observable in Fig. \ref{voidProp}, and the fact that our measurements are compatible with the $\mathrm{\Lambda}$CDM estimates for this zero-$\lambda_\mathrm{v}$ bins, we are assured of compatibility between our observed and simulated void populations.
\section{CONCLUSIONS}
\label{sec5}
In this paper, we present a series of new measurements of the cross-correlation between cosmic voids and \textit{Planck} CMB lensing (\citealt{Planck2020}, \citealt{PlanckCMB2020}). Our work is motivated by the opportunities offered by the exploitation of the new large photometric LRG catalog provided by the DESI Legacy Survey \citep{Legacy}, with the goal of investigating the tensions with the $\mathrm{\Lambda}$CDM simulation highlighted by some previous studies (\citealt{Vielzeuf2021}, \citealt{Hang2021}, \citealt{Kovacs2022}, \citealt{Camacho-Ciurana2023}).

We compared our observed signals with simulated templates derived from four realizations of the LRG samples from the Buzzard Mocks \citep{buzzard}, which are designed to mimic the DESI observations and provide realistic DESI-like photometric LRG samples. To address potential tensions arising from discrepancies between the observed and simulated galaxy catalogs, we corrected our mock realizations by calibrating the photometric redshifts of the mock galaxies, performed using a combination of photometric and spectroscopic redshifts provided by one million DESI spectra \citep{DESI1}, and matching the sparseness between the observed and simulated galaxy catalogs, as detailed in Sec. \ref{PhotCorrection} and Sec. \ref{sparsenessMatch}.

To create our 3D void samples, we utilized the modified \texttt{ZOBOV} algorithm \citep{zobov2008}, which is integrated into the \texttt{Revolver} void-finding code \citep{Nadathur2019}. The void identification process and the properties of the void catalogs are summarized in Sec. \ref{voidFinding}. The observed void catalog and the void catalogs from the calibrated mocks exhibit an almost perfect match across all considered properties ($r_\mathrm{v}$, $\bar{\delta}_\mathrm{v}$, $\lambda_\mathrm{v}$), as shown in Fig. \ref{voidProp}, allowing us to avoid the possible arising of tensions between our measurements and the $\mathrm{\Lambda}$CDM prediction, arising from the mismatch between the simulated and observed galaxy sample in which our void catalogues are identified. 

Once the void catalogs were produced, we evaluated their $\kappa$ imprint by cross-correlating them with the CMB lensing maps using a stacking methodology. We applied Gaussian smoothing with a FWHM of $0.5^\circ$ to the CMB lensing maps to mitigate noise and fluctuations at small scales.  We then utilized 1000 random CMB realizations to estimate the covariance matrices associated with our measurements. Through a template-fitting process, we assessed the consistency between the observed data and the simulated $\mathrm{\Lambda}$CDM predictions estimating the CMB lensing amplitude parameter $A_\kappa$ and its  uncertainty $\sigma_{A_\kappa}$.

We independently analyzed the North and South Caps to account for small differences in sparseness levels between the two regions. Afterwards, we merged the two void catalogs to cover the entire sky, correspondingly modifying our mocks to accurately reflect the properties of the full-sky observations. The result of the cross-correlation process are reported in Sec. \ref{sec4}. When considering the full void sample, we measured $A_\kappa = 1.088 \pm 0.081$ (S/N = 11.47) for the North region, $A_\kappa = 0.936 \pm 0.087$ (S/N = 7.83) for the South region, and $A_\kappa = 1.016 \pm 0.054$ (S/N = 14.06) for the full-sky observation. All three measurements are in perfect agreement with the simulated $\mathrm{\Lambda}$CDM predictions. Furthermore, the full-sky observation provides a new record detection of the CMB lensing signal from voids, with a signal-to-noise ratio slightly exceeding those reported in \citet{Hang2021} and \citet{Camacho-Ciurana2023}. 

To account for the different environments and evolution of our voids, following the procedure illustrated in \citet{nadathur2017} and \citet{Raghunathan2020}, we further subdivided the datasets according to the $\lambda_\mathrm{v}$ value of the voids, more precisely creating three non-equipopulated bins with $\lambda_\mathrm{v} \in (-\infty, -5)$, $\lambda_\mathrm{v} \in [-5, 5)$, and $\lambda_\mathrm{v} \in [5, \infty)$. These specific $\lambda_\mathrm{v}$-binning values were chosen to approximately maximize the signal-to-noise ratio. For the North, South, and full-sky observations, we measured void-CMB lensing cross-correlation signals in perfect agreement with the $\mathrm{\Lambda}$CDM expectations. The values of the CMB lensing amplitude parameter and corresponding signal-to-noise ratios are summarized in Tabs. \ref{NSsignificances} and \ref{ALLsignificances}.
This analysis underscores the enhanced signal significance obtained by subdividing the void sample into different populations based on their $\lambda_\mathrm{v}$ values. Specifically, for the full-sky observation, we measured $A_\kappa = 0.944 \pm 0.064$ (S/N = 16.94) for the \textit{void-in-voids} population and $A_\kappa = 0.975 \pm 0.060$ (S/N = 17.02) for the \textit{void-in-clouds}, significantly improving the signal-to-noise ratio compared to previous observations, despite the smaller number of voids in each sample, establishing a new record measurement for CMB lensing by voids. 

Lastly, we took advantage of the large size of our void catalog to perform a tomographic analysis, subdividing the redshift range into 4 different bins to explore the evolution of the void-CMB lensing signal across different cosmic epochs. Once again, we further subdivided our void samples into three distinct bins based on their $\lambda_\mathrm{v}$ values, allowing us to disentangle the varying lensing properties of different void populations across the redshift range. As shown in Fig. \ref{CCbins}, all the observed cross-correlations are consistent with the simulated $\mathrm{\Lambda}$CDM predictions at  $\sim1\sigma$ level. The corresponding signal-to-noise ratios and the measured $A_\kappa$ values, along with their associated uncertainties, are detailed in  \ref{CCamplBins}. Once again, this analysis highlights the enhanced power of the signal and the disentangling effect of the $\lambda_\mathrm{v}$ binning approach, revealing the evolution of the cross-correlation with redshift driven by the effect of the CMB lensing kernel, which would otherwise remain obscured by the overlapping contributions from different void populations to the total signal.

In summary, we presented a set of new cross-correlation measurements between 3D voids and CMB lensing with improved S/N, fully consistent with the $\mathrm{\Lambda}$CDM predictions from simulations, emphasizing the critical role of systematic management. In particular, we underscore the necessity for mock catalogs used in such analyses to precisely match the observed data in terms of sparseness and redshift error distribution. Even small deviations can lead to void populations with significantly different lensing properties, potentially causing apparent tensions between observations and simulations. This consideration should be prioritized in the development of any future mock catalogs aimed at the cosmological exploitation of voids, ensuring that the systematic alignment between mocks and observations is rigorously maintained to avoid introducing biases in the analysis.
Moreover, we demonstrated the importance of the $\lambda_\mathrm{v}$ binning approach, establishing it as essential for disentangling distinct void populations and enhancing the signal-to-noise ratio. This method reveals the redshift evolution of the CMB lensing signal, which would otherwise be obscured by the superposition of different void populations with differing physical and lensing properties, allowing a more precise comparison between observations and $\mathrm{\Lambda}$CDM predictions, thus significantly improving the robustness of the analysis.

In future analyses, it will be valuable to explore the fine-tuning of the \textit{void-in-voids} and \textit{void-in-clouds} populations to further improve the signal-to-noise ratio. This refined approach, combined with the new large spectroscopic datasets from DESI and other next-generation surveys (e.g., Euclid, Vera C. Rubin Observatory, Roman Space Telescope), which will provide high-quality data in the redshift range where the CMB lensing kernel is most sensitive, and with upcoming high-precision CMB observations (e.g., Simons Observatory, CMB-S4), which will significantly lower the noise level in convergence maps, will substantially enhance the cosmological constraining power of future measurements of the CMB lensing imprint of voids. These advancements will be critical for future efforts aimed at differentiating between cosmological models and constraining cosmological parameters (see \citealt{Vielzeuf2023} for details), offering a new and detailed perspective on cosmic voids and the nature of the Universe.

\section*{Data Availability}
The DESI Legacy Survey DR9 LRG catalogue and the Buzzard mocks are available upon request to respectively Rongpu Zhou and Joe DeRose. \textit{Planck} 2018 CMB lensing data are publicly available on the \textit{Planck} Legacy Archive at the following link: \url{http://pla.esac.esa.int/pla/#home}. Datapoint informations for all the figures of the publication can be found at \url{https://doi.org/10.5281/zenodo.14251132}.

\section*{Acknowledgements}
SS would like to thank Chris Blake and Joe DeRose for their efforts in the creation of the Buzzard galaxy mocks and their generous availability, as well as Rongpu Zhou for the creation and publication of the Legacy Survey DR9 sample, which forms the foundation of this work.
 
This material is based upon work supported by the U.S. Department of Energy (DOE), Office of Science, Office of High-Energy Physics, under Contract No. DE–AC02–05CH11231, and by the National Energy Research Scientific Computing Center, a DOE Office of Science User Facility under the same contract. Additional support for DESI was provided by the U.S. National Science Foundation (NSF), Division of Astronomical Sciences under Contract No. AST-0950945 to the NSF’s National Optical-Infrared Astronomy Research Laboratory; the Science and Technology Facilities Council of the United Kingdom; the Gordon and Betty Moore Foundation; the Heising-Simons Foundation; the French Alternative Energies and Atomic Energy Commission (CEA); the National Council of Humanities, Science and Technology of Mexico (CONAHCYT); the Ministry of Science, Innovation and Universities of Spain (MICIU/AEI/10.13039/501100011033), and by the DESI Member Institutions: \url{https://www.desi.lbl.gov/collaborating-institutions}.

The DESI Legacy Imaging Surveys consist of three individual and complementary projects: the Dark Energy Camera Legacy Survey (DECaLS), the Beijing-Arizona Sky Survey (BASS), and the Mayall z-band Legacy Survey (MzLS). DECaLS, BASS and MzLS together include data obtained, respectively, at the Blanco telescope, Cerro Tololo Inter-American Observatory, NSF’s NOIRLab; the Bok telescope, Steward Observatory, University of Arizona; and the Mayall telescope, Kitt Peak National Observatory, NOIRLab. NOIRLab is operated by the Association of Universities for Research in Astronomy (AURA) under a cooperative agreement with the National Science Foundation. Pipeline processing and analyses of the data were supported by NOIRLab and the Lawrence Berkeley National Laboratory. Legacy Surveys also uses data products from the Near-Earth Object Wide-field Infrared Survey Explorer (NEOWISE), a project of the Jet Propulsion Laboratory/California Institute of Technology, funded by the National Aeronautics and Space Administration. Legacy Surveys was supported by: the Director, Office of Science, Office of High Energy Physics of the U.S. Department of Energy; the National Energy Research Scientific Computing Center, a DOE Office of Science User Facility; the U.S. National Science Foundation, Division of Astronomical Sciences; the National Astronomical Observatories of China, the Chinese Academy of Sciences and the Chinese National Natural Science Foundation. LBNL is managed by the Regents of the University of California under contract to the U.S. Department of Energy. The complete acknowledgments can be found at \url{https://www.legacysurvey.org/}.

Any opinions, findings, and conclusions or recommendations expressed in this material are those of the author(s) and do not necessarily reflect the views of the U. S. National Science Foundation, the U. S. Department of Energy, or any of the listed funding agencies.

The authors are honored to be permitted to conduct scientific research on Iolkam Du’ag (Kitt Peak), a mountain with particular significance to the Tohono O’odham Nation.

The Large-Scale Structure (LSS) research group at Konkoly Observatory has been supported by a \emph{Lend\"ulet} excellence grant by the Hungarian Academy of Sciences (MTA). This project has received funding from the European Union’s Horizon Europe research and innovation programme under the Marie Skłodowska-Curie grant agreement number 101130774. Funding for this project was also available in part through the Hungarian Ministry of Innovation and Technology NRDI Office grant OTKA NN147550.

This work was supported by the French Space Agency, the Centre National d'Études Spatiales (CNES).

\bibliography{thebibliography}

\begin{thebibliography}{122}
\expandafter\ifx\csname natexlab\endcsname\relax\def\natexlab#1{#1}\fi

\bibitem[{{Alonso} {et~al.}(2018){Alonso}, {Hill}, {Hlo{\v{z}}ek}, \& {Spergel}}]{Alonso2018}
{Alonso}, D., {Hill}, J.~C., {Hlo{\v{z}}ek}, R., \& {Spergel}, D.~N. 2018, \prd, 97, 063514

\bibitem[{{Aubert} {et~al.}(2022){Aubert}, {Cousinou}, {Escoffier}, {Hawken}, {Nadathur}, {Alam}, {Bautista}, {Burtin}, {Chuang}, {de la Macorra}, {de Mattia}, {Gil-Mar{\'\i}n}, {Hou}, {Jullo}, {Kneib}, {Neveux}, {Rossi}, {Schneider}, {Smith}, {Tamone}, {Vargas Maga{\~n}a}, \& {Zhao}}]{Aubert2022}
{Aubert}, M., {Cousinou}, M.-C., {Escoffier}, S., {et~al.} 2022, \mnras, 513, 186

\bibitem[{{Baker} {et~al.}(2018){Baker}, {Clampitt}, {Jain}, \& {Trodden}}]{Baker2018}
{Baker}, T., {Clampitt}, J., {Jain}, B., \& {Trodden}, M. 2018, \prd, 98, 023511

\bibitem[{{Banerjee} \& {Dalal}(2016)}]{Banerjee2016}
{Banerjee}, A. \& {Dalal}, N. 2016, \jcap, 2016, 015

\bibitem[{{Barreira} {et~al.}(2015){Barreira}, {Cautun}, {Li}, {Baugh}, \& {Pascoli}}]{Barreira2015}
{Barreira}, A., {Cautun}, M., {Li}, B., {Baugh}, C.~M., \& {Pascoli}, S. 2015, \jcap, 2015, 028

\bibitem[{{Bennett} {et~al.}(2013){Bennett}, {Larson}, {Weiland}, {Jarosik}, {Hinshaw}, {Odegard}, {Smith}, {Hill}, {Gold}, {Halpern}, {Komatsu}, {Nolta}, {Page}, {Spergel}, {Wollack}, {Dunkley}, {Kogut}, {Limon}, {Meyer}, {Tucker}, \& {Wright}}]{Bennet2013wmap}
{Bennett}, C.~L., {Larson}, D., {Weiland}, J.~L., {et~al.} 2013, \apjs, 208, 20

\bibitem[{{Blanchard} \& {Schneider}(1987)}]{Blanchard1987}
{Blanchard}, A. \& {Schneider}, J. 1987, \aap, 184, 1

\bibitem[{{Blum} {et~al.}(2016){Blum}, {Burleigh}, {Dey}, {Schlegel}, {Meisner}, {Levi}, {Myers}, {Lang}, {Moustakas}, {Patej}, {Valdes}, {Kneib}, {Huanyuan}, {Nord}, {Olsen}, {Delubac}, {Saha}, {James}, {Walker}, \& {DECaLS Team}}]{DECALS}
{Blum}, R.~D., {Burleigh}, K., {Dey}, A., {et~al.} 2016, 228, 317.01

\bibitem[{{Bonici} {et~al.}(2023){Bonici}, {Carbone}, {Davini}, {Vielzeuf}, {Paganin}, {Cardone}, {Hamaus}, {Pisani}, {Hawken}, {Kovacs}, {Nadathur}, {Contarini}, {Verza}, {Tutusaus}, {Marulli}, {Moscardini}, {Aubert}, {Giocoli}, {Pourtsidou}, {Camera}, {Escoffier}, {Caminata}, {Di Domizio}, {Martinelli}, {Pallavicini}, {Pettorino}, {Sakr}, {Sapone}, {Testera}, {Tosi}, {Yankelevich}, {Amara}, {Auricchio}, {Baldi}, {Bonino}, {Branchini}, {Brescia}, {Brinchmann}, {Capobianco}, {Carretero}, {Castellano}, {Cavuoti}, {Cledassou}, {Congedo}, {Conversi}, {Copin}, {Corcione}, {Courbin}, {Cropper}, {Da Silva}, {Degaudenzi}, {Douspis}, {Dubath}, {Duncan}, {Dupac}, {Dusini}, {Ealet}, {Farrens}, {Ferriol}, {Fosalba}, {Frailis}, {Franceschi}, {Fumana}, {G{\'o}mez-Alvarez}, {Garilli}, {Gillis}, {Grazian}, {Grupp}, {Guzzo}, {Haugan}, {Holmes}, {Hormuth}, {Hornstrup}, {Jahnke}, {K{\"u}mmel}, {Kermiche}, {Kiessling}, {Kilbinger}, {Kunz}, {Kurki-Suonio}, {Laureijs}, {Ligori}, {Lilje}, {Lloro}, {Maiorano}, {Mansutti},
  {Marggraf}, {Markovic}, {Massey}, {Medinaceli}, {Melchior}, {Meneghetti}, {Meylan}, {Moresco}, {Munari}, {Niemi}, {Padilla}, {Paltani}, {Pasian}, {Pedersen}, {Percival}, {Pires}, {Polenta}, {Poncet}, {Popa}, {Raison}, {Rebolo}, {Renzi}, {Rhodes}, {Rossetti}, {Saglia}, {Sartoris}, {Scodeggio}, {Secroun}, {Seidel}, {Sirignano}, {Sirri}, {Stanco}, {Starck}, {Surace}, {Tallada-Cresp{\'\i}}, {Tavagnacco}, {Taylor}, {Tereno}, {Toledo-Moreo}, {Torradeflot}, {Valentijn}, {Valenziano}, {Wang}, {Weller}, {Zamorani}, {Zoubian}, \& {Andreon}}]{Bonici2023}
{Bonici}, M., {Carbone}, C., {Davini}, S., {et~al.} 2023, \aap, 670, A47

\bibitem[{Boschetti {et~al.}(2023)Boschetti, Vielzeuf, Cousinou, Escoffier, \& Jullo}]{boschetti2023}
Boschetti, R., Vielzeuf, P., Cousinou, M.-C., Escoffier, S., \& Jullo, E. 2023 [\eprint[arXiv]{2311.14586}]

\bibitem[{{Cai} {et~al.}(2014){Cai}, {Neyrinck}, {Szapudi}, {Cole}, \& {Frenk}}]{Cai2014}
{Cai}, Y.-C., {Neyrinck}, M.~C., {Szapudi}, I., {Cole}, S., \& {Frenk}, C.~S. 2014, \apj, 786, 110

\bibitem[{{Cai} {et~al.}(2015){Cai}, {Padilla}, \& {Li}}]{Cai2015}
{Cai}, Y.-C., {Padilla}, N., \& {Li}, B. 2015, \mnras, 451, 1036

\bibitem[{{Camacho-Ciurana} {et~al.}(2023){Camacho-Ciurana}, {Lee}, {Arsenov}, {Kov{\'a}cs}, {Szapudi}, \& {Csabai}}]{Camacho-Ciurana2023}
{Camacho-Ciurana}, G., {Lee}, P., {Arsenov}, N., {et~al.} 2023, arXiv e-prints, arXiv:2312.08483

\bibitem[{{Cautun} {et~al.}(2018){Cautun}, {Paillas}, {Cai}, {Bose}, {Armijo}, {Li}, \& {Padilla}}]{Cautun2018}
{Cautun}, M., {Paillas}, E., {Cai}, Y.-C., {et~al.} 2018, \mnras, 476, 3195

\bibitem[{{Chan} {et~al.}(2019){Chan}, {Hamaus}, \& {Biagetti}}]{Chan2019}
{Chan}, K.~C., {Hamaus}, N., \& {Biagetti}, M. 2019, \prd, 99, 121304

\bibitem[{{Chan} {et~al.}(2014){Chan}, {Hamaus}, \& {Desjacques}}]{Chan2014}
{Chan}, K.~C., {Hamaus}, N., \& {Desjacques}, V. 2014, \prd, 90, 103521

\bibitem[{Chen {et~al.}(2024)Chen, DeRose, Zhou, White, Ferraro, Blake, Lange, Wechsler, Aguilar, Ahlen, Brooks, Claybaugh, Dawson, de~la Macorra, Doel, Font-Ribera, Gaztañaga, Gontcho, Gutierrez, Honscheid, Howlett, Kehoe, Kirkby, Kisner, Kremin, Landriau, Guillou, Manera, Meisner, Miquel, Newman, Niz, Palanque-Delabrouille, Percival, Prada, Rossi, Sanchez, Schlegel, Schubnell, Sprayberry, Tarlé, \& Weaver}]{chen2024}
Chen, S., DeRose, J., Zhou, R., {et~al.} 2024, Not all lensing is low: An analysis of DESI$\times$DES using the Lagrangian Effective Theory of LSS

\bibitem[{{Clampitt} {et~al.}(2013){Clampitt}, {Cai}, \& {Li}}]{Clampitt2013}
{Clampitt}, J., {Cai}, Y.-C., \& {Li}, B. 2013, \mnras, 431, 749

\bibitem[{{Clampitt} {et~al.}(2016){Clampitt}, {Jain}, \& {S{\'a}nchez}}]{Clampitt2016}
{Clampitt}, J., {Jain}, B., \& {S{\'a}nchez}, C. 2016, \mnras, 456, 4425

\bibitem[{{Colberg} {et~al.}(2008){Colberg}, {Pearce}, {Foster}, {Platen}, {Brunino}, {Neyrinck}, {Basilakos}, {Fairall}, {Feldman}, {Gottl{\"o}ber}, {Hahn}, {Hoyle}, {M{\"u}ller}, {Nelson}, {Plionis}, {Porciani}, {Shandarin}, {Vogeley}, \& {van de Weygaert}}]{Colberg2008}
{Colberg}, J.~M., {Pearce}, F., {Foster}, C., {et~al.} 2008, \mnras, 387, 933

\bibitem[{{Cole} \& {Efstathiou}(1989)}]{Cole1989}
{Cole}, S. \& {Efstathiou}, G. 1989, \mnras, 239, 195

\bibitem[{{Contarini} {et~al.}(2023){Contarini}, {Pisani}, {Hamaus}, {Marulli}, {Moscardini}, \& {Baldi}}]{Contarini2023}
{Contarini}, S., {Pisani}, A., {Hamaus}, N., {et~al.} 2023, \apj, 953, 46

\bibitem[{{Contarini} {et~al.}(2024){Contarini}, {Pisani}, {Hamaus}, {Marulli}, {Moscardini}, \& {Baldi}}]{Contarini2024}
{Contarini}, S., {Pisani}, A., {Hamaus}, N., {et~al.} 2024, \aap, 682, A20

\bibitem[{{Contarini} {et~al.}(2019){Contarini}, {Ronconi}, {Marulli}, {Moscardini}, {Veropalumbo}, \& {Baldi}}]{Contarini2019}
{Contarini}, S., {Ronconi}, T., {Marulli}, F., {et~al.} 2019, \mnras, 488, 3526

\bibitem[{{Correa} \& {Paz}(2022)}]{Correa2022c}
{Correa}, C.~M. \& {Paz}, D.~J. 2022, Boletin de la Asociacion Argentina de Astronomia La Plata Argentina, 63, 193

\bibitem[{{Correa} {et~al.}(2022{\natexlab{a}}){Correa}, {Paz}, {Padilla}, {S{\'a}nchez}, {Ruiz}, \& {Angulo}}]{Correa2022a}
{Correa}, C.~M., {Paz}, D.~J., {Padilla}, N.~D., {et~al.} 2022{\natexlab{a}}, \mnras, 509, 1871

\bibitem[{{Correa} {et~al.}(2022{\natexlab{b}}){Correa}, {Paz}, {Padilla}, {S{\'a}nchez}, {Ruiz}, \& {Angulo}}]{Correa2022b}
{Correa}, C.~M., {Paz}, D.~J., {Padilla}, N.~D., {et~al.} 2022{\natexlab{b}}, \mnras, 509, 1871

\bibitem[{Crocce {et~al.}(2006)Crocce, Pueblas, \& Scoccimarro}]{Crocce2005}
Crocce, M., Pueblas, S., \& Scoccimarro, R. 2006, Mon. Not. Roy. Astron. Soc., 373, 369

\bibitem[{{Dawson} {et~al.}(2016){Dawson}, {Kneib}, {Percival}, {Alam}, {Albareti}, {Anderson}, {Armengaud}, {Aubourg}, {Bailey}, {Bautista}, {Berlind}, {Bershady}, {Beutler}, {Bizyaev}, {Blanton}, {Blomqvist}, {Bolton}, {Bovy}, {Brandt}, {Brinkmann}, {Brownstein}, {Burtin}, {Busca}, {Cai}, {Chuang}, {Clerc}, {Comparat}, {Cope}, {Croft}, {Cruz-Gonzalez}, {da Costa}, {Cousinou}, {Darling}, {de la Macorra}, {de la Torre}, {Delubac}, {du Mas des Bourboux}, {Dwelly}, {Ealet}, {Eisenstein}, {Eracleous}, {Escoffier}, {Fan}, {Finoguenov}, {Font-Ribera}, {Frinchaboy}, {Gaulme}, {Georgakakis}, {Green}, {Guo}, {Guy}, {Ho}, {Holder}, {Huehnerhoff}, {Hutchinson}, {Jing}, {Jullo}, {Kamble}, {Kinemuchi}, {Kirkby}, {Kitaura}, {Klaene}, {Laher}, {Lang}, {Laurent}, {Le Goff}, {Li}, {Liang}, {Lima}, {Lin}, {Lin}, {Lin}, {Long}, {Lundgren}, {MacDonald}, {Geimba Maia}, {Malanushenko}, {Malanushenko}, {Mariappan}, {McBride}, {McGreer}, {M{\'e}nard}, {Merloni}, {Meza}, {Montero-Dorta}, {Muna}, {Myers}, {Nandra}, {Naugle},
  {Newman}, {Noterdaeme}, {Nugent}, {Ogando}, {Olmstead}, {Oravetz}, {Oravetz}, {Padmanabhan}, {Palanque-Delabrouille}, {Pan}, {Parejko}, {P{\^a}ris}, {Peacock}, {Petitjean}, {Pieri}, {Pisani}, {Prada}, {Prakash}, {Raichoor}, {Reid}, {Rich}, {Ridl}, {Rodriguez-Torres}, {Carnero Rosell}, {Ross}, {Rossi}, {Ruan}, {Salvato}, {Sayres}, {Schneider}, {Schlegel}, {Seljak}, {Seo}, {Sesar}, {Shandera}, {Shu}, {Slosar}, {Sobreira}, {Streblyanska}, {Suzuki}, {Taylor}, {Tao}, {Tinker}, {Tojeiro}, {Vargas-Maga{\~n}a}, {Wang}, {Weaver}, {Weinberg}, {White}, {Wood-Vasey}, {Yeche}, {Zhai}, {Zhao}, {Zhao}, {Zheng}, {Ben Zhu}, \& {Zou}}]{eBOSS2016}
{Dawson}, K.~S., {Kneib}, J.-P., {Percival}, W.~J., {et~al.} 2016, \aj, 151, 44

\bibitem[{{Dawson} {et~al.}(2013){Dawson}, {Schlegel}, {Ahn}, {Anderson}, {Aubourg}, {Bailey}, {Barkhouser}, {Bautista}, {Beifiori}, {Berlind}, {Bhardwaj}, {Bizyaev}, {Blake}, {Blanton}, {Blomqvist}, {Bolton}, {Borde}, {Bovy}, {Brandt}, {Brewington}, {Brinkmann}, {Brown}, {Brownstein}, {Bundy}, {Busca}, {Carithers}, {Carnero}, {Carr}, {Chen}, {Comparat}, {Connolly}, {Cope}, {Croft}, {Cuesta}, {da Costa}, {Davenport}, {Delubac}, {de Putter}, {Dhital}, {Ealet}, {Ebelke}, {Eisenstein}, {Escoffier}, {Fan}, {Filiz Ak}, {Finley}, {Font-Ribera}, {G{\'e}nova-Santos}, {Gunn}, {Guo}, {Haggard}, {Hall}, {Hamilton}, {Harris}, {Harris}, {Ho}, {Hogg}, {Holder}, {Honscheid}, {Huehnerhoff}, {Jordan}, {Jordan}, {Kauffmann}, {Kazin}, {Kirkby}, {Klaene}, {Kneib}, {Le Goff}, {Lee}, {Long}, {Loomis}, {Lundgren}, {Lupton}, {Maia}, {Makler}, {Malanushenko}, {Malanushenko}, {Mandelbaum}, {Manera}, {Maraston}, {Margala}, {Masters}, {McBride}, {McDonald}, {McGreer}, {McMahon}, {Mena}, {Miralda-Escud{\'e}}, {Montero-Dorta},
  {Montesano}, {Muna}, {Myers}, {Naugle}, {Nichol}, {Noterdaeme}, {Nuza}, {Olmstead}, {Oravetz}, {Oravetz}, {Owen}, {Padmanabhan}, {Palanque-Delabrouille}, {Pan}, {Parejko}, {P{\^a}ris}, {Percival}, {P{\'e}rez-Fournon}, {P{\'e}rez-R{\`a}fols}, {Petitjean}, {Pfaffenberger}, {Pforr}, {Pieri}, {Prada}, {Price-Whelan}, {Raddick}, {Rebolo}, {Rich}, {Richards}, {Rockosi}, {Roe}, {Ross}, {Ross}, {Rossi}, {Rubi{\~n}o-Martin}, {Samushia}, {S{\'a}nchez}, {Sayres}, {Schmidt}, {Schneider}, {Sc{\'o}ccola}, {Seo}, {Shelden}, {Sheldon}, {Shen}, {Shu}, {Slosar}, {Smee}, {Snedden}, {Stauffer}, {Steele}, {Strauss}, {Streblyanska}, {Suzuki}, {Swanson}, {Tal}, {Tanaka}, {Thomas}, {Tinker}, {Tojeiro}, {Tremonti}, {Vargas Maga{\~n}a}, {Verde}, {Viel}, {Wake}, {Watson}, {Weaver}, {Weinberg}, {Weiner}, {West}, {White}, {Wood-Vasey}, {Yeche}, {Zehavi}, {Zhao}, \& {Zheng}}]{BOSS}
{Dawson}, K.~S., {Schlegel}, D.~J., {Ahn}, C.~P., {et~al.} 2013, \aj, 145, 10

\bibitem[{{Demirbozan} {et~al.}(2024){Demirbozan}, {Nadathur}, {Ferrero}, {Fosalba}, {Kovacs}, {Miquel}, {Davies}, {Pandey}, {Adamow}, {Bechtol}, {Drlica-Wagner}, {Gruendl}, {Hartley}, {Pieres}, {Ross}, {Rykoff}, {Sheldon}, {Yanny}, {Abbott}, {Aguena}, {Allam}, {Alves}, {Bacon}, {Bertin}, {Bocquet}, {Brooks}, {Carnero Rosell}, {Carretero}, {Cawthon}, {da Costa}, {Elidaiana da Silva Pereira}, {De Vicente}, {Desai}, {Doel}, {Everett}, {Flaugher}, {Friedel}, {Frieman}, {Gatti}, {Gaztanaga}, {Giannini}, {Gutierrez}, {Hinton}, {Hollowood}, {James}, {Jeffrey}, {Kuehn}, {Lahav}, {Lee}, {Marshall}, {Mena-Fern{\'a}ndez}, {Mohr}, {Myles}, {Ogando}, {Plazas Malag{\'o}n}, {Roodman}, {Sanchez}, {Sevilla}, {Smith}, {Soares-Santos}, {Suchyta}, {Swanson}, {Tarle}, {Weaverdyck}, {Weller}, \& {Wiseman}}]{Demirbozan2024}
{Demirbozan}, U., {Nadathur}, S., {Ferrero}, I., {et~al.} 2024, arXiv e-prints, arXiv:2404.18278

\bibitem[{{DeRose} {et~al.}(2022){DeRose}, {Becker}, \& {Wechsler}}]{DeRose2021}
{DeRose}, J., {Becker}, M.~R., \& {Wechsler}, R.~H. 2022, \apj, 940, 13

\bibitem[{{DeRose} {et~al.}(2019){DeRose}, {Wechsler}, {Becker}, {Busha}, {Rykoff}, {MacCrann}, {Erickson}, {Evrard}, {Kravtsov}, {Gruen}, {Allam}, {Avila}, {Bridle}, {Brooks}, {Buckley-Geer}, {Carnero Rosell}, {Carrasco Kind}, {Carretero}, {Castander}, {Cawthon}, {Crocce}, {da Costa}, {Davis}, {De Vicente}, {Dietrich}, {Doel}, {Drlica-Wagner}, {Fosalba}, {Frieman}, {Garcia-Bellido}, {Gutierrez}, {Hartley}, {Hollowood}, {Hoyle}, {James}, {Krause}, {Kuehn}, {Kuropatkin}, {Lima}, {Maia}, {Menanteau}, {Miller}, {Miquel}, {Ogando}, {Plazas Malag{\'o}n}, {Romer}, {Sanchez}, {Schindler}, {Serrano}, {Sevilla-Noarbe}, {Smith}, {Suchyta}, {Swanson}, {Tarle}, \& {Vikram}}]{buzzard}
{DeRose}, J., {Wechsler}, R.~H., {Becker}, M.~R., {et~al.} 2019, arXiv e-prints, arXiv:1901.02401

\bibitem[{{DeRose}(2024)}]{DeRose2024}
{DeRose}, J. e.~a. 2024, in prep

\bibitem[{{DESI Collaboration} {et~al.}(2016{\natexlab{a}}){DESI Collaboration}, {Aghamousa}, {Aguilar}, {Ahlen}, {Alam}, {Allen}, {Allende Prieto}, {Annis}, {Bailey}, {Balland}, {Ballester}, {Baltay}, {Beaufore}, {Bebek}, {Beers}, {Bell}, {Bernal}, {Besuner}, {Beutler}, {Blake}, {Bleuler}, {Blomqvist}, {Blum}, {Bolton}, {Briceno}, {Brooks}, {Brownstein}, {Buckley-Geer}, {Burden}, {Burtin}, {Busca}, {Cahn}, {Cai}, {Cardiel-Sas}, {Carlberg}, {Carton}, {Casas}, {Castander}, {Cervantes-Cota}, {Claybaugh}, {Close}, {Coker}, {Cole}, {Comparat}, {Cooper}, {Cousinou}, {Crocce}, {Cuby}, {Cunningham}, {Davis}, {Dawson}, {de la Macorra}, {De Vicente}, {Delubac}, {Derwent}, {Dey}, {Dhungana}, {Ding}, {Doel}, {Duan}, {Ealet}, {Edelstein}, {Eftekharzadeh}, {Eisenstein}, {Elliott}, {Escoffier}, {Evatt}, {Fagrelius}, {Fan}, {Fanning}, {Farahi}, {Farihi}, {Favole}, {Feng}, {Fernandez}, {Findlay}, {Finkbeiner}, {Fitzpatrick}, {Flaugher}, {Flender}, {Font-Ribera}, {Forero-Romero}, {Fosalba}, {Frenk}, {Fumagalli}, {Gaensicke},
  {Gallo}, {Garcia-Bellido}, {Gaztanaga}, {Pietro Gentile Fusillo}, {Gerard}, {Gershkovich}, {Giannantonio}, {Gillet}, {Gonzalez-de-Rivera}, {Gonzalez-Perez}, {Gott}, {Graur}, {Gutierrez}, {Guy}, {Habib}, {Heetderks}, {Heetderks}, {Heitmann}, {Hellwing}, {Herrera}, {Ho}, {Holland}, {Honscheid}, {Huff}, {Hutchinson}, {Huterer}, {Hwang}, {Illa Laguna}, {Ishikawa}, {Jacobs}, {Jeffrey}, {Jelinsky}, {Jennings}, {Jiang}, {Jimenez}, {Johnson}, {Joyce}, {Jullo}, {Juneau}, {Kama}, {Karcher}, {Karkar}, {Kehoe}, {Kennamer}, {Kent}, {Kilbinger}, {Kim}, {Kirkby}, {Kisner}, {Kitanidis}, {Kneib}, {Koposov}, {Kovacs}, {Koyama}, {Kremin}, {Kron}, {Kronig}, {Kueter-Young}, {Lacey}, {Lafever}, {Lahav}, {Lambert}, {Lampton}, {Landriau}, {Lang}, {Lauer}, {Le Goff}, {Le Guillou}, {Le Van Suu}, {Lee}, {Lee}, {Leitner}, {Lesser}, {Levi}, {L'Huillier}, {Li}, {Liang}, {Lin}, {Linder}, {Loebman}, {Luki{\'c}}, {Ma}, {MacCrann}, {Magneville}, {Makarem}, {Manera}, {Manser}, {Marshall}, {Martini}, {Massey}, {Matheson}, {McCauley},
  {McDonald}, {McGreer}, {Meisner}, {Metcalfe}, {Miller}, {Miquel}, {Moustakas}, {Myers}, {Naik}, {Newman}, {Nichol}, {Nicola}, {Nicolati da Costa}, {Nie}, {Niz}, {Norberg}, {Nord}, {Norman}, {Nugent}, {O'Brien}, {Oh}, {Olsen}, {Padilla}, {Padmanabhan}, {Padmanabhan}, {Palanque-Delabrouille}, {Palmese}, {Pappalardo}, {P{\^a}ris}, {Park}, {Patej}, {Peacock}, {Peiris}, {Peng}, {Percival}, {Perruchot}, {Pieri}, {Pogge}, {Pollack}, {Poppett}, {Prada}, {Prakash}, {Probst}, {Rabinowitz}, {Raichoor}, {Ree}, {Refregier}, {Regal}, {Reid}, {Reil}, {Rezaie}, {Rockosi}, {Roe}, {Ronayette}, {Roodman}, {Ross}, {Ross}, {Rossi}, {Rozo}, {Ruhlmann-Kleider}, {Rykoff}, {Sabiu}, {Samushia}, {Sanchez}, {Sanchez}, {Schlegel}, {Schneider}, {Schubnell}, {Secroun}, {Seljak}, {Seo}, {Serrano}, {Shafieloo}, {Shan}, {Sharples}, {Sholl}, {Shourt}, {Silber}, {Silva}, {Sirk}, {Slosar}, {Smith}, {Smoot}, {Som}, {Song}, {Sprayberry}, {Staten}, {Stefanik}, {Tarle}, {Sien Tie}, {Tinker}, {Tojeiro}, {Valdes}, {Valenzuela}, {Valluri},
  {Vargas-Magana}, {Verde}, {Walker}, {Wang}, {Wang}, {Weaver}, {Weaverdyck}, {Wechsler}, {Weinberg}, {White}, {Yang}, {Yeche}, {Zhang}, {Zhao}, {Zheng}, {Zhou}, {Zhou}, {Zhu}, {Zou}, \& {Zu}}]{DESI1}
{DESI Collaboration}, {Aghamousa}, A., {Aguilar}, J., {et~al.} 2016{\natexlab{a}}, arXiv e-prints, arXiv:1611.00036

\bibitem[{{DESI Collaboration} {et~al.}(2016{\natexlab{b}}){DESI Collaboration}, {Aghamousa}, {Aguilar}, {Ahlen}, {Alam}, {Allen}, {Allende Prieto}, {Annis}, {Bailey}, {Balland}, {Ballester}, {Baltay}, {Beaufore}, {Bebek}, {Beers}, {Bell}, {Bernal}, {Besuner}, {Beutler}, {Blake}, {Bleuler}, {Blomqvist}, {Blum}, {Bolton}, {Briceno}, {Brooks}, {Brownstein}, {Buckley-Geer}, {Burden}, {Burtin}, {Busca}, {Cahn}, {Cai}, {Cardiel-Sas}, {Carlberg}, {Carton}, {Casas}, {Castander}, {Cervantes-Cota}, {Claybaugh}, {Close}, {Coker}, {Cole}, {Comparat}, {Cooper}, {Cousinou}, {Crocce}, {Cuby}, {Cunningham}, {Davis}, {Dawson}, {de la Macorra}, {De Vicente}, {Delubac}, {Derwent}, {Dey}, {Dhungana}, {Ding}, {Doel}, {Duan}, {Ealet}, {Edelstein}, {Eftekharzadeh}, {Eisenstein}, {Elliott}, {Escoffier}, {Evatt}, {Fagrelius}, {Fan}, {Fanning}, {Farahi}, {Farihi}, {Favole}, {Feng}, {Fernandez}, {Findlay}, {Finkbeiner}, {Fitzpatrick}, {Flaugher}, {Flender}, {Font-Ribera}, {Forero-Romero}, {Fosalba}, {Frenk}, {Fumagalli}, {Gaensicke},
  {Gallo}, {Garcia-Bellido}, {Gaztanaga}, {Pietro Gentile Fusillo}, {Gerard}, {Gershkovich}, {Giannantonio}, {Gillet}, {Gonzalez-de-Rivera}, {Gonzalez-Perez}, {Gott}, {Graur}, {Gutierrez}, {Guy}, {Habib}, {Heetderks}, {Heetderks}, {Heitmann}, {Hellwing}, {Herrera}, {Ho}, {Holland}, {Honscheid}, {Huff}, {Hutchinson}, {Huterer}, {Hwang}, {Illa Laguna}, {Ishikawa}, {Jacobs}, {Jeffrey}, {Jelinsky}, {Jennings}, {Jiang}, {Jimenez}, {Johnson}, {Joyce}, {Jullo}, {Juneau}, {Kama}, {Karcher}, {Karkar}, {Kehoe}, {Kennamer}, {Kent}, {Kilbinger}, {Kim}, {Kirkby}, {Kisner}, {Kitanidis}, {Kneib}, {Koposov}, {Kovacs}, {Koyama}, {Kremin}, {Kron}, {Kronig}, {Kueter-Young}, {Lacey}, {Lafever}, {Lahav}, {Lambert}, {Lampton}, {Landriau}, {Lang}, {Lauer}, {Le Goff}, {Le Guillou}, {Le Van Suu}, {Lee}, {Lee}, {Leitner}, {Lesser}, {Levi}, {L'Huillier}, {Li}, {Liang}, {Lin}, {Linder}, {Loebman}, {Luki{\'c}}, {Ma}, {MacCrann}, {Magneville}, {Makarem}, {Manera}, {Manser}, {Marshall}, {Martini}, {Massey}, {Matheson}, {McCauley},
  {McDonald}, {McGreer}, {Meisner}, {Metcalfe}, {Miller}, {Miquel}, {Moustakas}, {Myers}, {Naik}, {Newman}, {Nichol}, {Nicola}, {Nicolati da Costa}, {Nie}, {Niz}, {Norberg}, {Nord}, {Norman}, {Nugent}, {O'Brien}, {Oh}, {Olsen}, {Padilla}, {Padmanabhan}, {Padmanabhan}, {Palanque-Delabrouille}, {Palmese}, {Pappalardo}, {P{\^a}ris}, {Park}, {Patej}, {Peacock}, {Peiris}, {Peng}, {Percival}, {Perruchot}, {Pieri}, {Pogge}, {Pollack}, {Poppett}, {Prada}, {Prakash}, {Probst}, {Rabinowitz}, {Raichoor}, {Ree}, {Refregier}, {Regal}, {Reid}, {Reil}, {Rezaie}, {Rockosi}, {Roe}, {Ronayette}, {Roodman}, {Ross}, {Ross}, {Rossi}, {Rozo}, {Ruhlmann-Kleider}, {Rykoff}, {Sabiu}, {Samushia}, {Sanchez}, {Sanchez}, {Schlegel}, {Schneider}, {Schubnell}, {Secroun}, {Seljak}, {Seo}, {Serrano}, {Shafieloo}, {Shan}, {Sharples}, {Sholl}, {Shourt}, {Silber}, {Silva}, {Sirk}, {Slosar}, {Smith}, {Smoot}, {Som}, {Song}, {Sprayberry}, {Staten}, {Stefanik}, {Tarle}, {Sien Tie}, {Tinker}, {Tojeiro}, {Valdes}, {Valenzuela}, {Valluri},
  {Vargas-Magana}, {Verde}, {Walker}, {Wang}, {Wang}, {Weaver}, {Weaverdyck}, {Wechsler}, {Weinberg}, {White}, {Yang}, {Yeche}, {Zhang}, {Zhao}, {Zheng}, {Zhou}, {Zhou}, {Zhu}, {Zou}, \& {Zu}}]{DESI2}
{DESI Collaboration}, {Aghamousa}, A., {Aguilar}, J., {et~al.} 2016{\natexlab{b}}, arXiv e-prints, arXiv:1611.00037

\bibitem[{{Dey} {et~al.}(2019){Dey}, {Schlegel}, {Lang}, {Blum}, {Burleigh}, {Fan}, {Findlay}, {Finkbeiner}, {Herrera}, {Juneau}, {Landriau}, {Levi}, {McGreer}, {Meisner}, {Myers}, {Moustakas}, {Nugent}, {Patej}, {Schlafly}, {Walker}, {Valdes}, {Weaver}, {Y{\`e}che}, {Zou}, {Zhou}, {Abareshi}, {Abbott}, {Abolfathi}, {Aguilera}, {Alam}, {Allen}, {Alvarez}, {Annis}, {Ansarinejad}, {Aubert}, {Beechert}, {Bell}, {BenZvi}, {Beutler}, {Bielby}, {Bolton}, {Brice{\~n}o}, {Buckley-Geer}, {Butler}, {Calamida}, {Carlberg}, {Carter}, {Casas}, {Castander}, {Choi}, {Comparat}, {Cukanovaite}, {Delubac}, {DeVries}, {Dey}, {Dhungana}, {Dickinson}, {Ding}, {Donaldson}, {Duan}, {Duckworth}, {Eftekharzadeh}, {Eisenstein}, {Etourneau}, {Fagrelius}, {Farihi}, {Fitzpatrick}, {Font-Ribera}, {Fulmer}, {G{\"a}nsicke}, {Gaztanaga}, {George}, {Gerdes}, {Gontcho}, {Gorgoni}, {Green}, {Guy}, {Harmer}, {Hernandez}, {Honscheid}, {Huang}, {James}, {Jannuzi}, {Jiang}, {Joyce}, {Karcher}, {Karkar}, {Kehoe}, {Kneib}, {Kueter-Young}, {Lan},
  {Lauer}, {Le Guillou}, {Le Van Suu}, {Lee}, {Lesser}, {Perreault Levasseur}, {Li}, {Mann}, {Marshall}, {Mart{\'\i}nez-V{\'a}zquez}, {Martini}, {du Mas des Bourboux}, {McManus}, {Meier}, {M{\'e}nard}, {Metcalfe}, {Mu{\~n}oz-Guti{\'e}rrez}, {Najita}, {Napier}, {Narayan}, {Newman}, {Nie}, {Nord}, {Norman}, {Olsen}, {Paat}, {Palanque-Delabrouille}, {Peng}, {Poppett}, {Poremba}, {Prakash}, {Rabinowitz}, {Raichoor}, {Rezaie}, {Robertson}, {Roe}, {Ross}, {Ross}, {Rudnick}, {Safonova}, {Saha}, {S{\'a}nchez}, {Savary}, {Schweiker}, {Scott}, {Seo}, {Shan}, {Silva}, {Slepian}, {Soto}, {Sprayberry}, {Staten}, {Stillman}, {Stupak}, {Summers}, {Sien Tie}, {Tirado}, {Vargas-Maga{\~n}a}, {Vivas}, {Wechsler}, {Williams}, {Yang}, {Yang}, {Yapici}, {Zaritsky}, {Zenteno}, {Zhang}, {Zhang}, {Zhou}, \& {Zhou}}]{Legacy}
{Dey}, A., {Schlegel}, D.~J., {Lang}, D., {et~al.} 2019, \aj, 157, 168

\bibitem[{{Eisenstein} {et~al.}(2005){Eisenstein}, {Zehavi}, {Hogg}, {Scoccimarro}, {Blanton}, {Nichol}, {Scranton}, {Seo}, {Tegmark}, {Zheng}, {Anderson}, {Annis}, {Bahcall}, {Brinkmann}, {Burles}, {Castander}, {Connolly}, {Csabai}, {Doi}, {Fukugita}, {Frieman}, {Glazebrook}, {Gunn}, {Hendry}, {Hennessy}, {Ivezi{\'c}}, {Kent}, {Knapp}, {Lin}, {Loh}, {Lupton}, {Margon}, {McKay}, {Meiksin}, {Munn}, {Pope}, {Richmond}, {Schlegel}, {Schneider}, {Shimasaku}, {Stoughton}, {Strauss}, {SubbaRao}, {Szalay}, {Szapudi}, {Tucker}, {Yanny}, \& {York}}]{Eisenstein2005bao}
{Eisenstein}, D.~J., {Zehavi}, I., {Hogg}, D.~W., {et~al.} 2005, \apj, 633, 560

\bibitem[{{Flaugher} {et~al.}(2015){Flaugher}, {Diehl}, {Honscheid}, {Abbott}, {Alvarez}, {Angstadt}, {Annis}, {Antonik}, {Ballester}, {Beaufore}, {Bernstein}, {Bernstein}, {Bigelow}, {Bonati}, {Boprie}, {Brooks}, {Buckley-Geer}, {Campa}, {Cardiel-Sas}, {Castander}, {Castilla}, {Cease}, {Cela-Ruiz}, {Chappa}, {Chi}, {Cooper}, {da Costa}, {Dede}, {Derylo}, {DePoy}, {de Vicente}, {Doel}, {Drlica-Wagner}, {Eiting}, {Elliott}, {Emes}, {Estrada}, {Fausti Neto}, {Finley}, {Flores}, {Frieman}, {Gerdes}, {Gladders}, {Gregory}, {Gutierrez}, {Hao}, {Holland}, {Holm}, {Huffman}, {Jackson}, {James}, {Jonas}, {Karcher}, {Karliner}, {Kent}, {Kessler}, {Kozlovsky}, {Kron}, {Kubik}, {Kuehn}, {Kuhlmann}, {Kuk}, {Lahav}, {Lathrop}, {Lee}, {Levi}, {Lewis}, {Li}, {Mandrichenko}, {Marshall}, {Martinez}, {Merritt}, {Miquel}, {Mu{\~n}oz}, {Neilsen}, {Nichol}, {Nord}, {Ogando}, {Olsen}, {Palaio}, {Patton}, {Peoples}, {Plazas}, {Rauch}, {Reil}, {Rheault}, {Roe}, {Rogers}, {Roodman}, {Sanchez}, {Scarpine}, {Schindler}, {Schmidt},
  {Schmitt}, {Schubnell}, {Schultz}, {Schurter}, {Scott}, {Serrano}, {Shaw}, {Smith}, {Soares-Santos}, {Stefanik}, {Stuermer}, {Suchyta}, {Sypniewski}, {Tarle}, {Thaler}, {Tighe}, {Tran}, {Tucker}, {Walker}, {Wang}, {Watson}, {Weaverdyck}, {Wester}, {Woods}, {Yanny}, \& {DES Collaboration}}]{Flaugher2015}
{Flaugher}, B., {Diehl}, H.~T., {Honscheid}, K., {et~al.} 2015, \aj, 150, 150

\bibitem[{{Hamaus} {et~al.}(2022){Hamaus}, {Aubert}, {Pisani}, {Contarini}, {Verza}, {Cousinou}, {Escoffier}, {Hawken}, {Lavaux}, {Pollina}, {Wandelt}, {Weller}, {Bonici}, {Carbone}, {Guzzo}, {Kovacs}, {Marulli}, {Massara}, {Moscardini}, {Ntelis}, {Percival}, {Radinovi{\'c}}, {Sahl{\'e}n}, {Sakr}, {S{\'a}nchez}, {Winther}, {Auricchio}, {Awan}, {Bender}, {Bodendorf}, {Bonino}, {Branchini}, {Brescia}, {Brinchmann}, {Capobianco}, {Carretero}, {Castander}, {Castellano}, {Cavuoti}, {Cimatti}, {Cledassou}, {Congedo}, {Conversi}, {Copin}, {Corcione}, {Cropper}, {Da Silva}, {Degaudenzi}, {Douspis}, {Dubath}, {Duncan}, {Dupac}, {Dusini}, {Ealet}, {Ferriol}, {Fosalba}, {Frailis}, {Franceschi}, {Franzetti}, {Fumana}, {Garilli}, {Gillis}, {Giocoli}, {Grazian}, {Grupp}, {Haugan}, {Holmes}, {Hormuth}, {Jahnke}, {Kermiche}, {Kiessling}, {Kilbinger}, {Kitching}, {K{\"u}mmel}, {Kunz}, {Kurki-Suonio}, {Ligori}, {Lilje}, {Lloro}, {Maiorano}, {Marggraf}, {Markovic}, {Massey}, {Maurogordato}, {Melchior}, {Meneghetti}, {Meylan},
  {Moresco}, {Munari}, {Niemi}, {Padilla}, {Paltani}, {Pasian}, {Pedersen}, {Pettorino}, {Pires}, {Poncet}, {Popa}, {Pozzetti}, {Rebolo}, {Rhodes}, {Rix}, {Roncarelli}, {Rossetti}, {Saglia}, {Schneider}, {Secroun}, {Seidel}, {Serrano}, {Sirignano}, {Sirri}, {Starck}, {Tallada-Cresp{\'\i}}, {Tavagnacco}, {Taylor}, {Tereno}, {Toledo-Moreo}, {Torradeflot}, {Valentijn}, {Valenziano}, {Wang}, {Welikala}, {Zamorani}, {Zoubian}, {Andreon}, {Baldi}, {Camera}, {Mei}, {Neissner}, \& {Romelli}}]{Hamaus2022}
{Hamaus}, N., {Aubert}, M., {Pisani}, A., {et~al.} 2022, \aap, 658, A20

\bibitem[{{Hamaus} {et~al.}(2014{\natexlab{a}}){Hamaus}, {Sutter}, {Lavaux}, \& {Wandelt}}]{Hamaus2014b}
{Hamaus}, N., {Sutter}, P.~M., {Lavaux}, G., \& {Wandelt}, B.~D. 2014{\natexlab{a}}, \jcap, 2014, 013

\bibitem[{{Hamaus} {et~al.}(2015){Hamaus}, {Sutter}, {Lavaux}, \& {Wandelt}}]{Hamaus2015}
{Hamaus}, N., {Sutter}, P.~M., {Lavaux}, G., \& {Wandelt}, B.~D. 2015, \jcap, 2015, 036

\bibitem[{{Hamaus} {et~al.}(2016){Hamaus}, {Sutter}, \& {Wandelt}}]{Hamaus2016}
{Hamaus}, N., {Sutter}, P.~M., \& {Wandelt}, B.~D. 2016, in The Zeldovich Universe: Genesis and Growth of the Cosmic Web, ed. R.~{van de Weygaert}, S.~{Shandarin}, E.~{Saar}, \& J.~{Einasto}, Vol. 308, 538--541

\bibitem[{{Hamaus} {et~al.}(2014{\natexlab{b}}){Hamaus}, {Wandelt}, {Sutter}, {Lavaux}, \& {Warren}}]{Hamaus2014a}
{Hamaus}, N., {Wandelt}, B.~D., {Sutter}, P.~M., {Lavaux}, G., \& {Warren}, M.~S. 2014{\natexlab{b}}, \prl, 112, 041304

\bibitem[{{Hang} {et~al.}(2021){Hang}, {Alam}, {Cai}, \& {Peacock}}]{Hang2021}
{Hang}, Q., {Alam}, S., {Cai}, Y.-C., \& {Peacock}, J.~A. 2021, \mnras, 507, 510

\bibitem[{{Hartlap} {et~al.}(2007){Hartlap}, {Simon}, \& {Schneider}}]{Hartlap2007}
{Hartlap}, J., {Simon}, P., \& {Schneider}, P. 2007, \aap, 464, 399

\bibitem[{{Hossen} {et~al.}(2022){Hossen}, {Ema}, {Bolejko}, \& {Lewis}}]{Hossen2022}
{Hossen}, M.~R., {Ema}, S.~A., {Bolejko}, K., \& {Lewis}, G.~F. 2022, \mnras, 513, 5575

\bibitem[{{Hu} \& {Okamoto}(2002)}]{HuOkamoto2002}
{Hu}, W. \& {Okamoto}, T. 2002, \apj, 574, 566

\bibitem[{{Ili{\'c}} {et~al.}(2013){Ili{\'c}}, {Langer}, \& {Douspis}}]{Ilic2013}
{Ili{\'c}}, S., {Langer}, M., \& {Douspis}, M. 2013, \aap, 556, A51

\bibitem[{{Jennings} {et~al.}(2013){Jennings}, {Li}, \& {Hu}}]{Jennings2013}
{Jennings}, E., {Li}, Y., \& {Hu}, W. 2013, \mnras, 434, 2167

\bibitem[{{Kashlinsky}(1988)}]{Kashlinsky1988}
{Kashlinsky}, A. 1988, \apjl, 331, L1

\bibitem[{{Komatsu} {et~al.}(2011){Komatsu}, {Smith}, {Dunkley}, {Bennett}, {Gold}, {Hinshaw}, {Jarosik}, {Larson}, {Nolta}, {Page}, {Spergel}, {Halpern}, {Hill}, {Kogut}, {Limon}, {Meyer}, {Odegard}, {Tucker}, {Weiland}, {Wollack}, \& {Wright}}]{Komatsu2011wmap}
{Komatsu}, E., {Smith}, K.~M., {Dunkley}, J., {et~al.} 2011, \apjs, 192, 18

\bibitem[{{Kov{\'a}cs}(2018)}]{Kovacs2018}
{Kov{\'a}cs}, A. 2018, \mnras, 475, 1777

\bibitem[{{Kov{\'a}cs} {et~al.}(2022){Kov{\'a}cs}, {Vielzeuf}, {Ferrero}, {Fosalba}, {Demirbozan}, {Miquel}, {Chang}, {Hamaus}, {Pollina}, {Bechtol}, {Becker}, {Carnero Rosell}, {Carrasco Kind}, {Cawthon}, {Crocce}, {Drlica-Wagner}, {Elvin-Poole}, {Gatti}, {Giannini}, {Gruendl}, {Porredon}, {Ross}, {Rykoff}, {Sevilla-Noarbe}, {Sheldon}, {Yanny}, {Abbott}, {Aguena}, {Allam}, {Annis}, {Bacon}, {Bernstein}, {Bertin}, {Bocquet}, {Brooks}, {Burke}, {Carretero}, {Castander}, {Costanzi}, {da Costa}, {Pereira}, {De Vicente}, {Desai}, {Diehl}, {Dietrich}, {Fert{\'e}}, {Flaugher}, {Frieman}, {Garcia-Bellido}, {Gazta{\~n}aga}, {Gerdes}, {Giannantonio}, {Gruen}, {Gschwend}, {Gutierrez}, {Hinton}, {Hollowood}, {Honscheid}, {Huterer}, {Kuehn}, {Lahav}, {Lima}, {March}, {Marshall}, {Melchior}, {Menanteau}, {Morgan}, {Muir}, {Ogando}, {Palmese}, {Paz-Chinchon}, {Pieres}, {Plazas Malag{\'o}n}, {Rodriguez Monroy}, {Roodman}, {Sanchez}, {Schubnell}, {Serrano}, {Smith}, {Suchyta}, {Tarle}, {Thomas}, {To}, {Varga}, {Weller}, \&
  {DES Collaboration}}]{Kovacs2022}
{Kov{\'a}cs}, A., {Vielzeuf}, P., {Ferrero}, I., {et~al.} 2022, \mnras, 515, 4417

\bibitem[{{Krause} {et~al.}(2013){Krause}, {Chang}, {Dor{\'e}}, \& {Umetsu}}]{Krause2013}
{Krause}, E., {Chang}, T.-C., {Dor{\'e}}, O., \& {Umetsu}, K. 2013, \apjl, 762, L20

\bibitem[{{Kreisch} {et~al.}(2019){Kreisch}, {Pisani}, {Carbone}, {Liu}, {Hawken}, {Massara}, {Spergel}, \& {Wandelt}}]{Kreisch2019}
{Kreisch}, C.~D., {Pisani}, A., {Carbone}, C., {et~al.} 2019, \mnras, 488, 4413

\bibitem[{{Lavaux} \& {Wandelt}(2010)}]{LavauxWandelt2010}
{Lavaux}, G. \& {Wandelt}, B.~D. 2010, \mnras, 403, 1392

\bibitem[{{Lavaux} \& {Wandelt}(2012)}]{Lavaux2012}
{Lavaux}, G. \& {Wandelt}, B.~D. 2012, \apj, 754, 109

\bibitem[{{Lee} \& {Park}(2009)}]{Lee2009}
{Lee}, J. \& {Park}, D. 2009, \apjl, 696, L10

\bibitem[{{Lewis} \& {Bridle}(2002)}]{Lewis2002}
{Lewis}, A. \& {Bridle}, S. 2002, \prd, 66, 103511

\bibitem[{{Li} {et~al.}(2024){Li}, {Ma}, {Tramonte}, \& {Li}}]{Li2024}
{Li}, G., {Ma}, Y.-Z., {Tramonte}, D., \& {Li}, G.-L. 2024, \mnras, 527, 2663

\bibitem[{{Li} {et~al.}(2020){Li}, {Ma}, \& {Nadathur}}]{Li2020}
{Li}, Y.-C., {Ma}, Y.-Z., \& {Nadathur}, S. 2020, arXiv e-prints, arXiv:2002.01689

\bibitem[{{Linder}(1990)}]{Linder1990}
{Linder}, E.~V. 1990, \mnras, 243, 353

\bibitem[{{Margon}(1999)}]{SDSS1999}
{Margon}, B. 1999, Philosophical Transactions of the Royal Society of London Series A, 357, 93

\bibitem[{{Massara} {et~al.}(2022){Massara}, {Percival}, {Dalal}, {Nadathur}, {Radinovi{\'c}}, {Winther}, \& {Woodfinden}}]{Massara2022}
{Massara}, E., {Percival}, W.~J., {Dalal}, N., {et~al.} 2022, \mnras, 517, 4458

\bibitem[{{Massara} {et~al.}(2015){Massara}, {Villaescusa-Navarro}, {Viel}, \& {Sutter}}]{Massara2015}
{Massara}, E., {Villaescusa-Navarro}, F., {Viel}, M., \& {Sutter}, P.~M. 2015, \jcap, 2015, 018

\bibitem[{{Mauland} {et~al.}(2023){Mauland}, {Elgar{\o}y}, {Mota}, \& {Winther}}]{Mauland2023}
{Mauland}, R., {Elgar{\o}y}, {\O}., {Mota}, D.~F., \& {Winther}, H.~A. 2023, \aap, 674, A185

\bibitem[{{Melchior} {et~al.}(2014){Melchior}, {Sutter}, {Sheldon}, {Krause}, \& {Wandelt}}]{Melchior2014}
{Melchior}, P., {Sutter}, P.~M., {Sheldon}, E.~S., {Krause}, E., \& {Wandelt}, B.~D. 2014, \mnras, 440, 2922

\bibitem[{Moresco {et~al.}(2022)Moresco, Amati, Amendola, Birrer, Blakeslee, Cantiello, Cimatti, Darling, Della~Valle, Fishbach, Grillo, Hamaus, Holz, Izzo, Jimenez, Lusso, Meneghetti, Piedipalumbo, Pisani, Pourtsidou, Pozzetti, Quartin, Risaliti, Rosati, \& Verde}]{Moresco2022}
Moresco, M., Amati, L., Amendola, L., {et~al.} 2022, Living Reviews in Relativity, 25

\bibitem[{{Nadathur} {et~al.}(2019){Nadathur}, {Carter}, {Percival}, {Winther}, \& {Bautista}}]{Nadathur2019b}
{Nadathur}, S., {Carter}, P.~M., {Percival}, W.~J., {Winther}, H.~A., \& {Bautista}, J.~E. 2019, Astrophysics Source Code Library, record ascl:1907.023

\bibitem[{{Nadathur} {et~al.}(2017){Nadathur}, {Hotchkiss}, \& {Crittenden}}]{nadathur2017}
{Nadathur}, S., {Hotchkiss}, S., \& {Crittenden}, R. 2017, \mnras, 467, 4067

\bibitem[{{Nadathur} {et~al.}(2012){Nadathur}, {Hotchkiss}, \& {Sarkar}}]{Nasathur2012}
{Nadathur}, S., {Hotchkiss}, S., \& {Sarkar}, S. 2012, \jcap, 2012, 042

\bibitem[{{Nadathur} \& {Percival}(2019)}]{Nadathur2019}
{Nadathur}, S. \& {Percival}, W.~J. 2019, \mnras, 483, 3472

\bibitem[{{Naidoo} {et~al.}(2024){Naidoo}, {Jaber}, {Hellwing}, \& {Bilicki}}]{Naidoo2024}
{Naidoo}, K., {Jaber}, M., {Hellwing}, W.~A., \& {Bilicki}, M. 2024, \prd, 109, 083511

\bibitem[{{Neyrinck}(2008)}]{zobov2008}
{Neyrinck}, M.~C. 2008, \mnras, 386, 2101

\bibitem[{{Paz} {et~al.}(2013){Paz}, {Lares}, {Ceccarelli}, {Padilla}, \& {Lambas}}]{Paz2013}
{Paz}, D., {Lares}, M., {Ceccarelli}, L., {Padilla}, N., \& {Lambas}, D.~G. 2013, \mnras, 436, 3480

\bibitem[{{Peebles}(1980)}]{Peebles1980}
{Peebles}, P.~J.~E. 1980, {The large-scale structure of the universe}

\bibitem[{{Pelliciari} {et~al.}(2023){Pelliciari}, {Contarini}, {Marulli}, {Moscardini}, {Giocoli}, {Lesci}, \& {Dolag}}]{Pelliciari2023}
{Pelliciari}, D., {Contarini}, S., {Marulli}, F., {et~al.} 2023, \mnras, 522, 152

\bibitem[{{Perlmutter} {et~al.}(1999){Perlmutter}, {Aldering}, {Goldhaber}, {Knop}, {Nugent}, {Castro}, {Deustua}, {Fabbro}, {Goobar}, {Groom}, {Hook}, {Kim}, {Kim}, {Lee}, {Nunes}, {Pain}, {Pennypacker}, {Quimby}, {Lidman}, {Ellis}, {Irwin}, {McMahon}, {Ruiz-Lapuente}, {Walton}, {Schaefer}, {Boyle}, {Filippenko}, {Matheson}, {Fruchter}, {Panagia}, {Newberg}, {Couch}, \& {Project}}]{Perlmutter1999}
{Perlmutter}, S., {Aldering}, G., {Goldhaber}, G., {et~al.} 1999, \apj, 517, 565

\bibitem[{Pisani {et~al.}(2019)Pisani, Massara, Spergel, Alonso, Baker, Cai, Cautun, Davies, Demchenko, Doré, Goulding, Habouzit, Hamaus, Hawken, Hirata, Ho, Jain, Kreisch, Marulli, Padilla, Pollina, Sahlén, Sheth, Somerville, Szapudi, van~de Weygaert, Villaescusa-Navarro, Wandelt, \& Wang}]{pisani2019cosmic}
Pisani, A., Massara, E., Spergel, D.~N., {et~al.} 2019, Cosmic voids: a novel probe to shed light on our Universe

\bibitem[{{Pisani} {et~al.}(2015){Pisani}, {Sutter}, {Hamaus}, {Alizadeh}, {Biswas}, {Wandelt}, \& {Hirata}}]{Pisani2015}
{Pisani}, A., {Sutter}, P.~M., {Hamaus}, N., {et~al.} 2015, \prd, 92, 083531

\bibitem[{{Planck Collaboration} {et~al.}(2020{\natexlab{a}}){Planck Collaboration}, {Aghanim}, {Akrami}, {Ashdown}, {Aumont}, {Baccigalupi}, {Ballardini}, {Banday}, {Barreiro}, {Bartolo}, {Basak}, {Battye}, {Benabed}, {Bernard}, {Bersanelli}, {Bielewicz}, {Bock}, {Bond}, {Borrill}, {Bouchet}, {Boulanger}, {Bucher}, {Burigana}, {Butler}, {Calabrese}, {Cardoso}, {Carron}, {Challinor}, {Chiang}, {Chluba}, {Colombo}, {Combet}, {Contreras}, {Crill}, {Cuttaia}, {de Bernardis}, {de Zotti}, {Delabrouille}, {Delouis}, {Di Valentino}, {Diego}, {Dor{\'e}}, {Douspis}, {Ducout}, {Dupac}, {Dusini}, {Efstathiou}, {Elsner}, {En{\ss}lin}, {Eriksen}, {Fantaye}, {Farhang}, {Fergusson}, {Fernandez-Cobos}, {Finelli}, {Forastieri}, {Frailis}, {Fraisse}, {Franceschi}, {Frolov}, {Galeotta}, {Galli}, {Ganga}, {G{\'e}nova-Santos}, {Gerbino}, {Ghosh}, {Gonz{\'a}lez-Nuevo}, {G{\'o}rski}, {Gratton}, {Gruppuso}, {Gudmundsson}, {Hamann}, {Handley}, {Hansen}, {Herranz}, {Hildebrandt}, {Hivon}, {Huang}, {Jaffe}, {Jones}, {Karakci},
  {Keih{\"a}nen}, {Keskitalo}, {Kiiveri}, {Kim}, {Kisner}, {Knox}, {Krachmalnicoff}, {Kunz}, {Kurki-Suonio}, {Lagache}, {Lamarre}, {Lasenby}, {Lattanzi}, {Lawrence}, {Le Jeune}, {Lemos}, {Lesgourgues}, {Levrier}, {Lewis}, {Liguori}, {Lilje}, {Lilley}, {Lindholm}, {L{\'o}pez-Caniego}, {Lubin}, {Ma}, {Mac{\'\i}as-P{\'e}rez}, {Maggio}, {Maino}, {Mandolesi}, {Mangilli}, {Marcos-Caballero}, {Maris}, {Martin}, {Martinelli}, {Mart{\'\i}nez-Gonz{\'a}lez}, {Matarrese}, {Mauri}, {McEwen}, {Meinhold}, {Melchiorri}, {Mennella}, {Migliaccio}, {Millea}, {Mitra}, {Miville-Desch{\^e}nes}, {Molinari}, {Montier}, {Morgante}, {Moss}, {Natoli}, {N{\o}rgaard-Nielsen}, {Pagano}, {Paoletti}, {Partridge}, {Patanchon}, {Peiris}, {Perrotta}, {Pettorino}, {Piacentini}, {Polastri}, {Polenta}, {Puget}, {Rachen}, {Reinecke}, {Remazeilles}, {Renzi}, {Rocha}, {Rosset}, {Roudier}, {Rubi{\~n}o-Mart{\'\i}n}, {Ruiz-Granados}, {Salvati}, {Sandri}, {Savelainen}, {Scott}, {Shellard}, {Sirignano}, {Sirri}, {Spencer}, {Sunyaev}, {Suur-Uski},
  {Tauber}, {Tavagnacco}, {Tenti}, {Toffolatti}, {Tomasi}, {Trombetti}, {Valenziano}, {Valiviita}, {Van Tent}, {Vibert}, {Vielva}, {Villa}, {Vittorio}, {Wandelt}, {Wehus}, {White}, {White}, {Zacchei}, \& {Zonca}}]{Planck2020}
{Planck Collaboration}, {Aghanim}, N., {Akrami}, Y., {et~al.} 2020{\natexlab{a}}, \aap, 641, A6

\bibitem[{{Planck Collaboration} {et~al.}(2020{\natexlab{b}}){Planck Collaboration}, {Aghanim}, {Akrami}, {Ashdown}, {Aumont}, {Baccigalupi}, {Ballardini}, {Banday}, {Barreiro}, {Bartolo}, {Basak}, {Benabed}, {Bernard}, {Bersanelli}, {Bielewicz}, {Bock}, {Bond}, {Borrill}, {Bouchet}, {Boulanger}, {Bucher}, {Burigana}, {Calabrese}, {Cardoso}, {Carron}, {Challinor}, {Chiang}, {Colombo}, {Combet}, {Crill}, {Cuttaia}, {de Bernardis}, {de Zotti}, {Delabrouille}, {Di Valentino}, {Diego}, {Dor{\'e}}, {Douspis}, {Ducout}, {Dupac}, {Efstathiou}, {Elsner}, {En{\ss}lin}, {Eriksen}, {Fantaye}, {Fernandez-Cobos}, {Finelli}, {Forastieri}, {Frailis}, {Fraisse}, {Franceschi}, {Frolov}, {Galeotta}, {Galli}, {Ganga}, {G{\'e}nova-Santos}, {Gerbino}, {Ghosh}, {Gonz{\'a}lez-Nuevo}, {G{\'o}rski}, {Gratton}, {Gruppuso}, {Gudmundsson}, {Hamann}, {Handley}, {Hansen}, {Herranz}, {Hivon}, {Huang}, {Jaffe}, {Jones}, {Karakci}, {Keih{\"a}nen}, {Keskitalo}, {Kiiveri}, {Kim}, {Knox}, {Krachmalnicoff}, {Kunz}, {Kurki-Suonio}, {Lagache},
  {Lamarre}, {Lasenby}, {Lattanzi}, {Lawrence}, {Le Jeune}, {Levrier}, {Lewis}, {Liguori}, {Lilje}, {Lindholm}, {L{\'o}pez-Caniego}, {Lubin}, {Ma}, {Mac{\'\i}as-P{\'e}rez}, {Maggio}, {Maino}, {Mandolesi}, {Mangilli}, {Marcos-Caballero}, {Maris}, {Martin}, {Mart{\'\i}nez-Gonz{\'a}lez}, {Matarrese}, {Mauri}, {McEwen}, {Melchiorri}, {Mennella}, {Migliaccio}, {Miville-Desch{\^e}nes}, {Molinari}, {Moneti}, {Montier}, {Morgante}, {Moss}, {Natoli}, {Pagano}, {Paoletti}, {Partridge}, {Patanchon}, {Perrotta}, {Pettorino}, {Piacentini}, {Polastri}, {Polenta}, {Puget}, {Rachen}, {Reinecke}, {Remazeilles}, {Renzi}, {Rocha}, {Rosset}, {Roudier}, {Rubi{\~n}o-Mart{\'\i}n}, {Ruiz-Granados}, {Salvati}, {Sandri}, {Savelainen}, {Scott}, {Sirignano}, {Sunyaev}, {Suur-Uski}, {Tauber}, {Tavagnacco}, {Tenti}, {Toffolatti}, {Tomasi}, {Trombetti}, {Valiviita}, {Van Tent}, {Vielva}, {Villa}, {Vittorio}, {Wandelt}, {Wehus}, {White}, {White}, {Zacchei}, \& {Zonca}}]{PlanckCMB2020}
{Planck Collaboration}, {Aghanim}, N., {Akrami}, Y., {et~al.} 2020{\natexlab{b}}, \aap, 641, A8

\bibitem[{{Platen} {et~al.}(2007){Platen}, {van de Weygaert}, \& {Jones}}]{platen2007}
{Platen}, E., {van de Weygaert}, R., \& {Jones}, B. J.~T. 2007, \mnras, 380, 551

\bibitem[{{Radinovi{\'c}} {et~al.}(2023){Radinovi{\'c}}, {Nadathur}, {Winther}, {Percival}, {Woodfinden}, {Massara}, {Paillas}, {Contarini}, {Hamaus}, {Kovacs}, {Pisani}, {Verza}, {Aubert}, {Amara}, {Auricchio}, {Baldi}, {Bonino}, {Branchini}, {Brescia}, {Camera}, {Capobianco}, {Carbone}, {Cardone}, {Carretero}, {Castellano}, {Cavuoti}, {Cimatti}, {Cledassou}, {Congedo}, {Conversi}, {Copin}, {Corcione}, {Courbin}, {Da Silva}, {Douspis}, {Dubath}, {Dupac}, {Farrens}, {Ferriol}, {Fosalba}, {Frailis}, {Franceschi}, {Fumana}, {Galeotta}, {Garilli}, {Gillard}, {Gillis}, {Giocoli}, {Grazian}, {Grupp}, {Haugan}, {Holmes}, {Hornstrup}, {Jahnke}, {K{\"u}mmel}, {Kiessling}, {Kilbinger}, {Kitching}, {Kurki-Suonio}, {Ligori}, {Lilje}, {Lloro}, {Maiorano}, {Mansutti}, {Marggraf}, {Markovic}, {Marulli}, {Massey}, {Mei}, {Melchior}, {Mellier}, {Meneghetti}, {Merlin}, {Meylan}, {Moresco}, {Moscardini}, {Niemi}, {Nightingale}, {Nutma}, {Padilla}, {Paltani}, {Pasian}, {Pedersen}, {Pettorino}, {Pires}, {Polenta}, {Poncet},
  {Popa}, {Pozzetti}, {Raison}, {Renzi}, {Rhodes}, {Riccio}, {Romelli}, {Roncarelli}, {Rosset}, {Saglia}, {Sapone}, {Sartoris}, {Schneider}, {Secroun}, {Seidel}, {Serrano}, {Sirignano}, {Sirri}, {Stanco}, {Starck}, {Surace}, {Tallada-Cresp{\'\i}}, {Tereno}, {Toledo-Moreo}, {Torradeflot}, {Tutusaus}, {Valentijn}, {Valenziano}, {Vassallo}, {Wang}, {Weller}, {Zamorani}, {Zoubian}, \& {Scottez}}]{Radinovic2023}
{Radinovi{\'c}}, S., {Nadathur}, S., {Winther}, H.~A., {et~al.} 2023, \aap, 677, A78

\bibitem[{{Raghunathan} {et~al.}(2020){Raghunathan}, {Nadathur}, {Sherwin}, \& {Whitehorn}}]{Raghunathan2020}
{Raghunathan}, S., {Nadathur}, S., {Sherwin}, B.~D., \& {Whitehorn}, N. 2020, \apj, 890, 168

\bibitem[{{Rezaei}(2020)}]{Rezaei2020}
{Rezaei}, Z. 2020, \apj, 902, 102

\bibitem[{{Riess} {et~al.}(1998){Riess}, {Filippenko}, {Challis}, {Clocchiatti}, {Diercks}, {Garnavich}, {Gilliland}, {Hogan}, {Jha}, {Kirshner}, {Leibundgut}, {Phillips}, {Reiss}, {Schmidt}, {Schommer}, {Smith}, {Spyromilio}, {Stubbs}, {Suntzeff}, \& {Tonry}}]{Riess1998}
{Riess}, A.~G., {Filippenko}, A.~V., {Challis}, P., {et~al.} 1998, \aj, 116, 1009

\bibitem[{{Ronconi} {et~al.}(2019){Ronconi}, {Contarini}, {Marulli}, {Baldi}, \& {Moscardini}}]{Ronconi2019}
{Ronconi}, T., {Contarini}, S., {Marulli}, F., {Baldi}, M., \& {Moscardini}, L. 2019, \mnras, 488, 5075

\bibitem[{{Ryden}(1995)}]{Ryden1995}
{Ryden}, B.~S. 1995, \apj, 452, 25

\bibitem[{{Sachs} \& {Wolfe}(1967)}]{ISW1967}
{Sachs}, R.~K. \& {Wolfe}, A.~M. 1967, \apj, 147, 73

\bibitem[{Sailer {et~al.}(2024)Sailer, Kim, Ferraro, Madhavacheril, White, Abril-Cabezas, Aguilar, Ahlen, Bond, Brooks, Burtin, Calabrese, Chen, Choi, Claybaugh, Dawson, de~la Macorra, DeRose, Dey, Dey, Doel, Dunkley, Embil-Villagra, Farren, Font-Ribera, Forero-Romero, Gaztañaga, Gluscevic, Gontcho, Honscheid, Howlett, Juneau, Kirkby, Kisner, Kremin, Landriau, Guillou, Levi, Manera, Meisner, Miquel, Moodley, Moustakas, Niemack, Niz, Palanque-Delabrouille, Percival, Prada, Qu, Rossi, Sanchez, Schaan, Schlafly, Schlegel, Schubnell, Sehgal, Seo, Sherwin, Sifón, Sprayberry, Staggs, Tarlé, Weaver, Yèche, Zhou, \& Zou}]{sailer2024}
Sailer, N., Kim, J., Ferraro, S., {et~al.} 2024, Cosmological constraints from the cross-correlation of DESI Luminous Red Galaxies with CMB lensing from Planck PR4 and ACT DR6

\bibitem[{{S{\'a}nchez} {et~al.}(2017){S{\'a}nchez}, {Clampitt}, {Kovacs}, {Jain}, {Garc{\'\i}a-Bellido}, {Nadathur}, {Gruen}, {Hamaus}, {Huterer}, {Vielzeuf}, {Amara}, {Bonnett}, {DeRose}, {Hartley}, {Jarvis}, {Lahav}, {Miquel}, {Rozo}, {Rykoff}, {Sheldon}, {Wechsler}, {Zuntz}, {Abbott}, {Abdalla}, {Annis}, {Benoit-L{\'e}vy}, {Bernstein}, {Bernstein}, {Bertin}, {Brooks}, {Buckley-Geer}, {Carnero Rosell}, {Carrasco Kind}, {Carretero}, {Crocce}, {Cunha}, {D'Andrea}, {da Costa}, {Desai}, {Diehl}, {Dietrich}, {Doel}, {Evrard}, {Fausti Neto}, {Flaugher}, {Fosalba}, {Frieman}, {Gaztanaga}, {Gruendl}, {Gutierrez}, {Honscheid}, {James}, {Krause}, {Kuehn}, {Lima}, {Maia}, {Marshall}, {Melchior}, {Plazas}, {Reil}, {Romer}, {Sanchez}, {Schubnell}, {Sevilla-Noarbe}, {Smith}, {Soares-Santos}, {Sobreira}, {Suchyta}, {Tarle}, {Thomas}, {Walker}, {Weller}, \& {DES Collaboration}}]{Sanchez2017}
{S{\'a}nchez}, C., {Clampitt}, J., {Kovacs}, A., {et~al.} 2017, \mnras, 465, 746

\bibitem[{{Sasaki}(1989)}]{Sasaki1989}
{Sasaki}, M. 1989, \mnras, 240, 415

\bibitem[{{Schlegel} {et~al.}(2021){Schlegel}, {Dey}, {Herrera}, {Juneau}, {Landriau}, {Lang}, {Meisner}, {Moustakas}, {Myers}, {Schlafly}, {Valdes}, {Weaver}, {Zhang}, {Zhou}, \& {DESI Legacy Imaging Surveys Team}}]{SchlegelDR9}
{Schlegel}, D., {Dey}, A., {Herrera}, D., {et~al.} 2021, 237, 235.03

\bibitem[{{Schmidt} {et~al.}(1998){Schmidt}, {Suntzeff}, {Phillips}, {Schommer}, {Clocchiatti}, {Kirshner}, {Garnavich}, {Challis}, {Leibundgut}, {Spyromilio}, {Riess}, {Filippenko}, {Hamuy}, {Smith}, {Hogan}, {Stubbs}, {Diercks}, {Reiss}, {Gilliland}, {Tonry}, {Maza}, {Dressler}, {Walsh}, \& {Ciardullo}}]{Schmidt1998}
{Schmidt}, B.~P., {Suntzeff}, N.~B., {Phillips}, M.~M., {et~al.} 1998, \apj, 507, 46

\bibitem[{{Schuster} {et~al.}(2023){Schuster}, {Hamaus}, {Dolag}, \& {Weller}}]{Schuster2023}
{Schuster}, N., {Hamaus}, N., {Dolag}, K., \& {Weller}, J. 2023, \jcap, 2023, 031

\bibitem[{{Schuster} {et~al.}(2024){Schuster}, {Hamaus}, {Dolag}, \& {Weller}}]{Schuster2024}
{Schuster}, N., {Hamaus}, N., {Dolag}, K., \& {Weller}, J. 2024, \jcap, 2024, 065

\bibitem[{{Schuster} {et~al.}(2019){Schuster}, {Hamaus}, {Pisani}, {Carbone}, {Kreisch}, {Pollina}, \& {Weller}}]{Schuster2019}
{Schuster}, N., {Hamaus}, N., {Pisani}, A., {et~al.} 2019, \jcap, 2019, 055

\bibitem[{{Sheth} \& {van de Weygaert}(2004)}]{Sheth2004}
{Sheth}, R.~K. \& {van de Weygaert}, R. 2004, \mnras, 350, 517

\bibitem[{{Silva} {et~al.}(2016){Silva}, {Blum}, {Allen}, {Dey}, {Schlegel}, {Lang}, {Moustakas}, {Meisner}, {Valdes}, {Patej}, {Myers}, {Sprayberry}, {Saha}, {Olsen}, {Safonova}, {Yang}, {Burleigh}, \& {MzLS Team}}]{MzLS}
{Silva}, D.~R., {Blum}, R.~D., {Allen}, L., {et~al.} 2016, 228, 317.02

\bibitem[{{Song} {et~al.}(2024){Song}, {Xiong}, {Gong}, {Deng}, {Chan}, {Chen}, {Guo}, {Han}, {Li}, {Li}, {Liu}, {Luo}, {Pei}, \& {Wei}}]{Song2024}
{Song}, Y., {Xiong}, Q., {Gong}, Y., {et~al.} 2024, arXiv e-prints, arXiv:2402.05492

\bibitem[{{Spolyar} {et~al.}(2013){Spolyar}, {Sahl{\'e}n}, \& {Silk}}]{Spolyar2013}
{Spolyar}, D., {Sahl{\'e}n}, M., \& {Silk}, J. 2013, \prl, 111, 241103

\bibitem[{{Springel}(2005)}]{Springel05}
{Springel}, V. 2005, \mnras, 364, 1105

\bibitem[{{Sunyaev} \& {Zeldovich}(1970)}]{SZ1970}
{Sunyaev}, R.~A. \& {Zeldovich}, Y.~B. 1970, \apss, 7, 3

\bibitem[{{Sutter} {et~al.}(2012){Sutter}, {Lavaux}, {Wandelt}, \& {Weinberg}}]{Sutter2012}
{Sutter}, P.~M., {Lavaux}, G., {Wandelt}, B.~D., \& {Weinberg}, D.~H. 2012, \apj, 761, 187

\bibitem[{{Sutter} {et~al.}(2014){Sutter}, {Pisani}, {Wandelt}, \& {Weinberg}}]{Sutter2014}
{Sutter}, P.~M., {Pisani}, A., {Wandelt}, B.~D., \& {Weinberg}, D.~H. 2014, \mnras, 443, 2983

\bibitem[{{The Dark Energy Survey Collaboration}(2005{\natexlab{a}})}]{thedarkenergysurveycollaboration2005dark}
{The Dark Energy Survey Collaboration}. 2005{\natexlab{a}}, The Dark Energy Survey

\bibitem[{{The Dark Energy Survey Collaboration}(2005{\natexlab{b}})}]{DES}
{The Dark Energy Survey Collaboration}. 2005{\natexlab{b}}, arXiv e-prints, astro

\bibitem[{{Thiele} {et~al.}(2023){Thiele}, {Massara}, {Pisani}, {Hahn}, {Spergel}, {Ho}, \& {Wandelt}}]{Thiele2023}
{Thiele}, L., {Massara}, E., {Pisani}, A., {et~al.} 2023, arXiv e-prints, arXiv:2307.07555

\bibitem[{Tomita \& Watanabe(1989)}]{Tomita1989}
Tomita, K. \& Watanabe, K. 1989, Progress of Theoretical Physics, 82, 563

\bibitem[{{Verza} {et~al.}(2024){Verza}, {Carbone}, {Pisani}, {Porciani}, \& {Matarrese}}]{Verza2024}
{Verza}, G., {Carbone}, C., {Pisani}, A., {Porciani}, C., \& {Matarrese}, S. 2024, arXiv e-prints, arXiv:2401.14451

\bibitem[{{Vielzeuf} {et~al.}(2023){Vielzeuf}, {Calabrese}, {Carbone}, {Fabbian}, \& {Baccigalupi}}]{Vielzeuf2023}
{Vielzeuf}, P., {Calabrese}, M., {Carbone}, C., {Fabbian}, G., \& {Baccigalupi}, C. 2023, \jcap, 2023, 010

\bibitem[{{Vielzeuf} {et~al.}(2021){Vielzeuf}, {Kov{\'a}cs}, {Demirbozan}, {Fosalba}, {Baxter}, {Hamaus}, {Huterer}, {Miquel}, {Nadathur}, {Pollina}, {S{\'a}nchez}, {Whiteway}, {Abbott}, {Allam}, {Annis}, {Avila}, {Brooks}, {Burke}, {Carnero Rosell}, {Carrasco Kind}, {Carretero}, {Cawthon}, {Costanzi}, {da Costa}, {De Vicente}, {Desai}, {Diehl}, {Doel}, {Eifler}, {Everett}, {Flaugher}, {Frieman}, {Garc{\'\i}a-Bellido}, {Gaztanaga}, {Gerdes}, {Gruen}, {Gruendl}, {Gschwend}, {Gutierrez}, {Hartley}, {Hollowood}, {Honscheid}, {James}, {Kuehn}, {Kuropatkin}, {Lahav}, {Lima}, {Maia}, {March}, {Marshall}, {Melchior}, {Menanteau}, {Palmese}, {Paz-Chinch{\'o}n}, {Plazas}, {Sanchez}, {Scarpine}, {Serrano}, {Sevilla-Noarbe}, {Smith}, {Suchyta}, {Tarle}, {Thomas}, {Weller}, {Zuntz}, {Zuntz}, \& {DES Collaboration}}]{Vielzeuf2021}
{Vielzeuf}, P., {Kov{\'a}cs}, A., {Demirbozan}, U., {et~al.} 2021, \mnras, 500, 464

\bibitem[{{Wechsler} {et~al.}(2022){Wechsler}, {DeRose}, {Busha}, {Becker}, {Rykoff}, \& {Evrard}}]{Wechsler2021}
{Wechsler}, R.~H., {DeRose}, J., {Busha}, M.~T., {et~al.} 2022, \apj, 931, 145

\bibitem[{{Wilson} \& {Bean}(2023)}]{Wilson2023}
{Wilson}, C. \& {Bean}, R. 2023, \prd, 107, 124008

\bibitem[{{Wright} {et~al.}(2010){Wright}, {Eisenhardt}, {Mainzer}, {Ressler}, {Cutri}, {Jarrett}, {Kirkpatrick}, {Padgett}, {McMillan}, {Skrutskie}, {Stanford}, {Cohen}, {Walker}, {Mather}, {Leisawitz}, {Gautier}, {McLean}, {Benford}, {Lonsdale}, {Blain}, {Mendez}, {Irace}, {Duval}, {Liu}, {Royer}, {Heinrichsen}, {Howard}, {Shannon}, {Kendall}, {Walsh}, {Larsen}, {Cardon}, {Schick}, {Schwalm}, {Abid}, {Fabinsky}, {Naes}, \& {Tsai}}]{WISE}
{Wright}, E.~L., {Eisenhardt}, P. R.~M., {Mainzer}, A.~K., {et~al.} 2010, \aj, 140, 1868

\bibitem[{{York} {et~al.}(2000){York}, {Adelman}, {Anderson}, {Anderson}, {Annis}, {Bahcall}, {Bakken}, {Barkhouser}, {Bastian}, {Berman}, {Boroski}, {Bracker}, {Briegel}, {Briggs}, {Brinkmann}, {Brunner}, {Burles}, {Carey}, {Carr}, {Castander}, {Chen}, {Colestock}, {Connolly}, {Crocker}, {Csabai}, {Czarapata}, {Davis}, {Doi}, {Dombeck}, {Eisenstein}, {Ellman}, {Elms}, {Evans}, {Fan}, {Federwitz}, {Fiscelli}, {Friedman}, {Frieman}, {Fukugita}, {Gillespie}, {Gunn}, {Gurbani}, {de Haas}, {Haldeman}, {Harris}, {Hayes}, {Heckman}, {Hennessy}, {Hindsley}, {Holm}, {Holmgren}, {Huang}, {Hull}, {Husby}, {Ichikawa}, {Ichikawa}, {Ivezi{\'c}}, {Kent}, {Kim}, {Kinney}, {Klaene}, {Kleinman}, {Kleinman}, {Knapp}, {Korienek}, {Kron}, {Kunszt}, {Lamb}, {Lee}, {Leger}, {Limmongkol}, {Lindenmeyer}, {Long}, {Loomis}, {Loveday}, {Lucinio}, {Lupton}, {MacKinnon}, {Mannery}, {Mantsch}, {Margon}, {McGehee}, {McKay}, {Meiksin}, {Merelli}, {Monet}, {Munn}, {Narayanan}, {Nash}, {Neilsen}, {Neswold}, {Newberg}, {Nichol}, {Nicinski},
  {Nonino}, {Okada}, {Okamura}, {Ostriker}, {Owen}, {Pauls}, {Peoples}, {Peterson}, {Petravick}, {Pier}, {Pope}, {Pordes}, {Prosapio}, {Rechenmacher}, {Quinn}, {Richards}, {Richmond}, {Rivetta}, {Rockosi}, {Ruthmansdorfer}, {Sandford}, {Schlegel}, {Schneider}, {Sekiguchi}, {Sergey}, {Shimasaku}, {Siegmund}, {Smee}, {Smith}, {Snedden}, {Stone}, {Stoughton}, {Strauss}, {Stubbs}, {SubbaRao}, {Szalay}, {Szapudi}, {Szokoly}, {Thakar}, {Tremonti}, {Tucker}, {Uomoto}, {Vanden Berk}, {Vogeley}, {Waddell}, {Wang}, {Watanabe}, {Weinberg}, {Yanny}, {Yasuda}, \& {SDSS Collaboration}}]{SDSS}
{York}, D.~G., {Adelman}, J., {Anderson}, John~E., J., {et~al.} 2000, \aj, 120, 1579

\bibitem[{{Zhou} {et~al.}(2023){Zhou}, {Dey}, {Newman}, {Eisenstein}, {Dawson}, {Bailey}, {Berti}, {Guy}, {Lan}, {Zou}, {Aguilar}, {Ahlen}, {Alam}, {Brooks}, {de la Macorra}, {Dey}, {Dhungana}, {Fanning}, {Font-Ribera}, {Gontcho}, {Honscheid}, {Ishak}, {Kisner}, {Kov{\'a}cs}, {Kremin}, {Landriau}, {Levi}, {Magneville}, {Manera}, {Martini}, {Meisner}, {Miquel}, {Moustakas}, {Myers}, {Nie}, {Palanque-Delabrouille}, {Percival}, {Poppett}, {Prada}, {Raichoor}, {Ross}, {Schlafly}, {Schlegel}, {Schubnell}, {Tarl{\'e}}, {Weaver}, {Wechsler}, {Y{\'e}che}, \& {Zhou}}]{Zhou2023}
{Zhou}, R., {Dey}, B., {Newman}, J.~A., {et~al.} 2023, \aj, 165, 58

\bibitem[{{Zivick} \& {Sutter}(2016)}]{Zivick2016}
{Zivick}, P. \& {Sutter}, P.~M. 2016, in The Zeldovich Universe: Genesis and Growth of the Cosmic Web, ed. R.~{van de Weygaert}, S.~{Shandarin}, E.~{Saar}, \& J.~{Einasto}, Vol. 308, 589--590

\bibitem[{{Zivick} {et~al.}(2015){Zivick}, {Sutter}, {Wandelt}, {Li}, \& {Lam}}]{zivick2015}
{Zivick}, P., {Sutter}, P.~M., {Wandelt}, B.~D., {Li}, B., \& {Lam}, T.~Y. 2015, \mnras, 451, 4215

\bibitem[{{Zou} {et~al.}(2017){Zou}, {Zhou}, {Fan}, {Zhang}, {Zhou}, {Nie}, {Peng}, {McGreer}, {Jiang}, {Dey}, {Fan}, {He}, {Jiang}, {Lang}, {Lesser}, {Ma}, {Mao}, {Schlegel}, \& {Wang}}]{BASS}
{Zou}, H., {Zhou}, X., {Fan}, X., {et~al.} 2017, \pasp, 129, 064101

\end{thebibliography}

\label{LastPage}
\end{document}